\pdfoutput=1
\documentclass[11pt]{article}

\usepackage[a4paper, total={7in, 9in}]{geometry}
\usepackage{setspace}
\usepackage{amsmath, amssymb, amsfonts, amsbsy, mathtools}
\usepackage{mathptmx} %
\usepackage{helvet}   %
\usepackage{courier}  %
\usepackage{type1cm}  %

\usepackage[authoryear,round]{natbib} %
\usepackage{etoolbox}

\makeatletter
\pretocmd{\NAT@citex}
  {\interlinepenalty=100} %
  {}{}
\makeatother

\usepackage{amsthm}
\newtheorem{assumption}{Assumption}[section]

\newtheorem{prop}{Proposition}[section]

\newtheorem{Thm}{Theorem}[section]
\newtheorem{Lemma}{Lemma}[section]

\usepackage{graphicx}
\usepackage{float}
\usepackage[caption=false,farskip=0pt]{subfig}
\usepackage{booktabs} %
\usepackage{colortbl} %
\usepackage[dvipsnames]{xcolor}

\usepackage{algorithm}
\usepackage[algo2e, boxed, ruled, vlined]{algorithm2e}

\usepackage{hyperref}
\usepackage{url}

\usepackage{adjustbox} %
\usepackage{moreverb}  %
\usepackage{microtype} %
\usepackage{nicefrac}  %
\usepackage{lmodern}   %

\usepackage{multicol}

\usepackage{amsmath}
\DeclareMathOperator*{\argmax}{arg\,max}
\DeclareMathOperator*{\argmin}{arg\,min}

\usepackage[normalem]{ulem}
\usepackage{paralist}

\providecommand{\bo}{\mathbf}

\newcommand{\pr}{^\prime}

\newcommand{\n}{^{(n)}}

\newcommand\bigfrown[2][\textstyle]{\ensuremath{%
  \array[b]{c}\text{\resizebox{14ex}{.7ex}{$#1\frown$}}\vspace{-2mm}\\[-1.3ex]#1#2\endarray}}

  \newcommand\biggfrown[2][\textstyle]{\ensuremath{%
  \array[b]{c}\text{\resizebox{16ex}{.7ex}{$#1\frown$}}\vspace{-2mm}\\[-1.3ex]#1#2\endarray}}

\newcommand\unaAdded[1]{{\color{black}{{#1}}}}

\numberwithin{equation}{section}

\title{\sc Directional Nonlinear Principal Components\\ a measure transportation approach}

\author{
Marc Hallin\footnote{Research supported by the Fonds Thelam for Mathematics of the King Baudouin Foundation and the Czech Science Foundation grant  GA24-10078S.}\\
\small Universit\'e libre de Bruxelles, Belgium and Czech Academy of Sciences, Czech Republic\\
\texttt{mhallin@ulb.ac.be}
\and
Klaus Nordhausen\footnote{Research supported by  the Research Council of Finland grant 363261.}\\
\small University of Helsinki, Finland\\
\texttt{klaus.nordhausen@helsinki.fi}
\and
Una Radojicic\footnote{Research supported by the Austrian Science Fund (FWF), grants [10.55776/I5799, 10.55776/PAT1440125].}\\
\small Vienna University of Technology, Austria\\
\texttt{una.radojicic@tuwien.ac.at}
}

\date{\today}

\begin{document}
\maketitle

\begin{abstract}
One of the earliest techniques in multivariate analysis, Principal Component Analysis (PCA) remains among the most popular ones in the area. PCA, however,  is a  {\it linear} and {\it bidirectional} technique (principal directions $\pm{\textbf u}$ are defined up to a sign), which is fine under Gaussian or elliptical assumptions but constitutes a major limitation outside that restrictive context. This has prompted the development of numerous nonlinear approaches.  {We show} here how recent results in measure transportation allow for a nonlinear and strictly directional (disentangling $\textbf u$ and $-{\textbf u}$) concept preserving the intuitive interpretation of the classical linear concept by maximizing expected squared arc lengths along data-driven principal curves rather than principal directions.
\end{abstract}


\section{Introduction}
One of the earliest techniques in multivariate analysis---it can be traced back to  {\cite{Pearson1901}} and  {\cite{Hotelling1933}}---Principal Component Analysis (PCA) probably remains the most popular and widespread tool in the area, with countless applications in all disciplines, ranging from biometrics or psychometrics to econometrics and quantitative finance. A major motivation for PCA is, of course, dimension reduction; but it also serves as an instrument in a variety of other data-analytical methods such as factor models, cluster and discriminant analysis, principal component regression, noise reduction, etc.

Principal components are defined, and PCA is usually carried out via the spectral decomposition of covariance matrices, under the assumption of finite second-order moments.  They achieve an important  {Karhunen--Loève} optimality property  {\citep{Gerbrands1981, Hallin2018}}, which provides a theoretical justification for their popularity.
We refer to the classical monograph by   {\cite{Joliffe2002}} for a thorough discussion of the subject.
PCA, however,  remains a  {\it linear} and {\it bidirectional} technique. This two-sidedness feature, which is fine under Gaussian or elliptical assumptions, constitutes a major limitation outside that restrictive context; let us examine it in more detail.

\subsection{Linearity}Classical PCA is a strictly {\it linear} technique. Its main objective is to identify {\it linear} combinations accounting for most of the variance of a $d$-dimensional variable of interest ${\textbf X}$. This is fine in an L$_2$ context, under Gaussian or more general elliptical assumptions, where   L$_2$ geometry %
 provides a natural and meaningful structure. Elliptical symmetry is a very restrictive assumption, though, and, in more general situations,   the inability of PCA  to detect nonlinear structures is a major limitation. This has  long been recognized, and a  variety of nonlinear versions of PCA have been proposed in the literature, extending some of the essential features of the classical concept.

 One of the earliest generalizations of classical PCA is %
 {\it polynomial} PCA  {\citep{GnanadesikanWilk1969}}, where, similar to polynomial regression modeling, higher-order polynomials based on the original variables and including interaction terms account for nonlinearities. Then, linear PCA is applied,  under adequate assumptions of finite higher-order moments,  to these polynomial transforms of the original variables. The disadvantage is that the dimensionality quickly grows as a function of the polynomial order.  Starting from the same idea,  {\citet{ScholkopfSmolaMuller1998}} propose  {\it kernel} PCA, where a kernel~$K$ is used to map a~$d$-dimensional random vector   $\bo X$ to some $d_k$-dimensional feature space  with~{$d_k > d$ (possibly, $d_k=\infty$)}, where linear PCA is again performed. The motivation in both cases is that a low-dimensional structure might be easier to identify when embedded in an appropriate high-dimensional space---now a possibly nonlinear one. Kernel PCA is quite popular but presents the challenge that the inverse kernel transformation might be difficult to perform.
   Both polynomial and kernel PCA thus rely on the same variance maximization principle as traditional PCA.

{\it Principal curves}, as introduced in  {\cite{hastie1989principal}}, were instead motivated by the least squares optimality of  a rank-one approximation. While a linear principal component is represented by  a straight line, a principal curve consists of a general (smooth) curve $\bo f$   parametrized by some $\lambda$.
Denoting by~${\bo X}\sim {\rm P}$  a $d$-variate random vector, let~$\lambda_{\bo f}(\bo x) \coloneqq  \sup_\lambda \{\lambda : \|\bo{x} - \bo{f}(\lambda)\| = \inf_{\mu}\|\bo{x} - \bo{f}(\mu)\|\}$ be the projection index  providing the value of $\lambda$ for which $\bo{f}(\lambda)$  is closest to $\bo x$: {the curve~$\bo f$} then is called a {\it principal curve of}~$\bo{X}$ if~$
\bo{f}(\lambda) = {\rm E_P}(\bo{X} | \lambda_{\bo{f}}(\bo{X}) = \lambda)
$  for all~$\lambda$.
This concept of a principal curve has been extended into a concept of $k$-dimen\-sional {\it principal surface} {\citep{Hastie1984, LeBlancTibshirani1994}: the} {\it principal surface}~$\bo{f}(\boldsymbol{\lambda})$,  $\boldsymbol{\lambda} \in \mathbb{R}^k$ {\it of}~$\bo{X}$  is a surface for which
$
\bo{f}(\boldsymbol{\lambda}) = {\rm E_P}(\bo{X} | \boldsymbol{\lambda}_{\bo{f}}(\bo{X}) = \boldsymbol{\lambda}),
$
where $\boldsymbol{\lambda}_{\bo{f}}$ is an appropriately defined projection index. In practice, visualization is usually the main goal, and the dimension of principal surfaces is limited~to~$k=2$. %

Closely related to principal curves is an %
approach based on a specific {\it autoencoder}---an artificial neural network   learning its own inputs; see, e.g.,   \cite{Kramer1991}, {\cite{ScholzKaplanGuyKopkaSelbig2005}}, and   {\cite{hinton2006reducing}}. Such autoencoders consist of two parts---the {\it encoder} and the {\it decoder}. The encoder maps the $d$-variate input~$\bo{x}$ to a $k$-dimensional (with~$k < d$) latent vector $\bo{z}$, or the decoder attempts to reconstruct
$\bo{x}$ from this~$\bo{z}$. The nonlinear mapping from $\bo{x}$ to~$\bo{z}$ is parameterized by a neural network using the encoder function~$
\bo{z} = \sigma_e(\bo{W}_e \bo{x} + \bo{b}_e)
$
where $\bo{W}_e$ and $\bo{b}_e$ are the weight matrix and bias vector, respectively,  of the encoder and  $\sigma_e$ is a nonlinear {\it activation function}. The decoder   reconstructing  $\bo{x}$ from $\bo{z}$ can be described as~$%
\hat{\bo{x}} = \sigma_d(\bo{W}_d \bo {z} + \bo{b}_d)
$,  %
where $\bo{W}_d$ and $\bo{b}_d$ are the weight matrix and bias vector, respectively,  of the decoder, and  $\sigma_d$  is again an activation function{. For simplicity, the encoder and the decoder are written here with a single layer; for nonlinear PCA, both contain at least one hidden layer with a nonlinear activation function, and the output layers are usually linear \citep{Kramer1991, ScholzKaplanGuyKopkaSelbig2005}}.
The goal is then to jointly learn $\bo{W}_e$, $\bo{W}_d$,~$\bo{b}_e$,\linebreak and  $\bo{b}_d$ by minimizing the difference between $\bo x$ and $\sigma_d(\bo{W}_d ( \sigma_e(\bo{W}_e \bo{x} + \bo{b}_e)) + \bo{b}_d)$ based on some loss function. In case of linear activation functions and using the $L_2$ loss function, it can be shown that the latent components $\bo z$ span the same subspace as the corresponding subspace obtained from linear PCA  {\citep{BaldiHornik1989,plaut2018principalsubspacesprincipalcomponents}}, which motivates the use of nonlinear activation functions to obtain a nonlinear variant of PCA.  For reviews of the topic, see, for example,  {\cite{LeeVerleysen07}},  {\cite{Mori16}},  {\cite{Mendez2023}}, or   {\cite{Wani2025}}.  For all these concepts, empirical and population versions are available, along with consistency results for kernel PCA {\citep[see, e.g.,][]{Blanchard2007}}, principal curves  {\citep[see, e.g.,][]{Kegl2000}}, and linear autoencoders  {\citep{BaldiHornik1989}}; consistency results for principal surfaces and nonlinear  {autoencoders are so far} limited.

\subsection{One-sidedness}
Another weakness of traditional PCA---which, unlike linearity,  {has seldom been} challenged---is its intrinsic {\it two-sidedness}. The definition of  %
the principal components of a random vector ${\textbf X}$, typically,  is based on the maximization, with respect to $\textbf u$ (a {\it direction} of ${\mathbb{R}^d}$, that is, a point on the unit hypersphere ${\mathcal S}_{d-1}$),  of the expected squared modulus $\langle {\textbf u},{{\textbf X}}
\rangle^2$ of the projection of  ${\textbf X}$ on the straight line with unit vector $\textbf u$. That squared modulus only depends on $\pm{\textbf u}$;  as a consequence, principal components are defined up to a sign---that is, they are associated with the linear subspace generated by $\pm \textbf u$  rather than with the direction~$\textbf u$. Again, this is fine under, e.g., elliptical symmetry, because the projection of a (centered) elliptical %
 ${\textbf X}$ on a direction~$\textbf u$ has the same distribution as its projection on the opposite direction $-\textbf u$. However, this is no longer the case in a nonsymmetric context, where better results can be achieved by disentangling  directions~$\textbf u$ and~$-\textbf u$ and maximizing the expected value of~$(\langle {\textbf u},{{\textbf X}}\rangle^+)^2$ (where~$\langle {\textbf u},{{\textbf X}}\rangle^+$ stands for the positive part of~$\langle {\textbf u},{{\textbf X}}\rangle$) rather than the expected value of~$\langle {\textbf u},{{\textbf X}}\rangle^2$: see Section~\ref{sec:lindirPCA}for details.

\subsection{A measure-transportation-based approach}
Measure transportation  {has recently emerged} as a novel and powerful tool in multivariate analysis. Based on measure transportation ideas,  {\cite{Chernozhukov2017},  {\cite{FaugRusch17}}},  {\cite{Hallin2022}}, and  {\cite{GHSHilbert}}  introduce concepts of {\it  {Monge--Kantorovich} depth, center-outward distribution} and {\it quantile functions, ranks}, and {\it signs} in ${\mathbb R}^d$ and Hilbert spaces;  {\cite{Carlier16},  {\cite{delBGonzHal24}},  {\cite{HLiu24}}}, and  {\cite{GHY2025}} develop multiple-output and manifold-valued quantile regression and autoregression methods; {\cite{GhoSen19},  {\cite{DebSen19}},  {\cite{HHH20,  HLL20, HLL19}},  {\cite{HALLINPortmanteau}}},   {\cite{ShiHallinAOS, HallinShiBernoulli}}, and {\cite{DebJRSS}}  propose distribution-free rank tests for location, regression, MANOVA, VARMA time series models, and vector independence, while  {\cite{HMordant}} define measure-transportation-based multi-attribute Lorenz curves and Gini indices; see {\cite{Hallin2022a}} for a review.

Measure transportation naturally offers fresh views on nonlinear concepts of principal components.
This novel perspective was first adopted in a paper by  {\cite{GunsiliusSchennach2023}}  who, under the terminology {\it  {independent nonlinear component analysis}}, {consider} an ingenious approach, %
{which can be seen as a nonlinear two-sided PCA method yielding mutually independent---not just uncorrelated---components.
Their idea consists of transporting the distribution of the variable~${\textbf X}$ of interest to that of a~$d$-variate  spherical normal distribution and defining nonlinear (but still ``bidirectional'') components based on the first~$k$ eigenvectors of  {the expected matrix logarithm of} the Jacobian of this transportation map   (see Section~\ref{SecG&S}). Below, we refer to their approach as {Gunsilius--Schennach PCA, in short GS-PCA}.}

In this paper, we  adopt a different approach combining the nonlinear features of measure transportation with the traditional arguments of maximal expected squared moduli of ``projections.''  However, our  projections are of a nonlinear and directional (one-sided) nature,  based on  transports to the spherical uniform on the unit ball (the {\it center-outward distribution function} ${\textbf F}_\pm$  and the so-called  {\it sign curves}  introduced in  {\cite{Hallin2022}}). These sign curves are the images, by the {\it center-outward quantile function}  ${\textbf Q}_\pm \coloneqq {\textbf F}_\pm^{-1}$, of the radii $\boldsymbol{\scriptstyle \mathcal U}\coloneqq \{r{\textbf u}\vert r\in [0, 1)\}$, where~${\textbf u}\in{\mathcal S}_{d-1}$ (the unit  sphere in ${\mathbb R}^d$). The mapping~${\textbf Q}_\pm$, in general, is  neither linear nor centrally symmetric (${\textbf Q}_\pm (-\boldsymbol{\scriptstyle \mathcal U})\neq - {\textbf Q}_\pm (\boldsymbol{\scriptstyle \mathcal U})$), and the sign curve ${\textbf Q}_\pm (\boldsymbol{\scriptstyle \mathcal U})$, in this context, bears no particular relation to the sign curve  ${\textbf Q}_\pm (-\boldsymbol{\scriptstyle \mathcal U})$; the mapping $(-1,1)\ni r\mapsto {\textbf Q}_\pm(r{\textbf u})$, moreover,  does not need to be  continuous at $r=0$, where it may be set-valued (see  {\cite{Fig18}}). This justifies the directional (one-sided) nature of our approach, selecting ``one-sided'' principal  curves~${\textbf Q}_\pm (\boldsymbol{{\scriptstyle \mathcal U}})$ instead of the two-sided ones ${\textbf Q}_\pm (\pm \boldsymbol{{\scriptstyle \mathcal U}})$, thereby capturing possible distributional anisotropies that remain invisible to a bidirectional formulation  and ending up, in dimension~$d$, with the definition of $2d$ {\it principal directions} and {\it curves}{---a definition which, moreover,  does not require finite second-order moments.}

{Now,  increased generality (nonlinearity, absence of moment assumptions, one-sidedness, ...) unavoidably comes at the cost of increased computational complexity. This is the case with all nonlinear PCA methods (kernel PCA, neural network learning, GS-PCA, ... ),  all of which are computer-intensive, and  our method is no exception.
}

The rest of the paper is organized as follows. In Section~\ref{Sec21}, we revisit the traditional definition of linear, bidirectional principal components  (Section~\ref{sec:linear_motivation}), define directional linear principal components (Section~\ref{sec:lindirPCA}), and propose (Section~\ref{Sec22}) an elliptical reformulation  that facilitates an extension to a nonlinear setting based on monotone transport maps. That extension is provided in Section~\ref{sec:nlpc}, the resulting dimension reduction (mapping the  observations  to a nonlinear manifold) in Section~\ref{sec:manifold}. Section~\ref{sec3} {proposes an empirical counterpart of that nonlinear extension,  %
 and (Section~\ref{SecG&S}) draws a comparison with the Gunsilius--Schennach approach.  Section~\ref{Sec:consist} deals with asymptotics and the consistency of our method.
Simulated and empirical examples are developed in Section~\ref{sec:examplesR2}. Section~\ref{ConclSec} concludes. Proofs are concentrated in  Appendix~\ref{sec:proofs}, and Appendix~\ref{sec:smooth} describes a regularized version of the empirical principal curves.
\color{Black}

\section{Principal components:  from  linear and bidirectional %
 to  nonlinear \\ and directional   %
}\label{Sec21}
\subsection{Traditional Principal Component Analysis (PCA)
} \label{sec:linear_motivation}
Assuming that $\mathbb{E}[{{\textbf X}}]={\boldsymbol 0}$ and that  the covariance matrix  $\boldsymbol \Sigma$ of ${\textbf X}$ is finite with full rank $d$ and $d$ distinct eigenvalues, the classical (bidirectional or two-sided)  linear {\it principal directions} of ${\textbf X}$ are defined~as\vspace{-1mm}
\begin{equation}\label{defU1}
\pm {\bf U}_1\coloneqq  \argmax_{{\textbf u}\in{\mathcal S}_{d-1}}\mathbb{E}\left[ \langle {\textbf u},{{\textbf X}}
\rangle ^2
  \right]\vspace{-2mm}
  \end{equation}
(where ${\mathcal S}_{d-1}$ stands for the unit sphere in ${\mathbb R}^d$) and\vspace{-2mm}
\begin{equation}\label{defUi}
\pm{\bf U}_i\coloneqq
\argmax\hspace{-20mm}_{\scriptscriptstyle{\begin{array}{c}$\,$\vspace{-8.5mm}\\ \scriptscriptstyle{{\textbf u}\in{\mathcal S}_{d-1} }\vspace{-5mm}\\  \scriptscriptstyle{{\textbf u}\perp {\bf U}_j,\ j=1,\ldots,i-1}\end{array}}}
\mathbb{E}\left[  \langle {\textbf u},{{\textbf X}}
\rangle ^2
 \right]\quad i=2,\ldots,d,\vspace{-2mm}
  \end{equation}
  hence coincide with the eigenvectors ${\bf P}_i$, $i=1,\ldots, d$ of $\boldsymbol\Sigma$. Call ${\mathcal L}_i\coloneqq  \{r{\bf U}_i\vert\, r\in {\mathbb R}\}$
   the $i$th {\it principal   line} of ${\textbf X}$.

  The classical (bidirectional or two-sided) $i$th {\it principal component} of ${\textbf X}$ then is the signed  modulus  \vspace{-2mm}%
  \begin{equation}\label{P_i}
  P_i\coloneqq\pm\langle{\bf U}_i ,{{\textbf X}} \rangle,\vspace{-3mm}
  \end{equation}
of the linear projection  onto ${\mathcal L}_i$ of $\textbf X$,     the sign of which depends  on the arbitrary labeling of $\pm {\bf U}_i$ as ${\bf U}_i$ and~$-{\bf U}_i$ (this choice is     irrelevant in most applications, and therefore receives little attention in practice). %

The first $k$ principal lines of $\textbf X$ span a $k$-dimensional linear subspace of ${\mathbb R}^d$\vspace{-2mm}
$${\mathfrak U}_k\coloneqq \{ ({\bf U}_1,\ldots,{\bf U}_k){\bf r} \vert\, {\bf r}\in{\mathbb R}^k\}.
\vspace{-1mm}$$
 A natural dimension reduction procedure then consists of replacing  $\textbf X$ with its projection\vspace{-2mm}
 $$\sum_{i=1}^k \langle {\bf U_i}, {\textbf X}
 \rangle {\bf U}_i = \sum_{i=1}^k P_i {\bf U_i}
\vspace{-2mm} $$
  onto   ${\mathfrak U}_k$ and constitutes the basis of traditional Principal Component Analysis (PCA).

  \subsection{Directional linear Principal Component Analysis}\label{sec:lindirPCA}

Recall that ${\textbf X}$ is assumed to have mean zero. Since $ \langle {\textbf u},{{\textbf X}}
\rangle =  \langle {\textbf u},{{\textbf X}}
\rangle^+ -  \langle {\textbf u},{{\textbf X}}
\rangle^- =  \langle {\textbf u},{{\textbf X}}
\rangle^+ -  \langle {-\textbf u},{{\textbf X}}
\rangle^+$ with $ \langle {\textbf u},{{\textbf X}}
\rangle^+  \langle {-\textbf u},{{\textbf X}} \rangle ^+ =0$ a.s., the criterion to be maxi\-mized in \eqref{defU1} and \eqref{defUi} is ``two-sided,''  hence ``bidirectional'' rather than ``directional.''  More precisely,~$\mathbb{E}\left[  \langle {\textbf u},{{\textbf X}}
\rangle^2
  \right]$ has the two-sided  form \vspace{-2mm}
\begin{equation}\label{2sided}
\mathbb{E}\left[  \langle {\textbf u},{{\textbf X}}
\rangle ^2
  \right]=\mathbb{E}\left[ (\langle {\textbf u},{{\textbf X}}
\rangle ^{+})^2
  \right] + \mathbb{E}\left[ (\langle {-\textbf u},{{\textbf X}}
\rangle ^{+})^2
  \right].\vspace{-2.5mm}
  \end{equation}
  This is fine in a symmetric situation, e.g., under  elliptical distributions:  then $\mathbb{E}\left[ (\langle {\textbf u},{{\textbf X}}
\rangle ^{+})^2
  \right] = \mathbb{E}\left[ (\langle {-\textbf u},{{\textbf X}}
\rangle ^{+})^2
  \right]$
 so that maximizing  $\mathbb{E}\left[  \langle {\textbf u},{{\textbf X}}
\rangle ^2
 \right]$, $\mathbb{E}\left[  (\langle {\textbf u},{{\textbf X}}
\rangle ^+)^2
 \right]$, or $\mathbb{E}\left[  (\langle -{\textbf u},{{\textbf X}}
\rangle ^+)^2
 \right]$ makes no difference---both ${\bf U}_i$ and~$-{\bf U}_i$ qualify as principal directions.  When the distribution of ${\textbf X}$ is asymmetric, however, $\mathbb{E}\left[  (\langle {\textbf u},{{\textbf X}}
\rangle ^+)^2
 \right]$ and~$\mathbb{E}\left[  (\langle -{\textbf u},{{\textbf X}}
\rangle ^+)^2\right] = \mathbb{E}\left[  (\langle {\textbf u},{{\textbf X}}
\rangle ^-)^2\right]$ may differ: ${\bf U}_i$ then maximizes their sum, or their average, %
not necessarily %
 any of them. %

We therefore propose to replace the classical {\it two-sided}  or {\it bidirectional} definitions  \eqref{defU1} and \eqref{defUi} of principal directions with the {\it one-sided} or {\it directional} ones \vspace{-2mm}
 \begin{equation}\label{defU1uni}
 {\bf U}_1^+\coloneqq  \argmax_{{\textbf u}\in{\mathcal S}_{d-1}}\mathbb{E}\left[( \langle {\textbf u},{{\textbf X}}
\rangle ^+) ^2
  \right]\vspace{-3mm}
  \end{equation}
and\vspace{-2mm}
\begin{equation}\label{defUiuni}
{\bf U}^+_i\coloneqq
\argmax_{%
{
\scriptscriptstyle{
\begin{array}{c} $\,$\vspace{-11mm}\\ \scriptscriptstyle{{\mathcal S}_{d-1}\ni {\textbf u}
\perp {\mathcal U}^+_{i-1}\vspace{-0mm}
}\vspace{-4mm}\\
 \text{\tiny{\rm or }}
 \scriptscriptstyle{
 {\textbf u}\in (-{\mathcal U}^+_{i-1})\setminus{\mathcal U}^+_{i-1} }
\end{array}}
\hspace{-3mm}
}
}
\mathbb{E}\left[ ( \langle {\textbf u},{{\textbf X}}
\rangle ^+) ^2
 \right]\quad i=2,\ldots,2d,\vspace{-5mm}
  \end{equation}
respectively, where
\begin{equation}\label{Uplus}{\mathcal U}^+_k\coloneqq \{{\bf U}^+_1,\ldots,{\bf U}^+_k\}\quad\text{and} \quad -{\mathcal U}^+_k\coloneqq \{-{\bf U}_1^+,\ldots,-{\bf U}^+_k\}\vspace{-0mm}\end{equation}
 (in case of  a tie of multiplicity $\kappa$ in \eqref{defU1uni} or \eqref{defUiuni},  tied values can be arbitrarily ordered and denoted as~${\bf U}^+_i, {\bf U}^+_{i+1},\ldots, {\bf U}^+_{i+\kappa -1}$), $i=1,\ldots,2d-\kappa +1$; in the sequel, we tacitly assume (in \eqref{defU1'}, \eqref{defUi'}, \eqref{defU1''}, and~\eqref{defUi''})  that this does not occur.  %
Call ${\bf U}^+_i$, $i=1,\ldots, 2d$, the $i$th {\it one-sided} or {\it directional linear principal direction} of $\textbf X$ and, defining the~$i$th {\it directional   principal halfline}  %
of ${\textbf X}$   as ${\cal L}^+_i\coloneqq\{r{\bf U}^+_i\vert\, r\in{\mathbb R}_+
\}$, %
denote~by   \begin{equation}\label{P_i+}P_i^+\coloneqq  \langle {\bf U}^+_i , {{\textbf X}}\rangle^+, \qquad i=1,\ldots,2d \vspace{-1mm}.\end{equation}
 the~$i$th {\it directional   principal component} of $\textbf X$, that is, the positive part of its projection onto ${\cal L}^+_i$.

  \renewcommand\thesubfigure{\roman{subfigure}}

\begin{figure}[b!]%
    \centering
    \includegraphics[width=\textwidth]{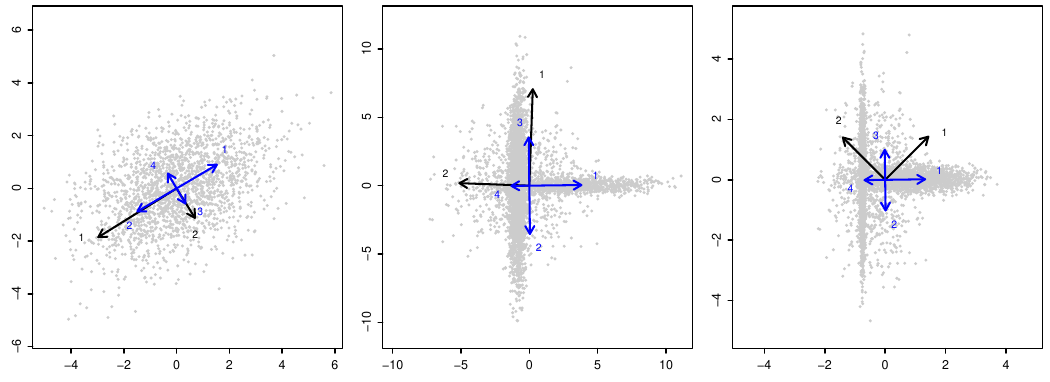}
    \caption{Linear principal components: directional (blue; ${\bf U}^+_i$, $i=1,\ldots,2d=4$) and classical (bidirectional, black; ${\bf U}_i$, $i=1,\ldots,d=2$). Left panel: an elliptical (Gaussian) distribution, for which the directional and classical linear principal components coincide. Middle panel: an asymmetric Gaussian mixture, for which the first two directional linear principal component coincides with the last two classical one. Right panel: an asymmetric Gaussian mixture, for which the directional principal directions differ from the classical ones by approximately~$\pi/4$.
    The numbers at the arrowheads give the order of the directions, and the length of each arrow is proportional to the empirical mean square of the corresponding principal component, $P_i^+$ or~$P_i$ (with the same factor for both approaches within a panel).}
    \label{fig:FIG1}
\end{figure}

\begin{figure}[b!]
    \vspace{-0.1cm}
    \centering
    \includegraphics[width=0.7\textwidth]{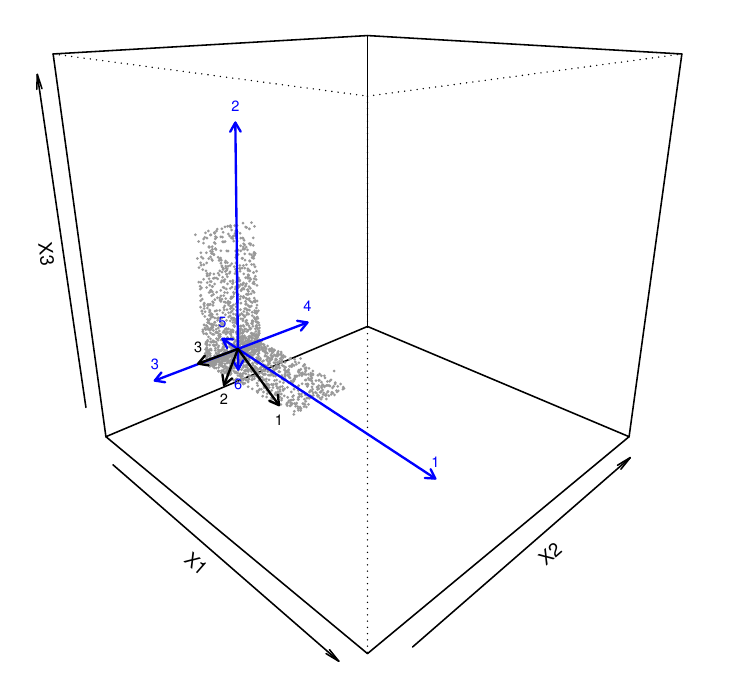}
    \caption{A three-dimensional scatterplot illustrating the difference between linear directional and classical principal directions. The arrows indicate the linear principal directions: the blue ones (${\bf U}^+_i$, $i=1,\ldots,2d=6$) correspond to the directional ones, the black ones (${\bf U}_i$, $i=1,\ldots,d=3$) to the traditional bidirectional ones. The numbers at the arrowheads give the order of the directions. The length of each arrow is proportional to the empirical mean square of the corresponding principal component, $P_i^+$ or~$P_i$; for better visibility, the black arrows are drawn at a quarter of the scale of the blue ones. The fifth and sixth directional components have small mean squares, so that their arrows are short and hardly visible near the origin; they are best seen in the interactive version available at \url{https://klauschn.github.io/DirectionalNonlinearPCA/linear_directions_3d.html}, which can be rotated and zoomed.
}

    \label{fig:FIG2}
\end{figure}

 \renewcommand\thesubfigure{\arabic{subfigure}}

\begin{figure}[t!]
     \centering
    \includegraphics[width=\textwidth]{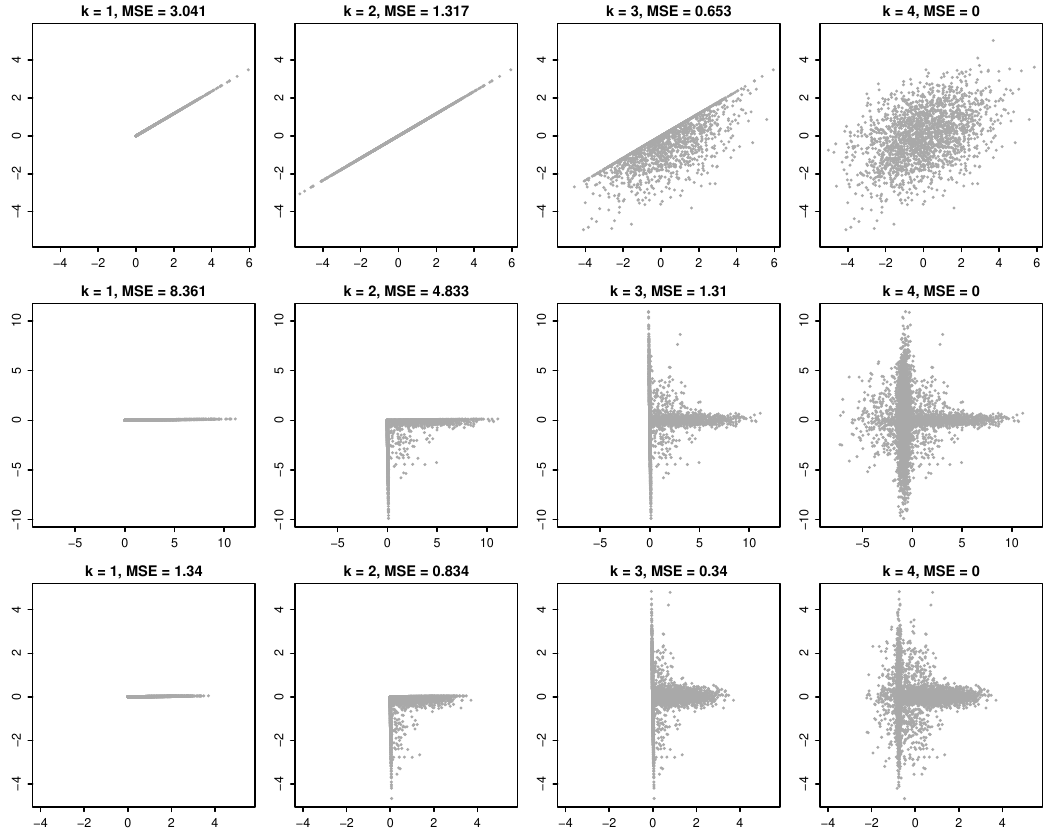}
    \caption{\small Directional  dimension reduction ($d=2$). Projections of the data from Figure~\ref{fig:FIG1} onto the positive spans ${\mathfrak U}_k^+$ of their first $k$ directional linear principal directions \eqref{defU1uni}--\eqref{defUiuni}, $k=1,\ldots,2d=4$. Top row:     the elliptical (Gaussian) data from the left panel of Figure~\ref{fig:FIG1}. Middle and bottom row:   the asymmetric Gaussian mixture data from the middle and right panels of the same figure. The MSE values at the top of each plot provide the mean squared reconstruction error associated with the projection onto the first $k$ one-sided principal directions. For~$k=4=2d$, this projection is identity, hence the reconstruction error and the MSE are zero. %
    }
    \label{fig:FIG3}
\end{figure}

The resulting   $2d$-tuple ${\mathcal U}^+\coloneqq \{{\bf U}^+_i \vert \ i=1,\ldots,{2d}\}={\mathcal U}^+_{2d}$ of directional principal directions   is such that if~${\bf U}^+_i\in {\mathcal U}^+$, then also~$-{\bf U}^+_i\in {\mathcal U}^+$ (with a distinct index $i$, though). However,  for $k<d$,
${\mathcal U}^+_{{2k}}$, in general, does not coincide with the $2k$-tuple
$ {\mathcal U}_k\coloneqq \{\pm{\bf U}_{{j}}\vert\ j=1,\ldots, k\}$   resulting from the classical bidirectional defini\-tions~\eqref{defU1}--\eqref{defUi}, while providing (in the sense of mean square distances) a better approximation of the $d$-dimensional distribution of ${{\textbf X}}$, hence a better dimension reduction.

This is illustrated in the following two- and three-dimensional examples. For a bivariate elliptically distributed dataset,  {the left panel} of Figure~\ref{fig:FIG1} shows that directional PCA and standard PCA yield identical results. In {the middle and right panels}, the dataset (obtained from Gaussian mixtures) gets more asymmetric, and the two approaches differ.  In  {the middle panel},  the differences between the two PCA variants look minor; however, the first standard PCA direction is less informative than the first two directional ones. The right panel further highlights the advantages of the directional approach, as it more effectively captures the dominant directions of variation in the data, whereas the traditional PCA directions are of questionable practical relevance.

The trivariate example shown in Figure~\ref{fig:FIG2} consists of a sample generated from  Gaussian distributions  concentrated around two intersecting planes---one   in the $x$--$y$ subspace ($z \approx 0$),  the other one in the $y$--$z$ subspace ($x \approx 0$).
	Despite its relatively simple   structure, this dataset is poorly represented by  classical PCA principal directions. In contrast, directional PCA clearly identifies the relevant directions of variation and orders them in a meaningful way.

If dimension reduction is the objective, it is naturally performed, in this directional context, by considering some chosen $k$-tuple ${\mathcal U}_k^+$~($k<2d$) of directional principal directions ${\bf U}_1^+,\ldots,{\bf U}_k^+$ ordered as in~\eqref{defU1uni}--\eqref{defUiuni}, that is, by decreasing values of $\mathbb{E}\left[ ( \langle {\bf U}_i^+,{{\textbf X}}
\rangle ^+) ^2\right]$. The positive span  ${\mathfrak{U}^+_k}\coloneqq \{ ({\bf U}_1^+,\ldots,{\bf U}_k^+){\bf r} \vert\, {\bf r}\in{\mathbb R}_+^k\}$ of these~$k$ directions either is a linear $k/2$-dimensional subspace (in case $k$ is even and ${\bf U}_i^+\in{\mathcal U}_k^+$ implies~$-{\bf U}_i^+\in{\mathcal U}_k^+$), or an orthant (in case ${\bf U}_i^+\in{\mathcal U}_k^+$ implies~$-{\bf U}^+_i\notin{\mathcal U}_k^+$), or a halfspace, or the product of an orthant and a linear subspace, etc. {Dimension reduction is naturally performed by projecting  $\textbf X$ onto $\mathfrak{U}^+_k$}, yielding
\begin{equation}\label{projmathfrakU+}
\sum_{i=1}^k \langle
{\bf U}_i^+,{\textbf X}
\rangle^+ {\bf U}_i^+ = \sum_{i=1}^k P_i^+ {\bf U}_i^+.
\end{equation}

\begin{figure}[t!]
    \centering
    \includegraphics[width=.9\textwidth]{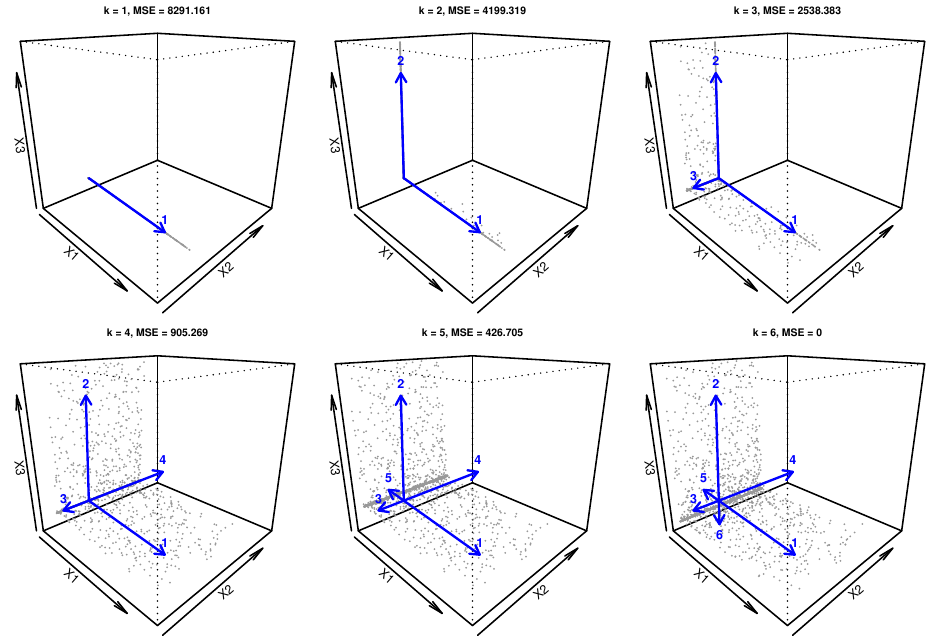}
    \caption{\small Directional  dimension reduction ($d=3$). Left to right, top to bottom: each panel gives the projection of the three-dimensional dataset from the mixture distribution shown in Figure~\ref{fig:FIG2} onto the positive span $\mathfrak{U}_k^+$ of the first~$k$ one-sided principal directions, $k=1,\dots,6$.
 The blue arrows are {proportional to the empirical  mean squares of}  the one-sided projections  (using the positive-part inner product~$\langle\cdot,\cdot\rangle^+$) of the data onto the $k$ first directional principal direc-\linebreak tions,~$k=1,\ldots,6=2d$, as in Figure~\ref{fig:FIG2} (the axes are, however, shorter, and arrows that would leave the box are shortened). The numbers displayed next to the arrowheads are the orders of the corresponding principal directions.
The MSE values at the top of each panel correspond to the projection onto the first $k$ one-sided directions. As in Figure~\ref{fig:FIG2}, the arrows of the fifth and sixth directions (panels $k=5$ and~$6$) are short, since the corresponding mean squares are small. An interactive version of this figure, in which the number~$k$ of directions can be chosen and the view can be rotated and zoomed, is available at \url{https://klauschn.github.io/DirectionalNonlinearPCA/linear_reconstructions_3d.html}.
}
    \label{fig:FIG4}
\end{figure}

The results, for the   data  shown in Figures~\ref{fig:FIG1} and~\ref{fig:FIG2}, are illustrated, in Figures~\ref{fig:FIG3} and~\ref{fig:FIG4},
with the corresponding dimension reductions. Note that even values of $k$ seem more meaningful in this context. To quantify the quality of the dimension reduction, mean squared error (MSE) values are reported in~each~figure.

For the elliptical data of the left panel of Figure~\ref{fig:FIG1}, the MSE values based on $k=2$ and $k=4$ directional components roughly coincide with those obtained from traditional PCA  with one and two components. However, for the  the middle and right panel data, the MSEs of traditional PCA with a single component are  $5.114$ and  $0.990$, respectively, which both  are larger than the   MSEs  obtained using the directional approach with $k=2$ (4.833 and 0.834). Thus, in addition to identifying more meaningful directions in the data, directional PCA also yields more accurate dimensionality reduction in terms of reconstruction error. MSE  improvements are even more pronounced in the three-dimensional case (Figure~\ref{fig:FIG4}), where the MSEs of traditional PCA are {$7{,}635.55$ and $3{,}293.80$}, respectively, for one and two bidirectional components, to be compared with {$4{,}199.32$} and~$905.27$ for two and four directional components, respectively.

\subsection{An elliptical reformulation of the classical concepts}\label{Sec22} Sections~\ref{sec:linear_motivation} and \ref{sec:lindirPCA} still are about linear (bidirectional and directional) principal directions and components concepts. Before turning to  nonlinear extensions,  let us reformulate their  definitions
 under a special form exploiting an assumption of elliptical symmetry---a reformulation that, after relaxing the ellipticity assumption, will naturally extend into a nonlinear and non-elliptical measure-transportation-based version.

The most natural context for PCA, indeed,  is the family of elliptically symmetric densities.  Denote\linebreak  by~${{\textbf X}}\coloneqq  (X_{1},\dots,X_{d})                      ^\prime$  a~$d$-dimensional elliptical random vector with location $\boldsymbol 0$,    full-rank  scatter matrix~$\boldsymbol\Sigma$,
and   {\it radial density}~$\mathfrak{f}$---the density of the modulus~$({{\textbf X}}^\prime{\boldsymbol\Sigma}^{-1}{{\textbf X}})^{1/2}\eqqcolon \Vert {\boldsymbol\Sigma}^{-1/2}{{\textbf X}}\Vert$. Denote  by ${\rm P}_{{\boldsymbol\Sigma},\mathfrak f}$ this elliptical distribution, by
$F_{\mathfrak f}:~\!t\mapsto F_{\mathfrak f}(t)\coloneqq\int_0^t{\mathfrak f}(u){\rm d}u={\rm P}_{{\boldsymbol\Sigma},\mathfrak f}\left[\Vert {\boldsymbol\Sigma}^{-1/2}{{\textbf X}}\Vert \leq t
\right]$,  $t\in{\mathbb R_+}$
 its {\it radial distribution function},
by~${\boldsymbol \Lambda}=\text{Diag}(\lambda_1,\ldots,\lambda_d)$   the diagonal matrix of eigenvalues of~$\boldsymbol\Sigma$ in decreasing order of magnitude (for simplicity, assume that they are all distinct),
by  ${\bf P}_1,\ldots,{\bf P}_d$ the  corresponding  eigenvectors, and by
${\bf P}$   the~$d\times d$ orthogonal matrix with  columns ${\bf P}_i$, $i=1,\ldots,d$; %
 let~${\boldsymbol\Sigma}^{1/2}\coloneqq  {{\bf P} {\boldsymbol \Lambda}^{1/2}{\bf P}^\prime}$.

   Since  ${\boldsymbol\Sigma}^{-1/2}{{\textbf X}}$ is spherical with radial density  $\mathfrak{f}$, it follows that\vspace{-2mm}
  \begin{equation}\label{defUell}
  {\bf U}\coloneqq F_{\mathfrak f}\left(\Vert {\boldsymbol\Sigma}^{-1/2}{{\textbf X}}\Vert \right)\frac
  {{\boldsymbol\Sigma}^{-1/2}{{\textbf X}}
}{\Vert {\boldsymbol\Sigma}^{-1/2}{{\textbf X}}
\Vert }\vspace{-2mm}
\end{equation}
 is {\it spherical uniform} over the unit ball ${\mathbb S}_d$---that is, elliptical with center $\boldsymbol 0$, scatter ${\bf I}$, and radial density~${\mathfrak f} (u)=1$ for $0\leq  u<1$ and ${\mathfrak f} (u)=0$ elsewhere---which we denote as ${\bf U}\sim {\rm U}_d$.
   In the terminology of measure transportation, the   transformation\vspace{-2mm}
 \begin{equation}\label{linT}
 \textbf{T}_{{\boldsymbol\Sigma},{\mathfrak f}}:\ {\bf x}
 \mapsto \textbf{T}_{{\boldsymbol\Sigma},{\mathfrak f}}({\bf x})
 \coloneqq F_{\mathfrak f}\left(\Vert {\boldsymbol\Sigma}^{-1/2}{\bf x}
\Vert \right)\frac{{\boldsymbol\Sigma}^{-1/2}{\bf x}
}{\Vert {\boldsymbol\Sigma}^{-1/2}{\bf x}
\Vert }\eqqcolon  F_{\mathfrak f}\left(\Vert {\boldsymbol\Sigma}^{-1/2}{\bf x}
\Vert \right)
\textbf{T}_{{\boldsymbol\Sigma}} ({\bf x})\vspace{-1mm}
 \end{equation}
  mapping ${\textbf X}$ to
  $\bf U$ is a {\it transport map} (in general, not the gradient of a convex function, hence not a cyclically monotone transport   in the sense of McCann, nor an optimal one in the sense of Monge and Kantorovich)
    {\it pushing} ${\rm P}_{{\boldsymbol\Sigma},\mathfrak f}$ {\it forward}
      to the spherical uniform   ${\rm U}_d$ (notation: $ \textbf{T}_{{\boldsymbol\Sigma},{\mathfrak f}}\# {\rm P}_{{\boldsymbol\Sigma},\mathfrak f}= {\rm U}_d$).
     Provided that $\boldsymbol\Sigma$ has $d$ distinct eigenvalues and $\mathfrak f$ is nonvanishing,  $ \textbf{T}_{{\boldsymbol\Sigma},{\mathfrak f}}$ is invertible, with inverse \vspace{-1mm}
    $$ \textbf{T}_{{\boldsymbol\Sigma},{\mathfrak f}}^{-1}: {\textbf u}\mapsto  \textbf{T}_{{\boldsymbol\Sigma},{\mathfrak f}}^{-1}({\textbf u}) =  F_{\mathfrak f}^{-1}(\Vert {\textbf u}\Vert){\boldsymbol\Sigma}^{1/2}\frac{{\textbf u}}{\Vert {\textbf u}\Vert}\quad {\textbf u}\in{\mathbb S}_d
   \vspace{-1mm} $$
  pushing ${\rm U}_d$ forward to ${\rm P}_{{\boldsymbol\Sigma},\mathfrak f}$: $ \textbf{T}_{{\boldsymbol\Sigma},{\mathfrak f}}^{-1}\# {\rm U}_d = {\rm P}_{{\boldsymbol\Sigma},\mathfrak f}$.

  This suggests an alternative description of ${\textbf X}$'s traditional principal components,   formally bypassing the recourse to the eigenvectors and eigenvalues of ${\boldsymbol\Sigma}$. Under ellipticity, both ${\bf U}_i$ and~$-{\bf U}_i$ qualify as  traditional~$i$th principal directions: actually, the ``one-sided'' definitions  \eqref{defU1uni} and \eqref{defUiuni}, at each step, characterize $\pm {\bf U}_i$ in a tie. This is also what we are doing in \eqref{defU1'} and \eqref{defUi'}  below. Denoting by ${\mathcal S}_{d-1}$ the unit hypersphere in ${\mathbb R}^d$ and assuming that ${\mathfrak f}$ is strictly positive over the positive real line ${\mathbb R}_+$ (this can be relaxed to strictly positive on an   interval~$(0,R)$ for some~$0<R<\infty$) so that $ \textbf{T}_{{\boldsymbol\Sigma},{\mathfrak f}}$ is invertible, define \vspace{-1mm}
\begin{equation}\label{defU1'}
\pm {\bf U}_{1;\, \text{\rm ell}}\coloneqq  \argmax_{{\textbf u}\in{\mathcal S}_{d-1}}\mathbb{E}\left[\left\Vert
   \textbf{T}_{{\boldsymbol\Sigma},{\mathfrak f}}^{-1}(\varrho {\textbf u})\right\Vert ^2
  \right]\vspace{-2mm}
  \end{equation}
and %
\begin{equation}\label{defUi'}
\pm{\bf U}_{i;\, \text{\rm ell}}\coloneqq\!\!  \argmax_{\scriptscriptstyle{\begin{array}{c}$\,$\vspace{-11mm}\\ \scriptscriptstyle{{\textbf u}\in{\mathcal S}_{d-1} }\vspace{-5mm}\\  \scriptscriptstyle{{\textbf u}\perp {\bf U}_j,\ j=1,\ldots,i-1}\end{array}}}
\mathbb{E}\left[\left\Vert
   \textbf{T}_{{\boldsymbol\Sigma},{\mathfrak f}}^{-1}(\varrho {\textbf u})\right\Vert ^2
  \right]\quad i=2,\ldots,d
  \end{equation}
 where $\varrho $ is uniformly distributed over $[0,1]$.
 Also
 define    the $i$th  {\it principal line} and {\it halflines} of ${\textbf X}$ (of ${\rm P}_{{\boldsymbol\Sigma},\mathfrak f}$) as
\begin{equation}\label{halflines}
{\mathcal L}_{i;\, \text{\rm ell}}\!\coloneqq\! \left\{\textbf{T}_{{\boldsymbol\Sigma},{\mathfrak f}}^{-1}(r{\bf U}_{i;\, \text{\rm ell}})\vert r\in{\mathbb R}\right\},\
{\mathcal L}^+_{i;\, \text{\rm ell}}\!\coloneqq\! \left\{\textbf{T}_{{\boldsymbol\Sigma},{\mathfrak f}}^{-1}(r{\bf U}_{i;\, \text{\rm ell}})\vert r\in{\mathbb R}_+\right\},\ \text{and}\ \ {\mathcal L}^-_{i;\, \text{\rm ell}}\!\coloneqq\! \left\{\textbf{T}_{{\boldsymbol\Sigma},{\mathfrak f}}^{-1}(-r{\bf U}_{i;\, \text{\rm ell}})\vert r\in{\mathbb R}_+\right\},
\end{equation}
respectively. The $i$th principal component of ${{\textbf X}}\sim {\rm P}_{{\boldsymbol\Sigma},\mathfrak f}$ then is the signed modulus  $P_{i;\, \text{\rm ell}} \coloneqq  \pm\langle{\bf U}_{i;\, \text{\rm ell}},{{\textbf X}}\rangle$  of the projection of~${\textbf X}$ on~$\pm {\bf U}_{i;\, \text{\rm ell}}$ (equivalently, on the principal line $\pm{\mathcal L}_{i;\, \text{\rm ell}}$).

Let us show that this, under ellipticity, is indeed a reformulation of the traditional concepts described in Section~\ref{sec:linear_motivation}. To see this, note that \vspace{-2mm}
\begin{align*}
 \pm{\bf U}_{1;\, \text{\rm ell}}&\coloneqq  \argmax_{{\textbf u}\in{\mathcal S}_{d-1}}\mathbb{E}\left[\left\Vert
   \textbf{T}_{{\boldsymbol\Sigma},{\mathfrak f}}^{-1}(\varrho {\textbf u})\right\Vert ^2
  \right] =\mathbb{E}\left[( F_{\mathfrak f}^{-1}(\rho ))^2  \right] \argmax_{{\textbf u}\in{\mathcal S}_{d-1}} \mathbb{E}\left[( F_{\mathfrak f}^{-1}(\rho ))^2  \right] \left({\textbf u}^\prime{\boldsymbol\Sigma}{\textbf u} \right)\\
&=\argmax_{{\textbf u}\in{\mathcal S}_{d-1}} \sigma^2_{\mathfrak f}\left({\textbf u}^\prime{\boldsymbol\Sigma}{\textbf u}\right)   =\argmax_{{\textbf u}\in{\mathcal S}_{d-1}} \left({\textbf u}^\prime{\boldsymbol\Sigma}{\textbf u}\right)=\pm{\bf P}_1 =   \pm{\bf U}_{1}
  \end{align*}
since $\pm {\bf P}_1$, the first eigenvector of $\boldsymbol\Sigma$,  coincides with  the principal directions $\pm{\bf U}_1$ previously defined in~\eqref{defU1}.
 The  case for $i=2,\ldots, d$ is entirely similar, and \eqref{defU1'}-\eqref{defUi'}, thus, are equivalent, under ellipticity, to~\eqref{defU1uni}-\eqref{defUiuni}:  %
   $ \pm{\bf U}_{i;\, \text{\rm ell}} =  \pm{\bf U}_{i}$ and, therefore,  also  $\pm{\mathcal L}_{i;\, \text{\rm ell}} =  \pm{\mathcal L}_{i}$ for $i=1,\ldots,d$, which implies
 \begin{equation}\label{mathfrak=}
 {\mathfrak{U}_{k;\, \text{\rm ell}}}\coloneqq \{ (\pm{\bf U}_{1;\, \text{\rm ell}},\ldots,\pm{\bf U}_{k;\, \text{\rm ell}}){\bf r} \vert\, {\bf r}\in{\mathbb R}^k\}
 = \{(\pm{\bf U}_1,\ldots,{\pm\bf U}_{k}) {\bf r} \vert\, {\bf r}\in{\mathbb R}^k\} \eqqcolon \mathfrak{U}_{k}.
 \end{equation}
 Elementary calculation yields
 $$ \textbf{T}_{{\boldsymbol\Sigma},{\mathfrak f}}\left( \sum_{i=1}^kr_i  {\bf U}_{i}\right) = F_{\mathfrak f}\left((\sum_{i=1}^kr_i^2\lambda_i^{-1})^{1/2}\right){\frac{\sum_{i=1}^k r_i\lambda_i^{-1/2}{\bf U}_i}{(\sum_{i=1}^kr_i^2\lambda_i^{-1})^{1/2}}}\in  \mathfrak{U}_{k}\bigcap{\mathbb S}_d \text{\ \ for any~$(r_1,\ldots,r_k)\in\mathbb{R}^k$,}$$
 hence %
\begin{equation}\label{mathfrakInv} \mathfrak{U}_{k}
 =
 \textbf{T}_{{\boldsymbol\Sigma},{\mathfrak f}}^{-1}\left({\mathfrak{U}_{k;\, \text{\rm ell}}}\bigcap\mathbb{S}_d\right) =  \textbf{T}_{{\boldsymbol\Sigma},{\mathfrak f}}^{-1}\left({\mathfrak{U}_{k}}\bigcap\mathbb{S}_d\right),\quad k=1,\ldots, d.
\end{equation}
The classical dimension reduction described in Section~\ref{sec:linear_motivation}, thus, can be performed by   projecting $\textbf X$ onto~$ \textbf{T}_{{\boldsymbol\Sigma},{\mathfrak f}}^{-1}\left({\mathfrak{U}_{k;\, \text{\rm ell}}}\bigcap\mathbb{S}_d\right)$.

Directional (one-sided) versions  ${\bf U}^+_{i;\, \text{\rm ell}} $, ${\mathcal L}^+_{i;\, \text{\rm ell}} $,  and ${\mathfrak{U}^+_{k;\, \text{\rm ell}}}$ of $\pm{\bf U}_{i;\, \text{\rm ell}} $, $\pm{\mathcal L}_{i;\, \text{\rm ell}} $,  and ${\mathfrak{U}_{k;\, \text{\rm ell}}}$ are straightforwardly obtained by combining the one-sided approach of Section~\ref{sec:lindirPCA} with the  elliptical definitions above; details are skipped.

 \subsection{Nonlinear directional principal components based on monotone transport maps}
 \label{sec:nlpc}

 In a general non-elliptical context, there is no such thing as a radial density $\mathfrak f$ and a transport map~$ \textbf{T}_{{\boldsymbol\Sigma},{\mathfrak f}}$ of the form \eqref{linT} pushing the distribution $\rm P$ of ${\textbf X}$ forward to the spherical uniform ${\rm U}_d$. A famous result by  {\cite{McCann1995}}, however, guarantees, without any moment assumptions,  the existence and almost sure uniqueness of a {\it monotone} transport map $\textbf{T}_\pm$, generally nonlinear,  pushing $\rm P$ forward to ${\rm U}_d$;  {\it monotone}, here,  means being\footnote{Actually, lying in the subdifferential of some convex potential function.} the gradient $\nabla\psi$ of some convex potential function~$\psi: {\mathbb R}^d\rightarrow {\mathbb R}$. This result by McCann is more general than the usual  {Monge--Kantorovich--Brenier} optimality results, which make no sense in the absence of finite second-order moments.  If, however, $\rm P$ admits finite second-order moments, $\textbf{T}_\pm$ coincides ($\rm P$-almost surely) with the  {Monge--Kantorovich--Brenier} optimal transport from $\rm P$ to ${\rm U}_d$---that is, mini\-mizes~$\int_{\mathbb{R}^d}  \Vert {\bf x}- {\bf T}(\mathbf{x}) \Vert^2_2
 \mathrm{dP}
\vspace{1mm}$
among all transport maps $\bf T$ pushing~$ {\rm P}$ forward to~${\rm U}_d$.
  Assuming,\footnote{These are the assumptions made in Figalli (2018); they can be relaxed---see  {\cite{BarrioGonzalezH}},   {\cite{delBarrio2023regularity}},  {\cite{ErausquinFig19}}.} moreover, that~$\rm P$ belongs to the family~${\mathcal P}^d_\pm$ of Lebesgue-absolutely continuous distributions  with support~$\mathbb{R}^d$ and density $f$ such that, for any compact subset $C$ of ${\mathbb R}^d$, there exist constants $0<b_C <B_C <\infty$ such that~$b_C \leq f({\bf x})\leq B_C$ for all ${\bf x}\in C$,  {\cite{Fig18}} shows that $\textbf{T}_\pm$ is  a  homeomorphism between  supp$({\rm P})\setminus \textbf{T}_\pm^{-1}(\{{\boldsymbol 0}\})$ (where supp$({\rm P})$ stands for the support of~$\rm P$) and  the punctured unit ball~${\mathbb S}_d\setminus\{{\boldsymbol 0}\}$, with continuous (over ${\mathbb S}_d\setminus\{{\boldsymbol 0}\}$)  inverse~$ \textbf{T}_\pm^{-1}$ (which, at~$\boldsymbol 0$, can be   set-valued). Both $\textbf{T}_\pm$ and  $\textbf{T}_\pm^{-1}$   entirely characterize~$\rm P$;   {\cite{Hallin2022}} define ${\textbf F}_\pm \coloneqq \textbf{T}_\pm$ and ${\textbf Q}_\pm\coloneqq  \textbf{T}_\pm^{-1}$ as the {\it center-outward distribution} and {\it quantile functions}, respectively, of $\rm P$ (equivalently, of~${{\textbf X}}\sim{\rm P}$).

 \subsubsection{A measure-transportation-based definition of  principal directions and   components}

 The basic idea behind our definition of nonlinear directional principal directions and principal components consists of  replacing the  map $ \textbf{T}_{{\boldsymbol\Sigma},{\mathfrak f}}$ in \eqref{linT} and its inverse $ \textbf{T}^{-1}_{{\boldsymbol\Sigma},{\mathfrak f}}$ with the \textcolor{black}{monotone} transport maps ${\bf T}_\pm = {\textbf F}_\pm$ and~${\bf T}^{-1}_\pm = {\textbf Q}_\pm$,  the definition, existence, and a.s.\ uniqueness of
 which require neither elliptical symmetry  nor finite variances. While the images by $ \textbf{T}^{-1}_{{\boldsymbol\Sigma},{\mathfrak f}}$ of the radii of the unit ball are straight halflines such that $
 \lim_{r\downarrow 0} \textbf{T}^{-1}_{{\boldsymbol\Sigma},{\mathfrak f}}(r{\textbf u}) = \lim_{r\downarrow 0} \textbf{T}^{-1}_{{\boldsymbol\Sigma},{\mathfrak f}}(-r{\textbf u})={\boldsymbol 0}$, their images by~${\bf T}^{-1}_\pm$, in general, are curves, and  there is no need for $\lim_{r\downarrow 0}{\bf T}^{-1}_\pm (r{\textbf u})$,  $\lim_{r\downarrow 0}{\bf T}^{-1}_\pm (-r{\textbf u})$, and ${\boldsymbol 0}$ to coincide.

More precisely,  assuming that ${\rm P}\in{\mathcal P}_\pm^d$  and   that the expectations in \eqref{defU1''} and \eqref{defUi''} below exist and are finite,  and   adopting, for notational convenience,  the notation ${\textbf F}_\pm$ and~${\textbf Q}_\pm$ where, after the arbitrary choice\footnote{This choice   has no impact on subsequent definitions.} of an orthonormal basis in ${\mathbb R}^d$, ${\textbf Q}_\pm$ takes the form
$({Q}_{\pm , 1},\ldots,{Q}_{\pm , d})^\prime$,
   define, parallel to~\eqref{defU1'}--\eqref{defUi'} and~\eqref{defU1uni}--\eqref{defUiuni},   the {\it first nonlinear directional  principal direction} of $\rm P$ (of~${{\textbf X}}\sim {\rm P}$) %
    as \vspace{-2mm}
\begin{equation}\label{defU1''}
 {\bf U}_{\pm ,1}\coloneqq  \argmax_{{\textbf u}\in{\mathcal S}_{d-1}} \ %
\mathbb{E}\left[L^2\left({\textbf Q}_\pm(\rho{\textbf u})\right)
  \right]%
  \end{equation}
 (recall that ${\mathcal S}_{d-1}$ stands for the unit sphere in ${\mathbb R}^d$). Similarly define      the {\it $i$th nonlinear directional principal direction} of $\rm P$ (of~${{\textbf X}}\sim {\rm P}$)
 as\vspace{-3mm}
\begin{equation}\label{defUi''}
 {\bf U}_{\pm , i}\coloneqq \argmax_{\scriptscriptstyle{\begin{array}{c}\, \vspace{-11mm}\\ \scriptscriptstyle{{\mathcal S}_{d-1}\ni {\textbf u} %
\perp {\mathcal U}^+_{i-1} }\vspace{-4mm}\\
 \text{\tiny{\rm or }} \scriptscriptstyle{{\textbf u}\in -{\mathcal U}^+_{i-1}\setminus{\mathcal U}^+_{i-1} }
\end{array}}}
\hspace{-3mm}
 \mathbb{E}\left[L^2\left({\textbf Q}_\pm(\rho{\textbf u})\right)
  \right]\vspace{1mm} %
\quad i=2,\ldots,2d\vspace{-4mm}
\end{equation}
with ${\mathcal U}^+_{i}$ defined in   \eqref{Uplus},  where $\rho$ is uniform over~$[0,1]$,  and\vspace{-1mm}
\begin{equation}\label{defL}
L^2\left({\textbf Q}_\pm(r{\textbf u})\right)\coloneqq\left[\int_0^r \left\Vert \frac{\rm d}{{\rm d}t}{\textbf Q}_\pm(t{\textbf u})\right\Vert  {\rm d}t\right]^2\! ,\ r\in (0,1)\ \ \text{with}\ \ \dfrac{\rm d}{{\rm d}t}{\textbf Q}_\pm(t{\textbf u})\coloneqq \Big(\dfrac{\rm d}{{\rm d}t}{Q}_{\pm , 1}(t{\textbf u}), \ldots, \dfrac{\rm d}{{\rm d}t}{Q}_{\pm , d}(t{\textbf u})\Big)\pr\! .\vspace{-1mm}
\end{equation}
  For the sake of simplicity in exposition, $ \mathbb{E}\left[L^2\left({\textbf Q}_\pm(\rho{\textbf u})\right)
  \right]$ is assumed to be bounded uniformly in~${\textbf u\in{\mathcal S}_{d-1}}$---that assumption, analogous to the traditional assumption of finite second-order moments, is dropped (see Assumption~\ref{AssA2})  in the consistency results of Section~\ref{Sec:consist}.  Similarly  to the classical PCA assumption of distinct eigenvalues, however, we assume that no ties occur in \eqref{defUi'} and~\eqref{defUi''}.

 Note that $L^2\left({\textbf Q}_\pm(r{\textbf u})\right)$  is the   squared length of the  image    by the center-outward quantile function~${\textbf Q}_\pm$ of the interval $({\boldsymbol 0},r {\textbf u}]$ (that image is an arc of the curve ${\textbf Q}_\pm((0,1){\textbf u})$). Maximizing the expectation of~$L^2\left({\textbf Q}_\pm(\rho {\textbf u})\right)$ therefore characterizes the direction~$\textbf u$ in which the quantile map ${\textbf Q}_\pm$  yields the greatest expected squared stretching---a nonlinear analogue of maximizing variance in classical bidirectional linear principal components.  This  criterion %
 is directional (``one-sided'')  since (unlike the ``two-sided'' criterion considered in~\eqref{defU1'}--\eqref{defUi'}) it is associated with $\textbf u$, not $\pm\textbf u$.  %
  The main difference between \eqref{defU1'}--\eqref{defUi'} and \eqref{defU1''}--\eqref{defUi''}, thus, consists in substituting the expected squared curve length~$\mathbb{E}[L^2 ({\textbf Q}_\pm(\rho{\textbf u}))]$ %
    for the expected squared modu\-lus~$\mathbb{E}[\Vert
   \textbf{T}_{{\boldsymbol\Sigma},{\mathfrak f}}^{-1}(\varrho {\textbf u})\Vert ^2] $ of the random segment $\textbf{T}_{{\boldsymbol\Sigma},{\mathfrak f}}^{-1}(\varrho {\textbf u})$.  Finally, also note that, while~${\textbf Q}_\pm$ in~\eqref{defL} under finite second-order moments  is an optimal transport, $\textbf{T}_{{\boldsymbol\Sigma},{\mathfrak f}}^{-1}$ in~{\eqref{defU1'}--\eqref{defUi'}} is a non-optimal one unless $\rm P$ is spherical---in which case ${\textbf Q}_\pm=\textbf{T}_{{\boldsymbol\Sigma},{\mathfrak f}}^{-1}$ irrespective of $\mathfrak f$.

  Associated with    the $i$th principal {\it direction} $ {\bf U}_{\pm , i}$
   define the $i$th  {\it principal curve} of~$\rm P$ (of  ${{\textbf X}}\sim {\rm P}$) as the image~${\mathcal L}_{\pm ,i}\coloneqq \Big\{{\textbf Q}_\pm(r{\bf U}_{\pm,i})\vert r\in (0,1)
  \Big\}$  of the %
  radius $\{r{\bf U}_{\pm,i}\vert r\in (0,1)\}$ by the quantile function ${\textbf Q}_\pm$---%
the nonlinear and directional counterpart of the principal halfline ${\mathcal L}^+_{i;\, \text{\rm ell}}$  in~\eqref{halflines}.

Finally, define  the $i$th {\it directional nonlinear principal component} of $\rm P$ (of $\textbf{X}\sim{\rm P}$) as\vspace{-1mm} %
\textcolor{black}{%
\begin{equation}\label{eq:nonlinear-projection+}
P_{\pm , i}({\textbf X})%
\coloneqq
{\mathop{\mathlarger{\mathlarger{\mathlarger{\int}}}}}_{0}^{\|{\textbf F}_\pm(\textbf{X})\|}%
\left\langle\frac{\frac{\mathrm{d}}{\mathrm{d}t}{\textbf Q}_\pm(t\textbf{U}_{\pm , i})}{\|\frac{\mathrm{d}}{\mathrm{d}t}{\textbf Q}_\pm(t\textbf{U}_{\pm , i})\|},\frac{\mathrm{d}}{\mathrm{d}t}{\textbf Q}_\pm\left(t\frac{{\textbf F}_\pm(\textbf{X})}{\|{\textbf F}_\pm(\textbf{X})\|}\right)\right\rangle^+%
\mathrm{d}t\quad i=1,\dots,2d .\vspace{-7mm}
\end{equation}
}

The integrand,  in \eqref{eq:nonlinear-projection+} %
is  the modulus of the {positive part of} the projection of the derivative with respect to $t$
 of   $t\mapsto {\textbf Q}_\pm(t{\textbf F}_\pm({{\textbf X}})/\Vert {\textbf F}_\pm({{\textbf X}})
\Vert )$  (the ${\textbf Q}_\pm$-image of a point in the unit ball moving %
 along the radius pointing to ${\textbf F}_\pm({{\textbf X}})$) onto the tangent   to the $i$th  {\it principal curve} ${\mathcal L}_{\pm ,i}$\vspace{-1mm} computed at~${\textbf Q}_\pm\left(t{\bf U}_{\pm , i}\right)$.\vspace{-3mm} Denote by$\bigfrown{({\textbf Q}_\pm({\boldsymbol 0}), {\bf P}_{\pm , i}]}$the arc  of ${\mathcal L}_{\pm ,i}$ (running from ${\textbf Q}_\pm({\boldsymbol 0})$ to ${\bf P}_{\pm , i}$, open at ${\textbf Q}_\pm({\boldsymbol 0})$ and closed at ${\bf P}_{\pm , i}$) of  length~$P_{\pm , i}({\textbf X})$. That arc can be interpreted as a ``curvilinear projection'' of $\textbf X$ onto the halfline ${\mathcal L}_{\pm ,i}$.

When $\textbf{X}\sim \rm P$ is spherical around $\boldsymbol 0$, ${\textbf F}_\pm$ and ${\textbf Q}_\pm$ are {monotone radial transformations}, their derivatives with respect to $t$ define the %
principal halflines ${\mathcal L}_i^+$,   and the same integrand reduces to the modulus of  a linear  projection  onto that halfline. Denoting by ${\textbf u}_{{\mathcal L}_{\pm ,i}}$ the unit vector of ${{\mathcal L}_{\pm ,i}}$ and taking into account the  fact that~${\textbf Q}_\pm$ is radial monotone, we obtain
\begin{align*}
P_{\pm , i}
&=\int_{0}^{\|{\textbf F}_\pm(\textbf X)\|}
\left\langle
\frac{\frac{\mathrm{d}}{\mathrm{d}t}{\textbf Q}_\pm(t\textbf u_{\pm , i})}
     {\left\|\frac{\mathrm{d}}{\mathrm{d}t}{\textbf Q}_\pm(t\textbf u_{\pm , i})\right\|},
\frac{\mathrm{d}}{\mathrm{d}t}{\textbf Q}_\pm\!\left(t\frac{{\textbf F}_\pm(\textbf X)}
{\|{\textbf F}_\pm(\textbf X)\|}\right)
\right\rangle^+ \mathrm{d}t
=\int_{0}^{\|{\textbf F}_\pm(\textbf{X})\|}\!%
\left\langle{\textbf u}_{{\mathcal L}_{\pm ,i}},\frac{\mathrm{d}}{\mathrm{d}t}{\textbf Q}_\pm\left(t\frac{{\textbf F}_\pm(\textbf{X})}{\|{\textbf F}_\pm(\textbf{X})\|}\right)\right\rangle^+\!\!%
\mathrm{d}t \\
&=\int_{0}^{\|{\textbf F}_\pm(\textbf X)\|}
\left\langle
\textbf u_{{\mathcal L}_{\pm ,i}},
\frac{1}{f_{\mathfrak f}(F_{\mathfrak f}^{-1}(t))}\,
\boldsymbol\Sigma^{1/2}\frac{{\textbf F}_\pm(\textbf X)}{\|{\textbf F}_\pm(\textbf X)\|}
\right\rangle^+ \mathrm{d}t
=
\left\langle
\textbf u_{{\mathcal L}_{\pm ,i}},
\boldsymbol\Sigma^{1/2}
\frac{{\textbf F}_\pm(\textbf X)}{\|{\textbf F}_\pm(\textbf X)\|}
\right\rangle^+
\!\!\int_{0}^{\|{\textbf F}_\pm(\textbf X)\|}\!\!\!
\frac{1}{f_{\mathfrak f}(F_{\mathfrak f}^{-1}(t))}\mathrm{d}t\\
&=
\left\langle
\textbf u_{{\mathcal L}_{\pm ,i}},
\boldsymbol\Sigma^{1/2}
\frac{{\textbf F}_\pm(\textbf X)}{\|{\textbf F}_\pm(\textbf X)\|}
\right\rangle^+
F_{\mathfrak f}^{-1}\!\left(\|{\textbf F}_\pm(\textbf X)\|\right)
=
\left\langle
\textbf u_{{\mathcal L}_{\pm ,i}},
\frac{\textbf X}{\|\boldsymbol\Sigma^{-1/2}\textbf X\|}
\right\rangle^+
\|\boldsymbol\Sigma^{-1/2}\textbf X\|
= \left\langle\textbf u_{{\mathcal L}_{\pm ,i}},\textbf X\right\rangle^+.
\end{align*}
The integral in \eqref{eq:nonlinear-projection+}, therefore, can be interpreted as the length of a form of nonlinear projection of ${\textbf X}$ onto ${\mathcal L}_{\pm ,i}$ reducing, in the linear case, to the linear  projection of ${\textbf X}$ onto a principal halfline,
  thus justifying for $P_{\pm , i}$
 the terminology $i$th {\it nonlinear directional principal component. \vspace{-0mm}
}
\begin{figure}[b!]
    \centering
     \centering
        \subfloat{
        \includegraphics[width=0.4\textwidth]{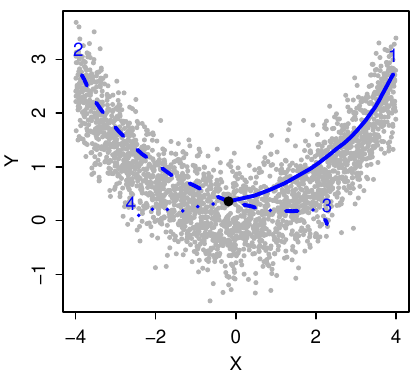}%
 }
    \vfill
     \subfloat{
        \includegraphics[width=0.65\textwidth]{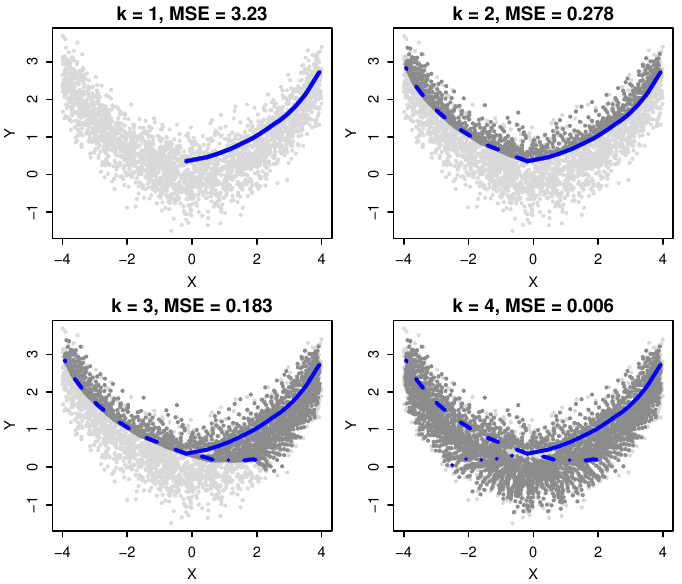}%
 }
    \caption{\small  Nonlinear directional PCA for an i.i.d.\ sample of size $n=2561$ from the banana-shaped distribution of~$({X},{Y})$, where ${X}\sim\mathrm{U}(-4,\, 4)$, ${Y}=(0.4X)^2+\varepsilon$, and $\varepsilon\sim\mathcal{N}(0,\, 0.25)$ is independent of~$X$. Top: the data with the four directional principal curves, numbered and distinguished by their line types (plain, dashed, ...) starting at the center (black dot). Bottom: the projections (dark grey) of the data (light grey) onto the manifolds~${\mathcal M}^\pm_k$, $k=1,\ldots, 4$ described in Section~\ref{sec:manifold}, together with the first~$k$ directional principal curves (blue). The corresponding MSEs are given for each~$k$; they indicate that a projection onto~$\mathcal{M}^\pm_2$ yields an excellent reconstruction of the data. The target grid has~$n_R=20$, $n_S=128$ and $n_0=1$, and the regularization parameters are as for Setting~1 in Section~\ref{simsec}.}
    \label{fig:non-linear-four-component-banana}
\end{figure}\vspace{-4mm}

\subsubsection{Nonlinear directional dimension reduction}\label{sec:manifold}

Once nonlinear directional principal directions and components have been defined, a natural idea consists in performing dimension reduction by dropping the less significant ones. As in the linear context of Section~\ref{sec:linear_motivation}, consider  some chosen number~$k<2d$ of principal directions ${\bf U}_{\pm ,1},\ldots,{\bf U}_{\pm , k}$ ordered   by decreasing values of~$\mathbb{E}[L^2 ({\textbf Q}_\pm(\rho{\textbf u}))]$ and denote by~${\mathfrak U}_k^\pm\coloneqq \{  ({\bf U}_{\pm,1},\ldots,{\bf U}_{\pm , k}){\bf r} \vert\, {\bf r}\in{\mathbb R}_+^k\}$  the positive span of these~$k$ directions.
 Since ${\textbf Q}_\pm$ is a diffeomorphism, the image by ${\textbf Q}_\pm$ of   ${\mathfrak U}_k^\pm\bigcap{\mathbb{S}}_d$    is a smooth  mani-\linebreak fold~${\mathcal M}_k^\pm\coloneqq {\textbf Q}_\pm\left({\mathfrak U}_k^\pm \bigcap{\mathbb{S}}_d\right)$. %
As in \eqref{projmathfrakU+}, dimension reduction is performed by projecting $\textbf X$ onto  ``the space of the $k$ first principal curves.'' That  ``space,'' however, now  is the manifold~${\mathcal M}_k^\pm$, which is no longer a linear space,  and the adequate projection onto ${\mathcal M}_k^\pm$ is no longer  a linear projection. For $k=1$, moreover,\vspace{-3mm} that projection  ({onto}~${\textbf Q}_\pm \big({\mathfrak U}_1^\pm\bigcap{\mathbb{S}}_d\big)$) should coincide with the arc $\bigfrown{({\textbf Q}_\pm({\boldsymbol 0}), {\bf P}_{\pm , 1}]}\!\!\!$. This is achieved\vspace{-2mm} by  the arc~$\biggfrown{({\textbf Q}_\pm({\boldsymbol 0}), {\bf P}_{\pm , {\mathcal M}_k}]}\!\!
$ of length
\begin{equation}\label{lengthcurv}
{\mathop{\mathlarger{\mathlarger{\mathlarger{\int}}}}}_{0}^{\|{\textbf F}_\pm(\textbf{X})\|}%
\left\langle\frac{\frac{\mathrm{d}}{\mathrm{d}t}{\textbf Q}_\pm(t{\bf U}_{_{\pm , {\mathcal M}_k}})}{\|\frac{\mathrm{d}}{\mathrm{d}t}{\textbf Q}_\pm(t{\bf U}_{_{\pm , {\mathcal M}_k}})\|},\frac{\mathrm{d}}{\mathrm{d}t}{\textbf Q}_\pm\left(t\frac{{\textbf F}_\pm(\textbf{X})}{\|{\textbf F}_\pm(\textbf{X})\|}\right)\right\rangle^+%
\mathrm{d}t
\end{equation}
of the halfcurve ${\textbf Q}_\pm \left(\{t{\bf U}_{_{\pm , {\mathcal M}_k}}\,
\vert\, t\in (0,1)\}\right)$, where
\begin{equation}\label{Argmaxlenth}
{\bf U}_{_{\pm , {\mathcal M}_k}}\coloneqq  \argmax_{{\textbf u}\in {\mathfrak U}_k^\pm \bigcap{\mathbb{S}}_d}{\mathop{\mathlarger{\mathlarger{\mathlarger{\int}}}}}_{0}^{\|{\textbf F}_\pm(\textbf{X})\|}%
\left\langle\frac{\frac{\mathrm{d}}{\mathrm{d}t}{\textbf Q}_\pm(t\textbf{u})}{\|\frac{\mathrm{d}}{\mathrm{d}t}{\textbf Q}_\pm(t\textbf{u})\|},\frac{\mathrm{d}}{\mathrm{d}t}{\textbf Q}_\pm\left(t\frac{{\textbf F}_\pm(\textbf{X})}{\|{\textbf F}_\pm(\textbf{X})\|}\right)\right\rangle^+%
\mathrm{d}t\vspace{-6mm}
\end{equation}
(the length \eqref{lengthcurv} of the projection $\biggfrown{({\textbf Q}_\pm({\boldsymbol 0}), {\bf P}_{\pm , {\mathcal M}_k}]}\!\!
$, thus, is the maximum achieved in  \eqref{Argmaxlenth}).

 Just as$\bigfrown{({\textbf Q}_\pm({\boldsymbol 0}), {\bf P}_{\pm , i}]}$is a ``curvilinear projection'' of $\textbf X$ onto ${\mathcal L}_{\pm ,i}= {\textbf Q}_\pm ({\bf U}_{\pm , i})$,$\biggfrown{({\textbf Q}_\pm({\boldsymbol 0}) , {\bf P}_{\pm , {\mathcal M}_k}]}\!\!
$is a ``curvilinear projection'' of~$\textbf X$  onto~$ {\mathcal M}_k^\pm =  {\textbf Q}_\pm \left({\mathfrak U}_k^\pm\bigcap{\mathbb{S}}_d\right)$.

A first illustration of nonlinear directional PCA is presented in Figure~\ref{fig:non-linear-four-component-banana}. The example considers a bivariate banana-shaped dataset obtained as an i.i.d.\ sample from the distribution of $(X,Y)$ with banana-shaped support, where $X \sim \mathrm{U}(-4,\,4)$ and $Y = (0.4X)^2 + \varepsilon$, with $\varepsilon \sim \mathcal{N}(0,\,0.25)$ independent of $X$. The corresponding directional components are associated with the manifolds $\mathcal{M}_k^{\pm}$ described in Section~\ref{sec:manifold}.
 {The top panel shows the data with the four directional principal curves. The first two curves lead from the center into the two arms of the banana and capture the overall shape of the data particularly well, while the third and fourth ones run from the center across the lower part of the distribution. The bottom panels display, for $k=1,\ldots,4$, the projections of the data onto~${\mathcal M}_k^\pm$ together with the corresponding reconstruction errors.} Based on $k=2$ components, $ {\mathcal M}_2^\pm$  provides a good representation of the banana-shaped structure, while the remaining two components mainly describe the thickness of the distribution around $ {\mathcal M}_2^\pm$, with modest contributions to the MSE. %

\subsubsection{Orthogonal invariance and equivariance} \label{sec::invariance}
Important properties of linear principal directions and components are their equivariance (for principal directions)   and invariance (for principal components) under orthogonal transformations. We show here that these properties are preserved for their nonlinear directional counterparts.

\begin{prop}\label{prop:invariance}
Denoting by  ${\boldsymbol{\boldsymbol{\mathcal O}}}$  a $d\times d$ orthogonal matrix\footnote{This matrix representation of orthogonal transformations presupposes the choice of a coordinate system for $\textbf X$---a choice which is, however, entirely arbitrary.},  let $\textbf{F}_\pm^{{\textbf X}}$ and $\textbf{F}_{\pm}^{\,\boldsymbol{\boldsymbol{\mathcal O}}\textbf{X}}$ be the \textcolor{black}{monotone}  transports pushing the distributions
 $\rm{P}^{\textbf X}\in{\mathcal P}^d_\pm$ of $\textbf{X}$ and~$\rm{P}^{{\boldsymbol{\mathcal O}}{\textbf X}}\in~\!{\mathcal P}^d_\pm$ of ${\boldsymbol{\mathcal O}}{\textbf X}$, respectively, to the uniform $\mathrm{U}_d$ over the unit ball ${\mathbb S}_d$. Using obvious notation, write  $\textbf{U}_{\pm,i}^{{\textbf X}}$,  $\textbf{U}_{\pm,i}^{\,\,\boldsymbol{\boldsymbol{\mathcal O}}\textbf{X}}$, ${\mathcal L}_{\pm ,i}^{\textbf X}$, ${\mathcal L}_{
\pm ,i}^{\,\boldsymbol{\boldsymbol{\mathcal O}}\textbf{X}}$, ${P}^{{\textbf X}}_{\pm,i}$, and ${P}_{\pm,i}^{\,\boldsymbol{\boldsymbol{\mathcal O}}\textbf{X}}$, $i=1,\dots,2d$ for the corresponding $i$th principal directions, curves, and components. Then, %
\begin{compactenum}
    \item[(i)] $\textbf{F}^{\,\boldsymbol{\boldsymbol{\mathcal O}}\textbf{X}}_{\pm}={\boldsymbol{\boldsymbol{\mathcal O}}}\textbf{F}_\pm^{\textbf X}{\boldsymbol{\boldsymbol{\mathcal O}}}^\prime$
    and $(\textbf{F}^{\,\boldsymbol{\boldsymbol{\mathcal O}}\textbf{X}}_{\pm})^{-1}
    ={\boldsymbol{\boldsymbol{\mathcal O}}}
    (\textbf{F}_\pm ^{\textbf X})^{-1}
    {\boldsymbol{\boldsymbol{\mathcal O}}}^\prime
   $; %
    \item[(ii)] $\textbf{U}_{\pm,i}^{\,\,\boldsymbol{\boldsymbol{\mathcal O}}\textbf{X}}={\boldsymbol{\boldsymbol{\mathcal O}}}\textbf{U}_{\pm,i}^{\textbf X}$, $i=1,\dots,2d$;
    \item[(iii)] $\mathcal{L}_{\pm,i}^{\,\boldsymbol{\boldsymbol{\mathcal O}}\textbf{X}}={\boldsymbol{\boldsymbol{\mathcal O}}}\mathcal{L}_{\pm,i}^{\textbf X}$, $i=1,\dots,2d$;%
    \item[(iv)] ${P}_{\pm,i}^{\,{\boldsymbol{\boldsymbol{\mathcal O}}}\textbf{X}}={P}_{\pm,i}^{\textbf{X}}$, $i=1,\dots,2d$.
\end{compactenum}
\end{prop}

 Nonlinear principal curves and principal components, just as their linear counterparts, are thus equivariant and invariant, respectively,  under orthogonal transformations. On the other hand, it is well known that classical linear principal components, despite their linear structure, are not affine-equivariant; unsurprisingly, the same conclusion applies to the nonlinear  ones.

\section{Empirical optimal-transport-based nonlinear directional principal components}\label{sec3}
\color{black} \subsection{Empirical  center-outward quantile functions %
 }\label{sec31}
Let  $\{\textbf{X}_1,\dots,\textbf{X}_n\}$ denote an i.i.d.\   sample with distribution ${\rm P}\in{\mathcal P}_\pm^d$, $d\geq 2$ and center-outward distribution and quantile functions $\textbf{F}_\pm$ and $\textbf{Q}_\pm$. We adopt the  definition of the empirical counterparts $\textbf{F}_\pm^{(n)}$ from  {\cite{Hallin2022}}. More precisely, factorizing $n$ into $n_R n_S+n_0$, where $n_R,\,n_S,\,n_0\in\mathbb{N}$ and  $0\leq n_0\leq\min\{n_R,n_S\}$, %
 we consider a {\it regular grid} ${\mathfrak G}^{(n)}$ of $n_Rn_S$ points obtained as the intersection between $n_S$ ``regularly distributed''  radii  $\{r\textbf{u}_j\vert\, r\in [0,1)\}$, $j=1,\ldots, n_S$, of ${\mathbb S}_d$ and $n_R$   hyperspheres  with radii $r_j\coloneqq\frac{j}{n_R+1}$, $j=1,\dots,n_R$,  centered at $\boldsymbol 0$, along with $n_0$ copies of the origin: see Figure~\ref{fig:banana}.

{The only requirement for the (strong) consistency of ${\textbf Q}^{(n)}_\pm$ and ${\textbf F}^{(n)}_\pm$ is the weak convergence, as $n_R$ and~$n_S$ both tend to infinity (hence~$n_0/n\to 0$), of the empirical distribution over the $n_Rn_S$ gridpoints\vspace{-8mm}\linebreak
}
 \begin{center}
\begin{figure}[h!]\hspace{10mm}
\includegraphics[trim=10mm 15mm 10mm 15mm, clip,width=7cm, height=7cm]{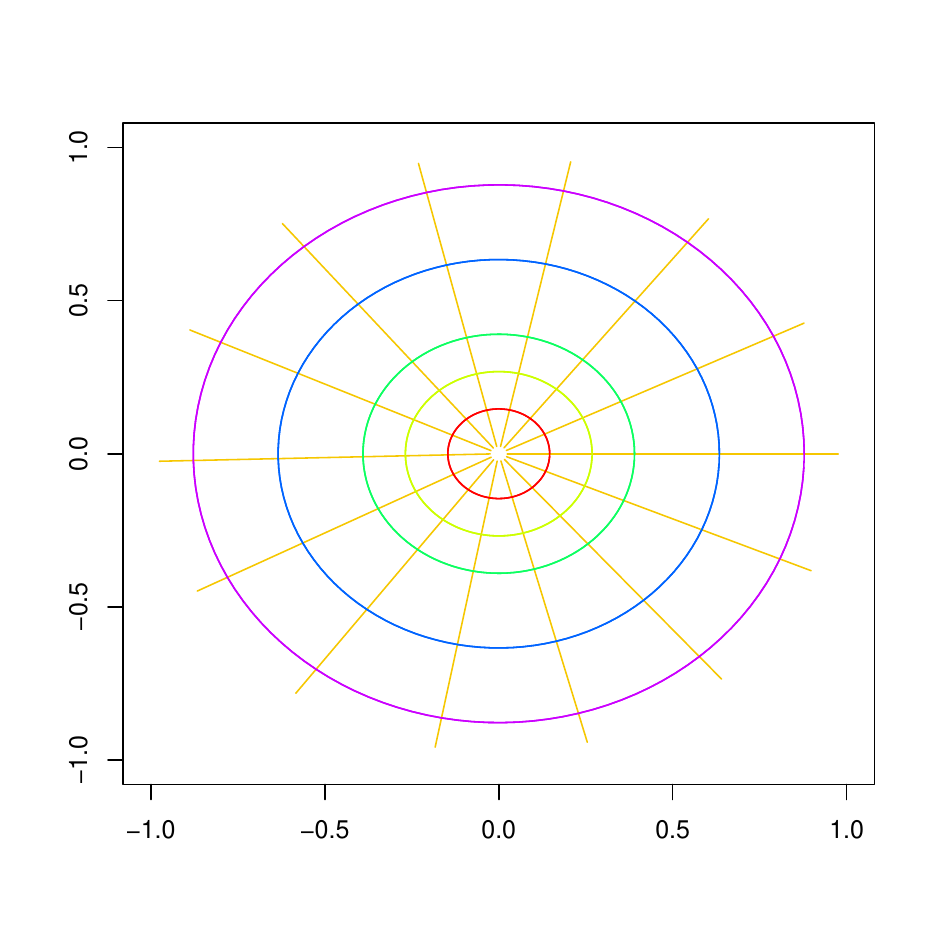}\hspace{5mm} %
\includegraphics[trim=10mm 15mm 10mm 15mm, clip,width=7cm, height=7cm]{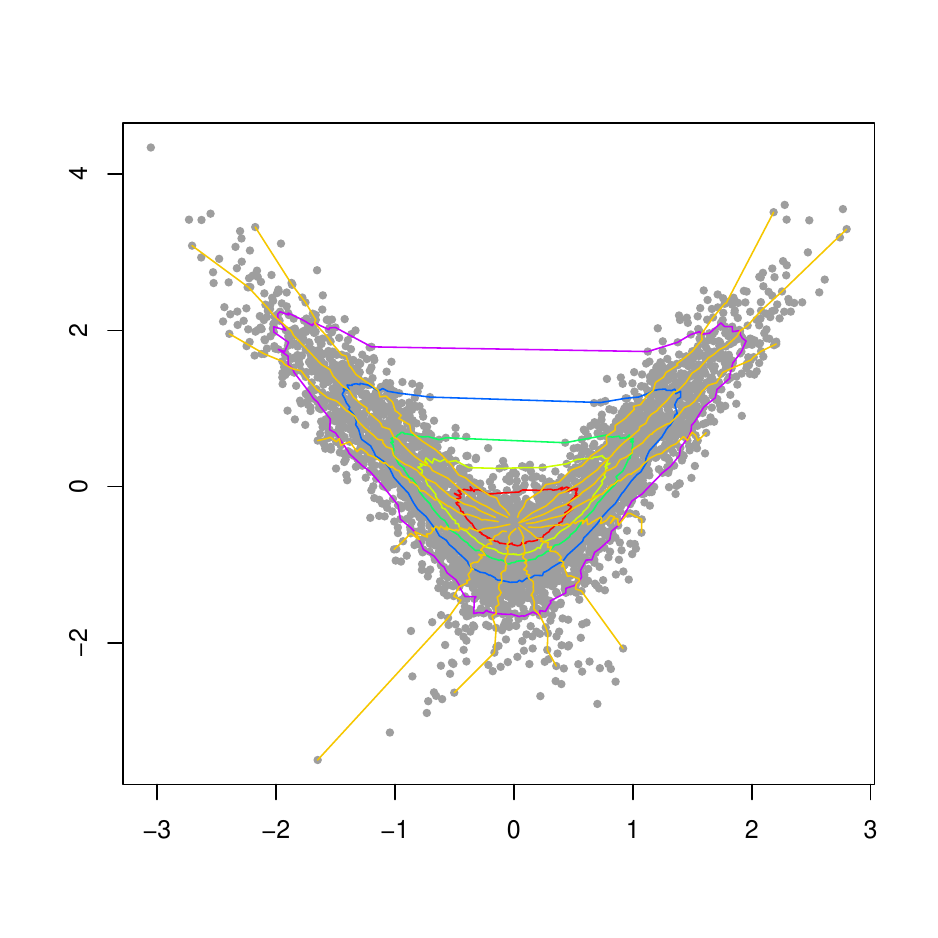} \\ \centering
 \definecolor{color1}{HTML}{FF0000}
   \definecolor{color2}{HTML}{CCFF00}
     \definecolor{color3}{HTML}{00FF66}
     \definecolor{color4}{HTML}{0066FF}
       \definecolor{color5}{HTML}{CC00FF}
\fbox{ \tiny $\tau =$  \textcolor{color1}{\textemdash} 0.146 \quad
 \textcolor{color2}{\textemdash} 0.268\quad
 \textcolor{color3}{\textemdash} 0.390 \quad
 \textcolor{color4}{\textemdash}  0.634 \quad
  \textcolor{color5}{\textemdash} 0.878\quad
 }\caption{The grid $\mathfrak G^{(n)}$ for $d=2$ (left-hand panel) and the empirical quantile contours (right-hand panel) of order $\tau =$ 0.146, 0.268, 0.390, 0.634, and~0.878  for a sample of size $n=5,000$ from a mixture of three bivariate Gaussians; $n_R=40$, $n_S=125$, and the smooth interpolation of  {\cite{Hallin2022}} described in Section~\ref{sec:smoothing} was implemented. Interpolated sign curves are shown in yellow.}
\label{fig:banana}\end{figure}
\end{center}\vspace{-12mm}

\noindent  {of~${\mathfrak G}^{(n)}$ to the  spherical uniform distribution~${\rm U}_d$ over the unit ball: ${\textbf F}\n_\pm$, then, is uniformly (over ${\mathbb R}^d$) consistent for ${\textbf F}_\pm$; while pointwise consistent (for ${\textbf Q}_\pm$) over   ${\mathbb S}_{d}$, ${\textbf Q}\n_\pm$ is   uniformly consistent over   compact subsets of ${\mathbb S}_d$.
 Consistency of ${\textbf Q}^{(n)}_\pm$ and ${\textbf F}^{(n)}_\pm$, however, is not sufficient for our purpose, and our consistency results (Theorem~\ref{Thm:Directions_Consistency}) require   an additional constraint on the relative growth rates of  $n_R$ and $n_S$ as $n\to\infty$, with~$n_S$ growing faster than $n_R$. Imposing this is not  a restriction of generality, though, as the choice of the factorization of $n$ into $n_Rn_S + n_0$ is left to the statistician.
A~similar  assumption is made by~\cite{GunsiliusSchennach2023} in their consistency results for  GS-PCA estimators; the intuitive reason is that $n_R$ is related to the ranks, which are univariate, while $n_S$ is related to the signs, which have dimension $(d-1)$. }

Now, for fixed $n$,  the gridpoints ideally should be distributed ``as regularly  as possible.''  For~$d=2$, this is achieved by equispacing the  $n_S$  points~$\textbf{u}_j$ on the unit circle. For~$d>2$, our construction involves, as suggested in  {\cite{HallinMordant2022}},   Halton sequences on the $d\times d$ unit hypercube. Without loss of generality, let~$n_S$ be even. Using the \texttt{halton} function in the~\texttt{SDraw} package~\citep{SDraw}, we generate a Halton sequence~$(v_{i,1}, \dots,v_{i,d})$, $i=1,\ldots,n_S/2$ and componentwise~perform the standard normal quantile transformation~$\Phi^{-1}$,~obtaining~$\textbf{w}_i=~\!(\Phi^{-1}(v_{i,1}),\dots,\Phi^{-1}(v_{i,d}))$.  %
 The   $n_S$  radii $\{r\textbf{u}_j\vert\, r\in [0,1)\}$ used in the construction of  ${\mathfrak G}^{(n)}$ then are determined by the $n_S$ unit vectors~$\pm\textbf{u}_i=\textbf{w}_i/\|\textbf{w}_i\|$, $i=1,\dots,n_S/2$.
 When\-ever~$n_0>0$, ${\mathfrak G}^{(n)}$ also includes $n_0$ additional  copies of the origin; call it an {\it augmented grid}. Since~$n_0/n\to0$ as~$n\to\infty$, the discrete distribution on ${\mathfrak G}^{(n)}$
 still converges weakly to  $\mathrm{U}_d$ as $n\to\infty$.

 The sample counterpart $\textbf{F}^{(n)}_\pm$ of $\textbf{F}_\pm$  then is defined at the sample points as $\textbf{F}_\pm^{(n)}(\textbf{X}_i)$, $i=1,\dots,n$ as a solution to the optimal coupling between the $n$ sample points   $\{\textbf{X}_1,\dots,\textbf{X}_n\}$ and the (augmented) grid ${\mathfrak G}^{(n)}$  just described.

   Let~$\cal T$ denote the set of all possible bijective mappings $T$ between $\{{\textbf X}_1,\ldots , {\textbf X}_n\}$ and  %
   ${\mathfrak G}^{(n)}$.
Call {\it empirical center-outward distribution function} any of the
 mappings\vspace{-2mm}
 $${\textbf F} _{\!{\pm}}^{(n)}\!\! :
 \big({\textbf X}_1,\ldots , {\textbf X}_n\big)\!\mapsto\! \big({\textbf F} _{\!{\pm}}^{(n)}({\textbf X}_1),\ldots ,{\textbf F} _{\!{\pm}}^{(n)}({\textbf X}_n)
\big) \vspace{-1mm}$$
 {\rm satisfying}
\vspace{-1mm}
$$\sum_{i=1}^n\big\Vert {\textbf X}_i - {\textbf F} _{\!{\pm}}^{(n)}({\textbf X}_i)
\big\Vert ^2
=
 \min_{T\in{\cal T}} \sum_{i=1}^n\big\Vert {\textbf X}_{i} - T({\textbf X}_i)
\big\Vert ^2
$$
{\rm and {\it empirical center-outward quantile function}    the corresponding inverse coupling $\textbf{Q}_\pm^{(n)}\coloneqq \left( \textbf{F}_\pm^{(n)}\right)^{-1}$  (which is set-valued at $\boldsymbol 0$ when $n_0>1$). }
Determining such   couplings is a standard optimal assignment problem, which takes the form of a linear program for which efficient algorithms are available (see   {\cite{PeyreCuturi19}} for a  survey of numerical methods).   The {\it empirical quantile contour} of order $\tau = r_j$ is the \linebreak collection $\{\textbf{Q}_\pm^{(n)}(r_j {\textbf u}_i)$, $i=1,\ldots,n_S\}$. The collections $\{\textbf{Q}_\pm^{(n)}(r_j {\textbf u}_i)$, $j=1,\ldots,n_R\}$ are  {\it  empirical sign curves}.

Figure~\ref{fig:banana} provides  a bivariate illustration where  the empirical quantile contours and the sign curves have been interpolated as explained in Section~\ref{sec:smoothing}.  The domain of $\textbf{Q}_\pm^{(n)}$, indeed, is the collection ${\mathfrak G}^{(n)}$ of~$n$ gridpoints, and the resulting   contours and sign curves, therefore, are collections of discrete observations.  {\cite{Hallin2022}} show how an interpolation ${\overline{\textbf{Q}}}_\pm^{(n)}$  yields a continuous extension of $\textbf{Q}_\pm^{(n)}$ with domain the open ball ${\mathbb S}_d$ which is the gradient of a convex function and still satisfies the same consistency properties as~$\textbf{Q}_\pm^{(n)}$; see Section~\ref{sec:smoothing}. The advantage of ${\overline{\textbf{Q}}}_\pm^{(n)}$ is that the interpolated sign curve $\big\{{\overline{\textbf{Q}}}_\pm^{(n)}(r{\textbf u})$, $r\in (0,1)\big\}$, unlike the sign curve~$\big\{{{\textbf{Q}}}_\pm^{(n)}(r{\textbf u})$, $r\in \{r_1=\frac{1}{n_R+1},\ldots, r_{n_R}=\frac{n_R}{n_R+1}\}\big\}$ (which is defined for ${\textbf u}\in\{{\textbf u}_1,\ldots,{\textbf u}_{n_S}\}$ only),  exists for all~${\textbf u}\in{\mathcal S}_{d-1}$ and $r\in (0,1)$. Our definition of empirical principal directions will be based on   polygonal evaluations of  arc lengths along such  curves.

These polygonal evaluations are based on grids on ${\mathbb S}_d$ of the same nature as ${\mathfrak G}^{(n)}$. However, being used for a measure-transportation-based estimator of~${\textbf Q}_\pm({\textbf X}_1),\ldots,{\textbf Q}_\pm({\textbf X}_n)$, ${\mathfrak G}^{(n)}$  consists of $n$ gridpoints while the grids to be used for the polygonal approximation of arc lengths along sign curves are not subjected to that constraint, and may involve $m\neq n$ points---typically, $m>\!>n$: the larger $m$, the better.

Starting with the definition of the  first \textit{empirical principal direction}  $ {\bf U}_{\pm,1}^{(n)}$, consider grids   ${\mathfrak G}^{(m)}_1$ of the same type as ${\mathfrak G}^{(n)}$, with $m$ factorizing {\it exactly}  into $m=m_Rm_S$  (no remainder here:  ${\mathfrak G}^{(m)}_1$ does not include the origin, where the continuity of ${\overline{\textbf{Q}}}_\pm^{(n)}$  is not guaranteed) with~$m_R=m_R(n)\to\infty$
  and $m_S=m_S(n)\to\infty$ as $n\to\infty$.   Letting $r_j\coloneqq j/(m_R+1)$, the $m$ gridpoints in  ${\mathfrak G}^{(m)}_1$   are of the form~$r_j{\textbf u}_k =
(j/(m_R+1)){\textbf u}_k$\linebreak with~$j=1,\ldots,m_R$ and $k= 1,\ldots,m_S$. Letting $\{{\textbf u}_1,\ldots,{\textbf u}_{m_S}\}\eqqcolon {\mathfrak G}^{\mathcal S}_m$,  the empirical distribution over~${\mathfrak G}^{\mathcal S}_m$ then converges weakly to the uniform over~${\mathcal S}_{d-1}$ as $n$ and hence also $m_S(n)$ converge to infinity.

 Next,
    for  $0<\delta\n<1$ with  $\delta\n\downarrow 0$ as $n\to\infty$,  %
    {define~$m_{\delta}\n \coloneqq \lfloor m_R(1-\delta\n)\rfloor$.}  This    $m_{\delta}\n $ will play the role of a sequence of trimming levels ensuring, for all $n$,  $r_{m_{\delta}\n }\leq (1-\delta\n)$ and~$r_{m_{\delta\n+1}}>(1-\delta\n)$.   This trimming is performed for two reasons: (i) it reduces the impact of  {\it extreme} quantile points~($ {\overline{\textbf{Q}}}_\pm^{(n)}(r_j{\textbf u})$ with~$r_j\approx 1$), and (ii) some amount of trimming is required for consistency in the absence of moment assumptions;   a similar assumption is also made by~\cite{GunsiliusSchennach2023} in their  asymptotics section.

Based on ${\overline{\textbf{Q}}}_\pm^{(n)}$, ${\mathfrak G}^{(m)}_1$,  and $m_{\delta}\n $,  let us  define the first \textit{empirical principal direction}  $ {\bf U}_{\pm,1}^{(n)}$ as the (trimmed) empirical version of   of the population concept \eqref{defU1''}:
   \begin{equation}\label{defU1n'}
 {\bf U}_{\pm,1}^{(n)}\coloneqq  \argmax_{{\textbf u}\in{\mathfrak G}^{\mathcal S}_m = \{\textbf{u}_1,\dots,\textbf{u}_{m_S}\}} \frac{1}{m_{\delta}\n } %
\sum_{j=\textcolor{black}{2}}^{m_{\delta}\n } L_n^2\left({\overline{\textbf{Q}}}_\pm^{(n)}(r_j{\textbf u})\right)
  \end{equation}
 with
     \begin{equation}\label{defLn}
L_n^2\left({\overline{\textbf{Q}}}_\pm^{(n)}(r_j{\textbf u})\right)\coloneqq \Bigg\{%
\sum_{l=\textcolor{black}{2}}^j\left\Vert
{\overline{\textbf{Q}}}_\pm^{(n)}(r_l{\textbf u}) - {\overline{\textbf{Q}}}_\pm^{(n)}(r_{l-1}{\textbf u})
\right\Vert\Bigg\} ^2,\quad j=1,\dots,m_{\delta}\n .
\end{equation}

We similarly define the $i$th empirical principal direction ($i=2,\ldots,2d-1$)  as a trimmed empirical version of~\eqref{defUi''}.
This  requires  grids ${\mathfrak G}_{i}^{(m^i)}$ and ${\mathfrak G}^{\mathcal S}_{m^i}$ of the same form as ${\mathfrak G}^{(m)}_1$
and ${\mathfrak G}^{\mathcal S}_m$, now over the subsets
$$
{\mathcal U}^{(n)\perp}_{\pm,\, i-1}\bigcup \Big(\!-{\mathcal U}^{(n)}_{\pm,\, i-1}\setminus {\mathcal U}^{(n)}_{\pm,\, i-1}\Big)\coloneqq \left\{ {\textbf u}\in{\mathcal S}_{d-1} \vert \, {\textbf u}\perp {\bf U}\n_{\pm ,\, j},\  j=1,\ldots, i-1
\right\} \bigcup \Big\{-{\mathcal U}^{(n)}_{\pm,\, i-1}\setminus {\mathcal U}^{(n)}_{\pm,\, i-1}\Big\}
$$
of ${\mathcal S}_{d-1}$ and the corresponding radii
of  ${\mathbb S}_d$,
with values of $m^i$, $m_R^i$,  and $m_S^i$ that can depend on $i$. \linebreak%
  Letting~$r_j\coloneqq j/(m_R^i+1)$, the $m^i$ gridpoints in  ${\mathfrak G}^{(m)}_i$   are of the form~$r_j{\textbf u}_k=
(j/(m_R^i+1)){\textbf u}_k$\linebreak  with~${\textbf u}_k\in  {\mathfrak G}^{\mathcal S}_{m^i}\subset {\mathcal U}^{(n)\perp}_{\pm,\, i-1}\bigcup \big(\!-{\mathcal U}^{(n)}_{\pm,\, i-1}\coloneqq  \{{\textbf u}_1,\ldots, {\textbf u}_{m^i_S}\}\setminus {\mathcal U}^{(n)}_{\pm,\, i-1}\big)$, $k= 1,\ldots,m_S^i$  and~$j=1,\ldots,m_R^i$. %
As in~\eqref{defU1n'}, further choose a sequence
 $m_{\delta ;i}\n $ %
  of trimming levels. %
The  $i$th ($i=2,\ldots, 2d$) {\it empirical principal direction} then~is \vspace{-5mm}
\begin{equation}\label{defUin'}
{\bf U}_{\pm, i}^{(n)}\coloneqq \hspace{0mm}
\argmax_{{\textbf u}\in {\mathfrak G}^{\mathcal S}_{m^i}=\{\textbf{u}_1,\dots,\textbf{u}_{m_S^i}\}}
\frac{1}{m_{\delta ;i}\n }\sum_{j=\textcolor{black}{2}}^{m_{\delta ;i}\n } L_n^2\left({\overline{\textbf{Q}}}_\pm^{(n)}(r_j{\textbf u})\right) ,
\end{equation}
where %
 $L_n^2\left({\overline{\textbf{Q}}}_\pm^{(n)}(r_j{\textbf u})\right)$ is defined as in \eqref{defU1n'}.

{In practice, the maximization in \eqref{defUin'} (with $i\geq 2$) requires the computation of $\overline{{\textbf Q}}\n_{\pm}$ over  $m^i_S$ grids  of~$m^i_R$ points each.
    These grids do not require any special structure but, if consistency is to be achieved, they should be asymptotically dense as $n\to\infty$: we thus assume that  $m^i_S$ and $m^i_R$  tend to infinity as $n\to\infty$, with the empirical distribution over  ${\mathfrak G}^{\mathcal S}_{m^i}$ converging weakly  to the uniform over~${\mathcal S}_{d-1}$. %
  }

\color{black}

Associated with ${\bf U}_{\pm, i}^{(n)}$ ($i=1,\ldots,2d$), define  the $i$th {\it empirical principal curve} %
as~${\mathcal L}_{\pm , i}^{(n)}\coloneqq {\mathcal L}_{\pm}^{(n)}({\bf U}_{\pm,i}^{(n)})$,   where\vspace{-1mm} %
   \begin{equation}\label{eq:PrincipalCurves_sample}
  {\mathcal L}_{\pm}^{(n)}({\textbf u})\coloneqq \left\{ \overline{\textbf{Q}}_\pm^{(n)}(r_j{\textbf u}): j\in \{1,\dots,m^i_R\}
  \right\},\quad {\textbf u}\in {\mathcal S}_{d-1}.
    \end{equation}
This empirical principal curve %
 is defined  at the $m^i_R$-tuple $\{r_j {\bf U}_{\pm,i}^{(n)}:j=1,\dots,n_R^i\}$  only (a discrete collection of $m_R^i$ points, thus) but naturally extends  to the image by $\overline{\textbf{Q}}_\pm^{(n)}$ of the whole~$(0,m^i_R/(m^i_R+1)]$ interval:%
\begin{equation}\label{eq:PrincipalCurves_smooth}
 \overline{{\mathcal L}}_{\pm , i}^{(n)}\coloneqq
 \overline{{\mathcal L}}_{\pm}^{(n)}({\bf U}_{\pm,i}^{(n)})\coloneqq  \overline{\textbf{Q}}_\pm^{(n)}\big( (0,m^i_R/(m^i_R+1)]{\bf U}_{\pm,i}^{(n)})
\big).
\end{equation}%

Finally, for $\textbf{X}_j\in\{\textbf{X}_1,\dots,\textbf{X}_n\}$ such that $\|\textbf{F}_\pm^{(n)}(\textbf{X}_{{j}})\| \eqqcolon  r_{l_{j}} >0 $, we define the empirical counterpart~$P_{\pm , k}^{(n)}({\textbf X}_{{j}})$ of~$\textbf{X}_j$'s $k$th principal component  {($k=1,\ldots,2d$)} introduced in \eqref{eq:nonlinear-projection+} as
\begin{equation}\label{eq:sample_nonlinear-projection+}
    P_{\pm , k}^{(n)}({\textbf X}_{{j}})%
\coloneqq
\sum_{l=1}^{l_{{j}}}%
\left\langle
\frac{\overline{\textbf Q}_\pm^{(n)}(r_{{{l}}}\textbf{U}_{\pm , k}^{(n)})
-
\overline{\textbf Q}_\pm^{(n)}(r_{{{l}}-1}\textbf{U}_{\pm , k}^{(n)})}{\|\overline{\textbf Q}_\pm^{(n)}(r_{{l}}\textbf{U}_{\pm , k}^{(n)})
-
\overline{\textbf Q}_\pm^{(n)}(r_{{{l}}-1}\textbf{U}_{\pm , k}^{(n)})\|}
,
{\textbf Q}_\pm^{(n)}
\Big(
r_{{l}}
\frac
{{\textbf F}_\pm^{(n)}
(
\textbf{X}_{
{j}
}
)
}
{
r_{l_{
{j}
}}}
\Big)
-
{\textbf Q}_\pm^{(n)}
\Big(r_{{{l}}-1}\frac{{\textbf F}_\pm^{(n)}(\textbf{X}_{{j}})}{r_{l_{{j}}}}
\Big)
\right\rangle^+\!\!\!\!\!
\end{equation}
whereas, for $\textbf{F}_\pm^{(n)}(\textbf{X}_j)=\textbf{0}$, we let $P_{\pm,k}^{(n)}(\textbf{X}_j)\coloneqq 0$, $k=1,\dots,2d$.

\subsection{Nonlinear dimension reduction}\label{dimreducemp}

Consider a chosen number $k<2d$ of empirical principal directions
${\bf U}_{\pm ,1}^{(n)},\ldots,{\bf U}_{\pm , k}^{(n)}$, ordered by decreasing values of $%
 {\frac{1}{m_{\delta ;i}\n }\sum_{j=2}^{m_{\delta ;i}\n }} L_n^2\big(\overline{\textbf Q}_\pm^{(n)}(r_j{\textbf u})\big)$---namely, $%
 {\frac{1}{m_{\delta; 1}\n}\sum_{j=2}^{m_{\delta; 1}\n}} L_n^2\big(\overline{\textbf Q}_\pm^{(n)}(r_j{\bf U}_{\pm,1}^{(n)})\big)
\geq \cdots \geq
{\frac{1}{m_{\delta;k}\n }\sum_{j=2}^{m_{\delta;k}\n }} L_n^2\big(\overline{\textbf Q}_\pm^{(n)}(r_j{\bf U}_{\pm,k}^{(n)})\big).$
We perform empirical dimension reduction by mimicking the population definitions of Section~\ref{sec:manifold}, that is, via a ``nonlinear projection'' onto the manifold ``spanned by''
$\overline{\textbf Q}_\pm^{(n)}({\bf U}_{\pm,1}^{(n)}),\dots,\overline{\textbf Q}_\pm^{(n)}({\bf U}_{\pm,k}^{(n)})$.

For $k=1$, this projection coincides with the arc $\bigfrown{\big(\overline{\textbf Q}_\pm^{(n)}({\boldsymbol 0}),\, {\bf P}_{\pm , 1}^{(n)}(\textbf{X}_i)\big]}$ of length $P_{\pm,1}^{(n)}(\textbf{X}_i)$ along the empirical principal curve $ \overline{{\mathcal L}}_{\pm , {1}}^{(n)}$. For $k\geq 2$, define the positive span of the first $k$ empirical principal directions
$$
{\mathfrak U}_{\pm,\, k}^{(n)}
\coloneqq \left\{ ({\bf U}_{\pm,1}^{(n)},\ldots,{\bf U}_{\pm , k}^{(n)}){\bf r} \;\middle|\; {\bf r}\in{\mathbb R}_+^k \right\}
\cap{\mathbb S}_d,
$$
and the associated nonlinear manifold
${\mathcal M}_{\pm,\, k}^{(n)}
\coloneqq
\overline{\textbf Q}_\pm^{(n)}\big({\mathfrak U}_{\pm ,\, k}^{(n)}\big).
$
 For each observation $\textbf{X}_i$, the nonlinear projection is determined by selecting a direction $
{\bf U}_{\pm , {\mathcal M}_k^{(n)}}^{(n)}(\textbf{X}_i)
\in {\mathfrak U}_{\pm,\, k}^{(n)}$ %
in the span ${\mathfrak U}_{\pm,\, k}^{(n)}$ of the first~$k$ principal directions for which the nonlinear projection score (in the sense of \eqref{eq:sample_nonlinear-projection+}) is maximal. More precisely,
\begin{equation}\label{Argmaxlenth_n}
{\bf U}_{\pm , {\mathcal M}_k^{(n)}}^{(n)}(\textbf{X}_i)
\coloneqq
\argmax_{{\textbf u}\in {\mathfrak U}_{\pm,\,k}^{(n)}}
\sum_{j=1}^{l_i}
\left\langle
\frac{\overline{\textbf Q}_\pm^{(n)}(r_{j}{\textbf u})-\overline{\textbf Q}_\pm^{(n)}(r_{j-1}{\textbf u})}
{\Big\|\overline{\textbf Q}_\pm^{(n)}(r_{j}{\textbf u})-\overline{\textbf Q}_\pm^{(n)}(r_{j-1}{\textbf u})\Big\|},
\overline{\textbf Q}_\pm^{(n)}\Bigg(r_j\frac{{\textbf F}_\pm^{(n)}(\textbf{X}_i)}{r_{l_i}}\Bigg)
-
\overline{\textbf Q}_\pm^{(n)}\Bigg(r_{j-1}\frac{{\textbf F}_\pm^{(n)}(\textbf{X}_i)}{r_{l_i}}\Bigg)
\right\rangle^+.
\end{equation}

Here, $l_i$ is the index such that
$\|{\textbf F}_\pm^{(n)}(\textbf{X}_i)\| = r_{l_i} > 0,$
so that the sequence $(r_j)_{j=\textcolor{black}{1}}^{l_i}$ discretizes the radial path associated with $\textbf{X}_i$. The corresponding projected point is then defined as
$\displaystyle {\bf P}_{\pm , {\mathcal M}_{\pm ,\, k}^{(n)}}^{(n)}(\textbf{X}_i)
\in {\mathcal M}_{\pm,\, k}^{(n)},$
namely, the point on the curve
$\overline{\textbf Q}_\pm^{(n)}\big(\{t\,{\bf U}_{\pm , {\mathcal M}_k^{(n)}}^{(n)}(\textbf{X}_i) \mid t\in (0,1)\}\big)$ such that the length of the arc running   from~$\overline{\textbf Q}_\pm^{(n)}\Big({\textcolor{black}{\frac{1}{m^k_R +1}{\bf U}_{\pm , {\mathcal M}_k^{(n)}}^{(n)}(\textbf{X}_i) }}\Big)$ to~${\bf P}_{\pm , {\mathcal M}_{\pm ,\, k}^{(n)}}^{(n)}(\textbf{X}_i)$ equals the maximal nonlinear projection score. This arc length is %
\begin{equation}\label{lengthcurv_n}
\sum_{j=1}^{l_i}
\left\langle
\frac{\overline{\textbf Q}_\pm^{(n)}(r_{j}{\bf U}_{\pm , {\mathcal M}_k^{(n)}}^{(n)}(\textbf{X}_i))
-
\overline{\textbf Q}_\pm^{(n)}(r_{j-1}{\bf U}_{\pm , {\mathcal M}_k^{(n)}}^{(n)}(\textbf{X}_i))}
{\Big\|\overline{\textbf Q}_\pm^{(n)}(r_{j}{\bf U}_{\pm , {\mathcal M}_k^{(n)}}^{(n)}(\textbf{X}_i))
-
\overline{\textbf Q}_\pm^{(n)}(r_{j-1}{\bf U}_{\pm , {\mathcal M}_k^{(n)}}^{(n)}(\textbf{X}_i))\Big\|},
\,
\overline{\textbf Q}_\pm^{(n)}\!\Bigg(r_j\frac{{\textbf F}_\pm^{(n)}(\textbf{X}_i)}{r_{l_i}}\Bigg)
-
\overline{\textbf Q}_\pm^{(n)}\!\Bigg(r_{j-1}\frac{{\textbf F}_\pm^{(n)}(\textbf{X}_i)}{r_{l_i}}\Bigg)
\right\rangle^+.
\end{equation}

As in the population case, the value in \eqref{lengthcurv_n} coincides with the maximum in \eqref{Argmaxlenth_n} and represents the nonlinear projection score of $\textbf{X}_i$ onto the manifold ${\mathcal M}_{\pm,\, k}^{(n)}$.

\medskip

{The selection criterion, in \eqref{defU1n'} and \eqref{defUin'}, is based on a polygonal evaluation of empirical principal curve lengths, that is, involves a linear interpolation of  the $m_R^i$  $\overline{\textbf{Q}}_\pm^{(n)}(r_j\textbf{u})$ values for $m_S^i$ values of $\textbf u$.   Depending on the sample size $n$, that approximation can be poor and quite wiggly, yielding over\-estimated~$L_n^2\left(\overline{\textbf{Q}}_\pm^{(n)}(r_j{\textbf u})\right)$ values in  {\eqref{defLn}}. Some regularization, then, is desirable, and a Tikhonov regularization strategy is discussed in Section~\ref{sec:smooth} of the Appendix.  }

\color{black}

\subsection{Smooth extensions of  principal curves}\label{sec:smoothing}
For  ${\rm P}\in\mathcal{P}_\pm^d$,
$\textbf{Q}_\pm$ is a {homeomorphism}
  between the punctured unit ball~${\mathbb S}_d\setminus\{{\boldsymbol 0}\}$ and  supp$({\rm P})\setminus \textbf{Q}_\pm(\{{\boldsymbol 0}\})$  {\citep{Fig18}}, hence is smooth in ${\mathbb S}_d\setminus\{{\boldsymbol 0}\}$. This  implies that the mappings $(0,1)\ni r\mapsto {\textbf Q}_\pm(r{\bf U}_i)$, $i=1,\ldots, 2d$ are  {homeomorphisms}  as well, so that the corresponding principal curves ${\mathcal L}_{\pm ,i}$  are smooth.

The empirical center-outward distribution function, however, is only defined at the observed\linebreak  points~$\textbf{X}_1,\dots,\textbf{X}_n$, and the empirical principal curves $ {\mathcal L}_{\pm}^{(n)}({\bf U}\n_{\pm ,\, i})$ defined in~\eqref{eq:PrincipalCurves_sample} (as well as their regularized counterparts in Section~\ref{sec:smooth})  are just collections of discrete points. These collections can be turned  into a smooth curve~$\overline{{\mathcal L}}_{\pm}^{(n)}({\bf U}\n_{\pm ,\, i})$  by interpolating   the empirical center-outward quantile function into a smooth function $\overline{\textbf{Q}}_\pm^{(n)}$,  naturally defining the smooth extensions \eqref{eq:PrincipalCurves_smooth} of the discrete, empirical principal curves~\eqref{eq:PrincipalCurves_sample}.%

The interpolation $\overline{\textbf{Q}}_\pm^{(n)}$ of   $\textbf{Q}_\pm^{(n)}$ proposed in~\cite{Hallin2022}  is based on  a smooth interpolation~$\overline{\textbf{F}}_\pm^{(n)}$ of the empirical center-outward distribution function ${\textbf{F}}_\pm^{(n)}$ within the class of all gradients of convex functions mapping $\mathbb{R}^d$ to $\mathbb{S}_d$. Gradients of convex functions are characterized by the property of \textit{cyclical monotonicity} (see~\cite{Hallin2022} or  {\cite{Rockafellar1}} for definitions) and  such interpolations are, in some sense, the $d$-dimensional and center-outward counterpart of a smooth and monotone  non-decreasing interpolation of the $n$ points~$(\text{X}_i, {F}^{(n)}(\text{X}_i))$, $i=1,\dots,n$ in a univariate empirical process (with $F^{(n)}$ the traditional univariate empirical distribution function). %
Their construction   in~\cite{Hallin2022} proceeds
in two steps. First,~$\textbf{F}_\pm^{(n)}$ is extended to a piecewise constant cyclically monotone map deﬁned a.e.\ in $\mathbb{R}^d$. Being piecewise constant, that map cannot be smooth, and a regularization   based on  {\it Moreau envelopes} and {\it Yoshida regularization}~\citep{Yoshida} is performed,  yielding the required smoothness while preserving  interpolation features. The resulting $\overline{\textbf{F}}_\pm^{(n)}$ is also Lipschitz-continuous and enjoys Glivenko--Cantelli consistency, viz.
$$
\sup_{\textbf{x}\in\mathbb{R}^{d}}\|\overline{\textbf{F}}_\pm^{(n)}(\textbf{x})-\textbf{F}_\pm(\textbf{x})\| \to 0,\,\, \text{a.s.\ as }n\to\infty.
$$
The resulting $\overline{\textbf{F}}_\pm^{(n)}$ admits a smooth inverse $\overline{\textbf{Q}}_\pm^{(n)}$ on ${\mathbb S}_d\setminus\{{\boldsymbol 0}\}$, thus yielding a smooth interpolation  of ${\textbf{Q}}_\pm^{(n)}$. Such interpolation is strongly consistent for ${\textbf{Q}}_\pm$, but   uniform consistency typically only holds over compact subsets of ${\mathbb S}_d\setminus\{{\boldsymbol 0}\}$.

\subsection{{Comparison with GS-PCA  {\citep{GunsiliusSchennach2023}}}}\label{SecG&S}

Although the GS-PCA approach proposed by Gunsilius and Schennach (2023) is also based on measure transportation, it significantly differs from ours.

At  population level, GS-PCA  {\citep{GunsiliusSchennach2023}} first considers the optimal\footnote{Optimality requires finite second-order moments, and so do  {\citep{GunsiliusSchennach2023}}. However, that assumption can be dispensed with if optimality is relaxed to (cyclical) monotonicity: see   {\cite{McCann1995}}.} trans\-port~$\textbf{T}_{\text{\tiny GS}}$ from~$\mathbb{R}^d$ to $\mathbb{R}^d$  pushing the distribution $\rm P$ of the random vector $\textbf{X}$ with density $f$ forward to the spherical standard normal distribution~${\cal N}(\boldsymbol{0}, {\bf I}_d)$. Next, in order to obtain a $k$-dimensional ($k\leq d$) representation of~$\textbf{X}$, they select a $d$-tuple of mutually orthogonal unit vectors $({\textbf u}_{\text{\tiny GS};1},\ldots, {\textbf u}_{\text{\tiny GS};d})$  by sequentially maximizing an entropy criterion and show that these vectors coincide with the $d$  eigenvectors of the\linebreak  matrix~$\overline{\bf J}\coloneqq -\int_{{\mathbb R}^d}f({\bf x})\ln ({\bf J}({\bf x})){\rm d}{\bf x}$,
 where $\ln ({\bf J}({\bf x}))$
 is the matrix logarithm\footnote{Recall   {\citep{HornJohnson91}} that
 $ \ln ({\bf P}
  {\text{diag}}
 (\lambda_1,\ldots,\lambda_d){\bf P}^{-1})
 \coloneqq  {\bf P}{\text{diag}}(\ln \lambda_1,\ldots,\ln\lambda_d){\bf P}^{-1}$.} of the Jacobian of $\textbf{T}_{\text{\tiny GS}}$ computed at ${\bf x}$. This defines $d$  {\it principal curves} ${\mathcal L}_{\text{\tiny GS};i}\coloneqq\big\{\textbf{T}_{\text{\tiny GS}}^{-1}\left( r{\textbf u}_{\text{\tiny GS};i}
 \right)\vert r\in (-1,1)
  \big\}$, $i=1,\ldots,d$, to be compared with our $2d$ principal curves ${\mathcal L}_{\pm ,i}$, $i=1,\ldots, 2d$, and $d$ manifolds ${\mathcal M}_k^{\text{\tiny GS}}\coloneqq \textbf{T}_{\text{\tiny GS}}^{-1}\left(
   \{  ({\textbf u}_{\text{\tiny GS};1},\ldots,{\textbf u}_{\text{\tiny GS};k}){\bf r} \vert\, {\bf r}\in{\mathbb R}^k\}
  \right)$ to be compared with our $2d$ manifolds ${\mathcal M}_k^\pm$, $k=1,\ldots,2d$. The  $k$-dimensional (${\mathcal M}_k^{\text{\tiny GS}}$-valued) representation of $\textbf X$ they propose  is
 $\textbf{T}_{\text{\tiny GS}}^{-1}\left(\sum_{j=1}^k \langle {\textbf X}, {\textbf u}_{\text{\tiny GS};j}
 \rangle  {\textbf u}_{\text{\tiny GS};j}
 \right).
 $

At sample level (based on a sample ${\textbf X}_1,\ldots, {\textbf X}_n$  of $n$ independent copies of ${\textbf X}\sim{\rm P}\in{\mathcal P}^d_\pm$), Gunsilius and Schennach first compute a kernel-based  estimator $f^{(n)}$ of the density $f$ of $\rm P$. That  $f^{(n)}$ allows them   to obtain an estimate   of  $\textbf{T}_{\text{\tiny GS}}$   that is defined at every point of the support of $\rm P$, and is required in the  evaluation of the entropy criterion used in the selection of  empirical versions of $({\textbf u}_{\text{\tiny GS};1},\ldots, {\textbf u}_{\text{\tiny GS};d})$.  Kernel-based density estimation, however, is not without its pitfalls,  particularly as $d$ increases.

A similar strategy could be adopted within our framework. Since the quality of our estimators depends on the target grid being sufficiently dense, one could generate a larger sample of size $n_{\mathrm{new}} \gg n$ from the estimated density $f^{(n)}$, with $n_{\mathrm{new}} = n_{\mathrm{new},\, S} n_{\mathrm{new},\, R} + n_{\mathrm{new},\, 0}$ for $n_{\mathrm{new},\, S}$ and $n_{\mathrm{new},\, R}$  sufficiently large. However, {in addition to the pitfalls noted above,} kernel density estimation is itself computationally intensive, and we therefore sample from the empirical distribution instead. In the real-data examples of Section~\ref{sec::realdata}, GS-PCA and the other competing methods are applied to the original observations, and this resampling is used for our method only (for the Old Faithful and Gvessel data).

Although their approach  to nonlinear dimension reduction and ours both yield a representation of~$\textbf X$  on a lower-dimensional manifold,  they also differ on several crucial points.  First, while the GS-PCA transports~$\textbf{T}_{\text{\tiny GS}}$ and~$\textbf{T}_{\text{\tiny GS}}^{-1}$ are to and from the spherical Gaussian, ours (${\textbf T}_\pm={\textbf F}_\pm$ and ${\textbf T}_\pm^{-1}={\textbf Q}_\pm$) are to and from the spherical uniform, which better matches the properties expected from a distribution and a quantile  function (see  {\cite{Hallin2022a}}).  Second, being based on the {eigenvectors} of $\overline{\bf J}$,  GS-PCA remains bidirectional rather than directional.
The main difference, however, lies in the   selection of the principal  curves ${\mathcal L}_{\pm ,i}$ and ${\mathcal L}_{\text{\tiny GS};i}$. Our  cri\-terion~$\mathbb{E}\left[L^2\left({\textbf Q}_\pm(\rho{\textbf u})\right)\right]$  is the expected squared length of a curve, yielding an evaluation of the expected strain of the quantile mapping ${\textbf T}_\pm^{-1}={\textbf Q}_\pm$ along~${\textbf u}$---an expectation  that only involves  the points in $\{r{\textbf u}\vert\, r\in [0,1)\}$ and their image by ${\textbf T}_\pm^{-1}={\textbf Q}_\pm$, that is,
the corresponding   sign curve.
This is very much in the spirit of the  traditional concepts based on variance maximization  along~${\textbf u}$, a variance which is also an  expected  squared length.
In contrast, Gunsilius and Schennach are selecting directions based on a form of expected log-strain of  $\textbf{T}_{\text{\tiny GS}}^{-1}$ that involves  the entire space.

\section{Large-sample asymptotics}\label{Sec:consist}

In this section, we are dealing with  the consistency of the %
 empirical quantities introduced in Section \ref{sec31}.

Throughout this section, the following assumptions are made.

\begin{assumption}\label{AssA1}{\rm The observations ${\textbf X}_1,\ldots,{\textbf X}_n$ are i.i.d., with distribution ${\rm P}\in{\mathcal P}^d_\pm$ and density $f\in C^{0,\alpha}_{\mathrm{loc}}(\mathbb{R}^d)$ for some $\alpha\in(0,1)$.\footnote{Recall that a function $f$ belongs to the class $C_\mathrm{loc}^{k,\alpha}(\mathbb{R}^d)$,   $k\geq 0$ and $\alpha\in(0,1)$,  if $f\in C^k(\mathbb{R}^d)$ and its $k$th derivative $\nabla^k f$ is locally $\alpha$-Hölder-continuous, i.e.\  $
\sup_{x,y\in \mathbb{S}(R),\, x\neq y} {\|\nabla^kf(x)-\nabla^kf(y) \|}/{\|x-y\|^\alpha}<\infty
$
for every $R>0$. \textcolor{black}{For $k=0$, this reduces to  local $\alpha$-Hölder-continuity of $f$.}}
}
\end{assumption}

\begin{assumption}\label{AssA2}{\rm  There exist~$\varepsilon_0,\delta_0 >0$ such that, for all $\delta_1\in (0,\varepsilon_0]$, $\delta_2 \in (0,\delta_0]$, ${\textbf u}^\prime$, and ${\textbf u}^{\prime\prime}$ in ${\mathcal S}_{d-1}$,
  $$\mathbb{E}\left[\left(L^{(\delta_1)}\left({\textbf Q}_\pm(\rho{\textbf u}^\prime)\right)\right)^2I[\delta_1\leq\rho\leq (1-\delta_2)]\right]
  \leq
  \mathbb{E}\left[\left(L^{(\delta_1)}\left({\textbf Q}_\pm(\rho{\textbf u^{\prime\prime}})\right)\right)^2I[\delta_1\leq\rho\leq (1-\delta_2)]\right]
  $$
  iff
$$\mathbb{E}\left[\left(L^{(\varepsilon_0)}\left({\textbf Q}_\pm(\rho{\textbf u}^\prime)\right)\right)^2I[\varepsilon_0\leq\rho\leq (1-\delta_0)]\right]
  \leq
\mathbb{E}\left[\left(L^{(\varepsilon_0)}\left({\textbf Q}_\pm(\rho{\textbf u^{\prime\prime}})\right)\right)^2I[\varepsilon_0\leq\rho\leq (1-\delta_0)]\right]
,$$
where \textcolor{black}{$\rho$ is uniformly distributed  over $[0,1]$} and
$$
L^{(\delta_1)}(\textbf{Q}_\pm(r\textbf{u})) \coloneqq  \int_{\delta_1}^r \left\|\frac{\mathrm{d}}{\mathrm
dt}\textbf{Q}_\pm(t\textbf{u})%
\right\|\,\mathrm{d}t,
\qquad r\in[\delta_1,1).
$$
is the trimmed-at-zero length of $r\mapsto \textbf{Q}_\pm(r\textbf{u})$.}
\end{assumption}

Under this assumption, the ordering of the expected squared trimmed arc lengths over the truncated region $\delta_1\leq \rho \leq (1-\delta_2)$, with $\delta_1\leq \varepsilon_0$ and $\delta_2\leq \delta_0$, is completely determined by the ordering corresponding to the fixed trimming and truncation levels $\varepsilon_0$ and $\delta_0$.
Replacing in~\eqref{defU1''}--\eqref{defUi''} the non-trimmed and non-truncated expectations with the expectations based on the trimmed-at-zero arc lengths~$
\mathbb{E}\left[\left(L^{(\varepsilon_0)}\left({\textbf Q}_\pm(\rho{\textbf u})\right)\right)^2
I[\varepsilon_0\leq\rho\leq (1-\delta_0)]\right]
$
yields a ``new'' definition of the population principal directions which avoids requiring boundedness of~$
\sup_{{\textbf u}\in{\mathcal S}_{d-1}}
\mathbb{E}\left[L^2\left({\textbf Q}_\pm(\rho{\textbf u})\right)\right],
$
while still coinciding with the ``old'' definition whenever boundedness holds. Bounded $
\sup_{{\textbf u}\in{\mathcal S}_{d-1}}
\mathbb{E}\left[L^2\left({\textbf Q}_\pm(\rho{\textbf u})\right)\right]
$, of course, is sufficient for Assumption~\ref{AssA2} to hold.

\color{black}

The following theorems establish the consistency of the sample principal directions computed from a sample of size $n$. These directions are obtained by maximizing  average squared trimmed polygonal lengths over the interval $[\delta_1^{(n)},\,1-\delta_2^{(n)}]\subset (0,1),$
using a radial partition with\footnote{For simplicity, we are dropping here the index $i$ used in the definitions of Section~\ref{sec31}.} mesh size $m_R^{-1}\to 0$ and trimming levels satisfying
$\delta_1^{(n)}\downarrow \varepsilon_0$ and
$\delta_2^{(n)}\downarrow \delta_0,$ where the constants $\varepsilon_0$ and $\delta_0>0$ are as in Assumption~\ref{AssA2}. Observe that the sequences $\delta_1^{(n)}$ and $\delta_2^{(n)}$ could equally well be chosen to converge to arbitrary constants $0<\varepsilon_1\leq \varepsilon_0<1$ and~$0<\delta_1\leq \delta_0<1,$ respectively. However, to avoid %
 additional notation, we restrict %
  to %
$\delta_1^{(n)}\downarrow \varepsilon_0$ and~$\delta_2^{(n)}\downarrow \delta_0$ as $n\to\infty$.

Consistency holds provided that the mesh size $m_R^{-1}$ tends to zero sufficiently slowly relative to the convergence rate of the empirical center-outward distribution function. More precisely, we require the a.s.\ convergence to zero of
$m_R s_n$ where %
\begin{equation}\label{GCrate}
s_n\coloneqq\sup_{\textbf{x}\in \mathbb{R}^d}\|\textbf{F}_\pm(\textbf{x})-\overline{\textbf{F}}_\pm^{(n)}(\textbf{x})\| .
\end{equation}
  provides the  Glivenko--Cantelli uniform consistency rate\footnote{See  {\citet{delBarrio2018}}.} of $\overline{\textbf{F}}_\pm^{(n)}$.
  Although the precise dependence of $s_n$ on $\textbf{F}_\pm$ and the radial and spherical discretization parameters $n_R$ and~$n_S$ is not explicitly available in general, assumptions of this type also can be found in, e.g.,  %
  {\cite[Theorem 4]{GunsiliusSchennach2023}}.%

We also emphasize that the lower trimming condition
$\delta_1^{(n)}\downarrow\varepsilon_0>0$ is primarily technical. It is needed to establish the uniform convergence of the sample objective in Lemma~\ref{lem:uniform_convergence_M_n}. Since the reference distribution is radial uniform on the unit ball, uniform regularity fails near the origin; see  {\cite[Theorem 1.1]{Fig18}}. In particular, the derivative and Lipschitz bounds for the center-outward quantile map are only guaranteed on compact annuli bounded away from zero; see Lemma~\ref{lem:bounds}. The lower trimming, therefore, ensures that all quantities are evaluated away from the origin, where the relevant regularity constants remain uniformly bounded. This discontinuity at the origin also provides a compelling reason for considering a directional, rather than bi-directional, approach.

Finally, Assumption~\ref{AssA2} allows $\varepsilon_0$ to be chosen arbitrarily small, provided it remains strictly positive. In our implementation, %
 we simply take $\delta_1^{(n)}=(m_R+1)^{-1}$, corresponding to
the radial gridpoints closest to the origin. For sufficiently large $m_R$, this yields a very small lower trimming level, so that the resulting estimator coincides, for practical purposes, with \eqref{defU1n'} with upper trimming level $\delta^{(n)}=\delta_2^{(n)}\to\delta_0$.

\color{black}

The following three theorems establish the consistency as $n\to\infty$ of our empirical concepts; see Appendix~A for the proofs.  We first establish the consistency  of the first  empirical principal direction.

\begin{Thm}\label{Thm:Directions_Consistency}
Let Assumptions \ref{AssA1} and \ref{AssA2} hold, the latter with constants $\varepsilon_0$ and $\delta_0$.
Let    $\delta_1^{(n)}\downarrow\varepsilon_0$,~$\delta_2^{(n)}\downarrow\delta_0$, and $m=m(n)=m_Rm_S\geq n$ be such that $m_R,m_S\to\infty$ as $n\to\infty$. With $s_n$ defined in \eqref{GCrate}, assume that~$m_Rs_n\to 0$ {\rm a.s.}~as $n\to\infty$.  For $m_{\delta_1^{(n)}}\coloneqq \lfloor \delta_1^{(n)}(m_R+1)\rfloor$, $m_{\delta_2^{(n)}}\coloneqq \lfloor(1-\delta_2^{(n)})(m_R+1)\rfloor$, let~$r_j\coloneqq j/(m_R+1)$, $j=m_{\delta_1^{(n)}},\dots,m_{\delta_2^{(n)}}$ and   %
denote by   $\textbf{U}_{\pm,1}^{(n)}$  the  unique maximizer of \vspace{-1mm}%
\begin{equation}\label{MEmpirical}
\textbf{u}\mapsto M_n^{\delta_1^{(n)}\!,\delta_2^{(n)}}(\textbf{u})\coloneqq \frac{1}{m_R+1}
\sum_{j=m_{\delta_1^{(n)}}}^{m_{\delta_2^{(n)}}}\Big( L_n^{(\delta_1^{(n)})}(\overline{\textbf{Q}}_\pm^{(n)}(r_j{\textbf u})\Big)^2 \qquad n\in\mathbb N
\end{equation}
 over the set of  directions $\{\textbf{u}_1,\dots,\textbf{u}_{m_S}\}$   considered in \eqref{defU1n'}.\footnote{\textcolor{black}{The factor $\frac{1}{m_R+1}$ in \eqref{MEmpirical} could be replaced with  $\frac{1}{m_{\delta_2^{(n)}}-m_{\delta_1^{(n)}}}$ without affecting asymptotic results. }}
 Then, \vspace{-1mm}
 $$
\big\|\textbf{U}_{\pm, 1}^{(n)}-\textbf{U}_{\pm, 1}\big\|= o_{\rm P}(1) \quad\text{as }n\to\infty ,
$$
where   $\textbf{U}_{\pm, 1}$ stands\footnote{For the sake of simplicity, we keep the same notation $\textbf{U}_{\pm, 1}$ as in \eqref{defU1''}.}  for the unique  maximizer of  the trimmed version \vspace{-1mm}
\begin{equation}\label{MTrimmed}
 {\textbf u}\mapsto M^{\varepsilon_0,\delta_0}(\textbf{u})\coloneqq \int_{\varepsilon_0}^{1-\delta_0}
\left(L^{(\varepsilon_0)}(\textbf Q_\pm(r\textbf u))\right)^2{\mathrm d} r
\end{equation}
of \eqref{defU1''}.
\end{Thm}

The second theorem establishes the consistency of the subsequent ($2\leq k \leq 2d$) empirical principal directions~$\textbf{U}_{\pm,k}^{(n)}$, $k=1,\ldots,2d$.

{\begin{Thm}\label{Thm:Directions_Consistency_i}
Let the assumptions of Theorem~\ref{Thm:Directions_Consistency} hold.\footnote{The sequences $\delta_1^{(n)}\downarrow \varepsilon_0$,
$\delta_2^{(n)}\downarrow\delta_0$ may depend on
$i=2,\dots,2d$; for simplicity, however, this dependence is not reflected in the notation.} For every $k=2,\dots,2d$ and $n\in\mathbb N$, \linebreak let $m_{\delta_1;k}^{(n)}\coloneqq \lfloor \delta_1^{(n)}(m_R^i+~\!1)\rfloor$, $m_{\delta_2;k}^{(n)}\coloneqq \lfloor(1-\delta_2^{(n)})(m_R^i+1)\rfloor$, $r_j\coloneqq  j/(m_R^k+1)$, $j=m_{\delta_1;k}^{(n)},\dots,m_{\delta_2;k}^{(n)}$,  %
and denote by   $\textbf{U}_{\pm,k}^{(n)}$ the unique maximizer~of
\begin{equation}\label{MEmpirical_i}
\textbf{u}\mapsto M_n^{\delta_1^{(n)}\delta_2^{(n)}}(\textbf{u})\coloneqq \frac{1}{m_R^k+1}
\sum_{j=m_{\delta_1;k}^{(n)}}^{m_{\delta_2;k}^{(n)}}\Big( L_n^{(\delta_1^{(n)})}(\overline{\textbf{Q}}_\pm^{(n)}(r_j{\textbf u})\Big)^2, \qquad n\in\mathbb N
\end{equation}
 over the set of sign directions $(\textbf{u}_1,\dots,\textbf{u}_{m_S^k})$ considered in \eqref{defUin'}.\footnote{\textcolor{black}{The factor $\frac{1}{m_R+1}$ in \eqref{MEmpirical_i} could be replaced with  $\frac{1}{m_{\delta_2;k}^{(n)}-m_{\delta_1;k}^{(n)}}$ without affecting asymptotic results. }}
Then,
$$
\big\|\textbf{U}_{\pm,k}^{(n)}-\textbf{U}_{\pm,k}\big\| = o_{\rm P}(1) \quad\text{as }n\to\infty,
$$
where $\textbf{U}_{\pm,k}$ is the unique maximizer, over the same feasible set as in \eqref{defUi''}, of the trimmed version  \eqref{MTrimmed}
 of~\eqref{defUi''}.
\end{Thm}

The main technical issue in    Theorem~\ref{Thm:Directions_Consistency_i} is the fact that the feasible set over which optimization is carried out in \eqref{MEmpirical_i} now depends on $\textbf{U}_{\pm,1}^{(n)},\ldots, \textbf{U}_{\pm,i-1}^{(n)}$, hence   constitutes a set-valued random element; see, e.g.,  {\cite{molchanov2018random}} for a precise formulation and measurability considerations.  As a consequence,  standard M-estimation consistency results %
are not sufficient to prove Theorem~\ref{Thm:Directions_Consistency_i}.

We therefore develop an extension (Theorem~\ref{thm:Argmax-compact-random} in Appendix~A) of the classical consistency result for M-estimators~\citep[Corollary~3.2.3]{vanderVaart1996} which covers the case in which the sample objective is optimized over random feasible sets that converge, in probability and with respect to the Hausdorff distance, to the feasible set of the population objective.\footnote{A similar extension has been proposed by~\cite{cox2022}, covering the case of a sample objective that is optimized over  feasible sets that converge, in the Painlevé--Kuratowski distance, to the feasible set of the population optimization problem. However, this result does not cover the case  of  random feasible sets, with convergence in probability and the set distance.  This result, thus, does not apply to our case.} \footnote{Recall that, in compact spaces, Painlevé--Kuratowski convergence is equivalent to convergence in the Hausdorff distance: see, e.g., Chapter 5 in \cite{rockafellar2009variational}.}

 The extension given in Theorem~\ref{thm:Argmax-compact-random} is of independent interest, since many deflation-based and projection-pursuit methods proceed sequentially by optimizing given objective functions subject to constraints induced by previous steps. Prominent examples include PCA and ICA, in which each new direction must be orthogonal to those previously selected.   In this context, our result shows that the consistency of subsequent directions follows directly from the consistency of the initial (unconstrained) one, provided that the corresponding sequence of feasible sets converges in the above-described sense.  In particular, Lemma~\ref{lem:orthogonality_convergence} establishes that orthogonality constraints on the unit sphere with respect to previously estimated principal directions satisfy this condition.

Finally, a third theorem yields the convergence in probability of the interpolated principal curves, uniformly over any compact subset of $(0,1)$.

\begin{Thm}\label{Thm:principal_curves}
Let the assumptions of Theorems~\ref{Thm:Directions_Consistency} and  \ref{Thm:Directions_Consistency_i} hold.
Then, letting  $\mathcal{L}_{\pm, k}(r)\coloneqq \textbf{Q}_\pm(r\textbf{U}_{\pm, k})$ and~$\overline{\mathcal{L}}_{\pm, k}^{(n)}(r)\coloneqq \overline{\textbf{Q}}_\pm^{(n)}(r\textbf{U}_{\pm, k}^{(n)})$,  for any $\varepsilon>0$,
\begin{equation}\label{first statement}
\sup_{r\in [\varepsilon,1-\varepsilon]}\big\Vert \mathcal{L}_{\pm,k}(r)-\overline{\mathcal{L}}_{\pm, k}^{(n)}(r)\big\Vert  = o_{\rm P}(1) \quad\text{$k=1\dots,2d$ \quad as }n\to\infty ,
\end{equation}
hence,
for any  $n_{R_0} < n_{R_1}\in\{1,\dots,n_R\}$ such that $\varepsilon\leq r_{n_{R_0}}<r_{n_{R_1}}\leq 1-\varepsilon$,
\begin{equation}\label{second statement}
\sup_{ j\in \{n_{R_0},\dots,n_{R_1}\}}\big\|\mathcal{L}_{\pm,k}(r_j)-{\mathcal{L}}_{\pm, k}^{(n)}(r_j)\big\| = o_{\rm P}(1) \quad\text{$k=1\dots,2d$ \quad as }n\to\infty
\end{equation}
where %
 ${\mathcal{L}}_{\pm, k}^{(n)}(r)\coloneqq {\textbf{Q}}_\pm^{(n)}(r\textbf{U}_{\pm, k}^{(n)})$.
\end{Thm}

\section{Numerical illustrations} \label{sec:examplesR2}
\subsection{Simulated examples} \label{simsec}

This section presents several simulation-based applications of the proposed method.
Principal directions are selected based on the regularized curve lengths introduced in Section~\ref{sec:smooth},
while the resulting principal curves are, for simplicity, obtained via linear interpolation of the regularized discrete curves \eqref{eq:PrincipalCurves_sample}.
We compare the proposed approach to standard (two-sided linear) PCA, GS-PCA  {\citep{GunsiliusSchennach2023}},
principal curves as introduced by  {\cite{hastie1989principal}}, and nonlinear PCA implemented via an autoencoder  {\citep{ScholzKaplanGuyKopkaSelbig2005}}.
Principal curves are fitted using the \textsf{R} package \texttt{princurve}  {\citep{princurve}},
and the autoencoder is implemented using the \textsf{R} package \texttt{pcaMethods}  {\citep{pcaMethods}},
both with their default settings.
The \textsf{R} package \texttt{ospca}, which implements nonlinear directional PCA and GS-PCA, and the script \texttt{paper\_examples.R}, which reproduces the examples of Sections~\ref{Sec21} and~\ref{sec:examplesR2}, are available at \url{https://github.com/klauschN/DirectionalNonlinearPCA}.

We compare competing nonlinear dimension reduction methods across six bivariate settings, each specified by an explicit data-generating mechanism and constructed to represent a distinct form of nonlinear structure.  These settings are summarized below. Throughout, the sample size is fixed at $n = 2561$, and the target grid is defined by $n_R = 20$, $n_S = 128$, and $n_0 = 1$. In each setting, the regularization parameter $\varepsilon_n$ is selected based on visual assessment of the smoothness of the estimated principal curves; the chosen values are reported below.  These values are used in the criterion that selects the principal directions, which leaves out the outer ten percent of the radii ($\delta_2=0.1$). The principal curves shown in Figure~\ref{fig:2D} are regularized with $\varepsilon_n=1$. In Setting~2, a stronger regularization with $\varepsilon_n=8$ is used, in which the penalty also includes the step from the center to the first point of the curve, so that the regularized curves still start at the center. In all settings, the second parameter of the normal distribution denotes its variance.}\footnote{Note that all methods considered here are based on different objective functions; hence, there is no universally accepted gold standard for their evaluation.  Additionally, as these methods impose no common constraint on the shape or length of the  curves they produce, mean-squared reconstruction error is not a meaningful measure of the quality of  dimension reduction. For instance, any curve passing through all of the points would provide zero reconstruction error.
Accordingly, our comparisons are primarily based on qualitative visual assessment.}

\emph{Setting~1 (Banana-shaped distribution).}
Let $X_1 \sim \mathcal{U}(-4,4)$ and define $X_2 = (0.4 X_1)^2 + \varepsilon$,\linebreak
where~$\varepsilon \sim~\!\mathcal{N}(0,\,0.25)$ is independent of $X_1$. The resulting distribution exhibits a single curved one-dimensional manifold with additive noise orthogonal to the manifold. The regularization parameter is set to~$\varepsilon_n=0.5$.

\emph{Setting~2 (Noisy banana).}
{The sample consists of $2001$ i.i.d.\ copies of $(X_1,X_2)$ and $560$ i.i.d.\ copies of~$(Y_1,Y_2)$, where\vspace{-3mm}
$$X_1 \sim \mathcal{U}(-6,6),\ \ X_2 = (0.4 X_1)^2 + \varepsilon,\  \ \varepsilon \sim \mathcal{N}(0,\,0.25),\
\ Y_1 \sim \mathcal{U}(-7,7),\  \ Y_2 \sim \mathcal{U}(-2,7),\vspace{-3mm}$$
 and all variables are mutually independent. The observations are standardized componentwise and then multiplied by~${\bf S}^{-1/2}$, where~${\bf S}$ denotes their sample covariance matrix. This standardization removes second-order structure while preserving nonlinear features. The regularization parameter is set to $\varepsilon_n=1.5$.}

\emph{Setting~3 (Exponential transformation of a Gaussian vector).}
Letting $(X_1,X_2) \sim \mathcal{N}(\boldsymbol 0,\,\mathbf I_2)$, de-\linebreak fine~${\bf Y}=~\!(Y_1,Y_2)$ with
$Y_2 = \exp(X_1 + X_2)$ and $Y_1 = Y_2 + X_1$.
This setting induces a smooth but strongly nonlinear dependence structure and was also considered in  {\cite{GunsiliusSchennach2023}}.  The regularization parameter is set to $\varepsilon_n={1.5}$.

\emph{Setting~4 (Sigmoid-shaped distribution).}
 {Letting $X_1 \sim \mathcal{U}(0,1)$ and $T\sim t_5$ independent of~$X_1$, define $X_2 = \log\{X_1/(1-X_1)\} + 0.4\,T$; the observations are centered at their means. The resulting distribution lies near the graph of the logit function, a sigmoid curve with the roles of the two coordinates exchanged, with homoscedastic heavy-tailed noise. The regularization parameter is set to $\varepsilon_n=1.5$.

\emph{Setting~5 (Two-component Gaussian mixture with nonlinear transformation).}
The data are drawn from a two-component mixture distribution. In the first component ($1200$ observations),
$(X_1,X_2) \sim \mathcal{N}_2\!\big((3,3),\,\boldsymbol\Sigma_1\big)$,\linebreak
with~$\boldsymbol\Sigma_1=\Bigg(\!\begin{array}{cc} 1.5&1\vspace{-2mm} \\ 1& 1.5
\end{array}\!
\Bigg)$,
followed by the componentwise transformation
$(X_1,X_2) \mapsto (X_1^2, X_2^2)$.\vspace{-4mm}
In the second  component ($1361$ observations),
$(X_1,X_2) \sim \mathcal{N}_2\!\big((5,-5),\,\boldsymbol\Sigma_2\big)$,
with $\boldsymbol\Sigma_2=\Bigg(\!\begin{array}{cc} 10& {-10}\vspace{-2mm} \\ {-10}& 45
\end{array}\!
\Bigg)$; the observations are standardized componentwise.
This setting combines multimodality with nonlinear distortions.  {The regularization} parameter is set to $\varepsilon_n=1.5$.}

{\emph{Setting~6 (Noisy bird-shaped distribution).}
Define the random vector $
\mathbf W = B\,(X_1,X_2) + (1-B)\,(Y_1,Y_2)$,
where $B \sim \mathrm{Bernoulli}(\pi)$ with $\pi = 2000/2561$. Conditional on $B=1$, $(X_1,X_2) \sim \mathrm{P}_{\mathrm{bird}}$, where $\mathrm{P}_{\mathrm{bird}}$ denotes the empirical distribution induced by the silhouette of a flying bird.\footnote{Based on the image available at \url{https://creazilla.com/media/silhouette/63833/flying-bird}.} This distribution is obtained by thresholding a grayscale image to identify foreground pixels corresponding to the bird shape, sampling these pixels uniformly at random, and mapping their pixel indices to Cartesian coordinates in the unit square $[0,1]^2$, with the vertical axis reversed.}

Conditional on $B=0$, $(Y_1,Y_2) \sim \mathcal{U}([0,1]^2)$. All random variables are mutually independent. This construction combines a highly nonlinear, image-derived low-dimensional structure with uniform background noise in the ambient space. The regularization parameter is set to $\varepsilon_n = 1$.

\begin{figure}[p]%
\centering
\includegraphics[width=\linewidth, height=0.84\textheight, keepaspectratio]{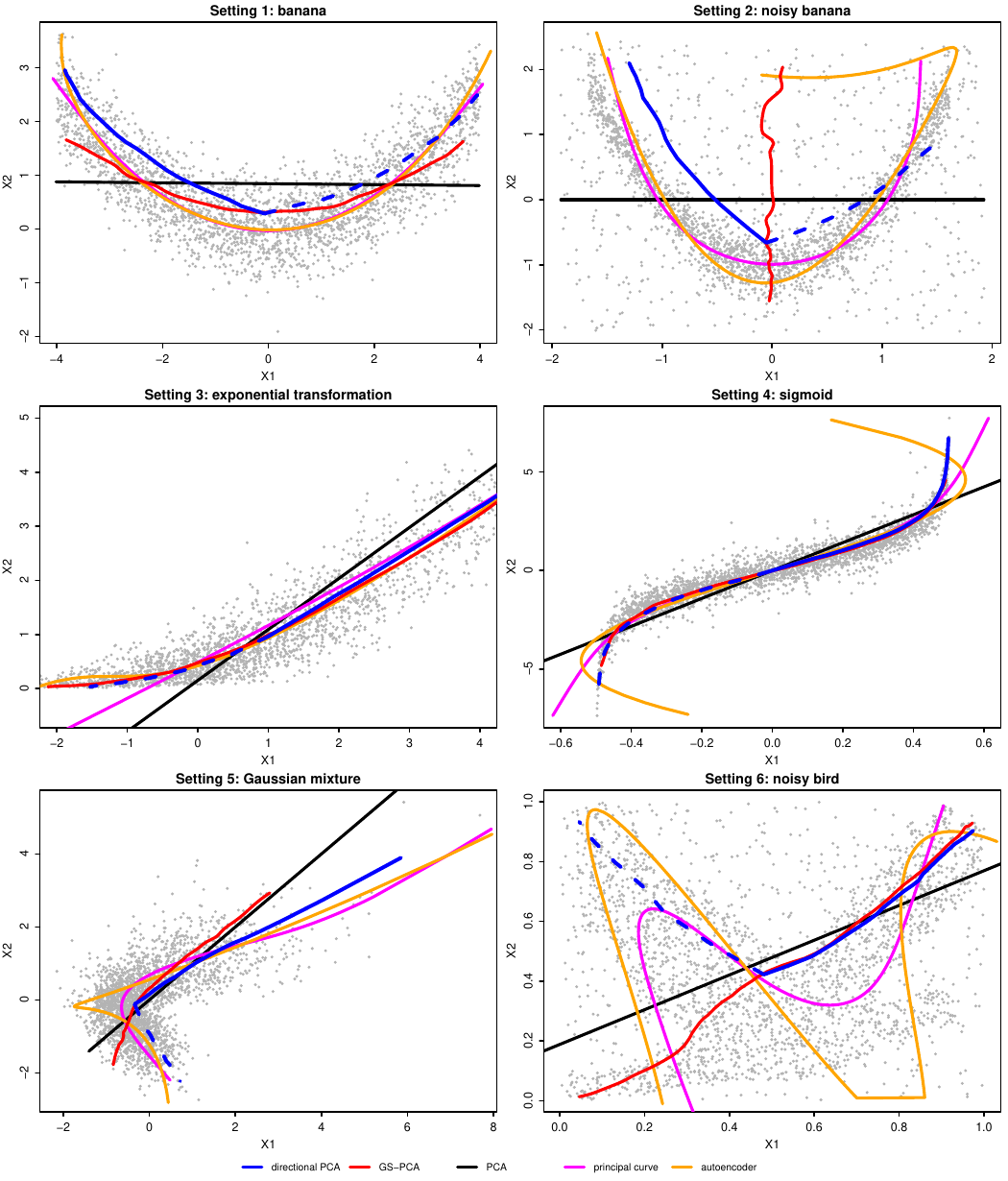}

\caption{Comparison of nonlinear dimension reduction methods across six bivariate settings.
Shown are the first two regularized nonlinear directional PCA curves (\textcolor{blue}{blue}), the first GS-PCA curve (\textcolor{red}{red}),
the first classical linear principal component (black), the Hastie--Stuetzle principal curve (\textcolor{magenta}{magenta}),
and an autoencoder-based reconstruction (\textcolor{orange}{orange}).
For directional PCA, solid and dashed (\textcolor{blue}{blue}) lines correspond to the first and second directional components, respectively. {In Settings~3 and~4, the plot windows are restricted to the bulk of the data.}}
\label{fig:2D}

\end{figure}

Figure~\ref{fig:2D} displays the leading direction obtained from the standard methods, together with the first two directions produced by the directional approach. Across all settings, it is evident that linear PCA fails to provide a meaningful representation of the underlying data structure. {The autoencoder performs reasonably well in Settings~1, 3 and~5, but it forms loops in Settings~2 and~6 and turns back at both ends of the sigmoid in Setting~4. The principal curve accounts for the banana quite well in Settings~1 and~2, but it is almost linear in Setting~3, does not fully capture the steep ends of the sigmoid in Setting~4, and is adversely affected by the noise in Setting~6.} In Setting~6, most methods focus on quite different features of the data; interestingly, PCA and GS-PCA are quite similar and focus on a combination of the right wing and the tail of the bird, whereas the directional PCA uses one direction for each of the two wings.

Overall, the directional and GS-PCA approaches produce similar results in Settings~1, 3 and~4, and in Setting~4 they are the only methods that account for the sigmoid up to its steep ends. In the noisy banana of Setting~2, the GS-PCA curve is distorted by the noise, whereas the two directional curves correctly capture the two ``arms'' of the banana; the first one, however, lies inside the left arm, and the principal curve follows the banana more closely. In Setting~5, the first directional curve accounts for the elongated component of the mixture further than GS-PCA, and the second one leads into the other component. In Setting~6, finally, the directional approach focuses on features of the data (the wings) that are perhaps more natural than those found by GS-PCA. These observations suggest that, at least for the settings considered here, the directional approach delivers consistently competitive performance, while each of the other methods fails in at least one of the settings.

\subsection{Real data examples}\label{sec::realdata}

\subsubsection{\texorpdfstring{{Old Faithful}}
{Old Faithful}}

\begin{figure}[b!]%
\centering
\includegraphics[width = 0.7\linewidth]{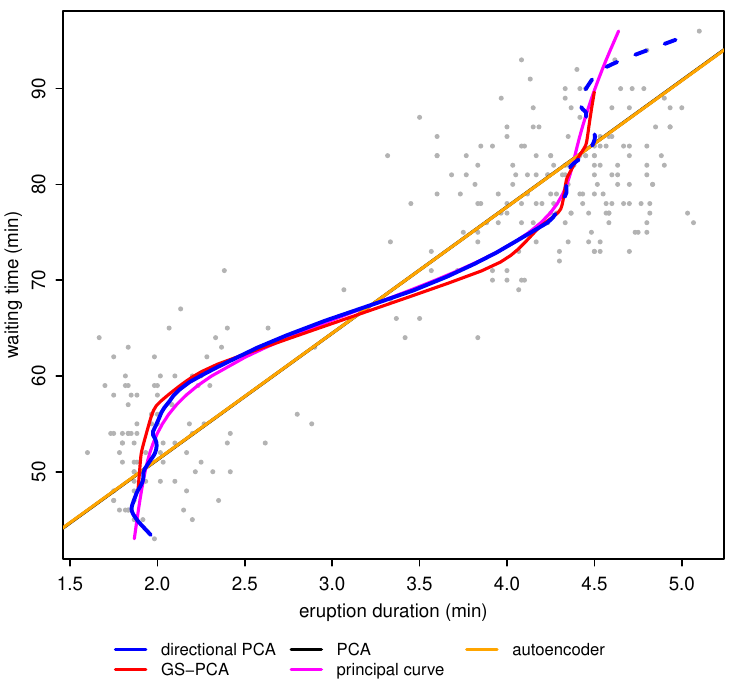}
\caption{%
Univariate approximations of the Old Faithful data.
The first two regularized nonlinear directional PCA curves (\textcolor{blue}{blue}),
the first GS-PCA curve (\textcolor{red}{red}), the first classical linear principal component ({black}), the  {Hastie--Stuetzle} principal curve (\textcolor{magenta}{magenta}), and the autoencoder (\textcolor{orange}{orange}). For directional PCA, solid and dashed (\textcolor{blue}{blue}) lines correspond to the first and second directional components, respectively.  {The autoencoder curve almost coincides with the classical principal component. The points are the $272$ original observations.}
}
\label{fig:OldFaithful}
\end{figure}

The Old Faithful dataset consists of bivariate measurements of geyser activity at Old Faithful in Yellowstone National Park. Each observation comprises the eruption duration (in minutes) and the subsequent waiting time until the next eruption (in minutes). The data exhibit a pronounced nonlinear and bimodal structure, reflecting distinct eruption regimes, and have been widely used in the statistical literature as a benchmark example for mixture modeling, clustering, and nonlinear dimension-reduction methods.

For the numerical experiments, the original dataset {(272 observations)} was resampled with replacement to obtain a bivariate sample of size $n=2561${, and the ties created by the resampling were broken by adding a small amount of noise}. This resampling allows nonlinear directional PCA  to be applied based on {a target grid with} $n_R = 20$, $n_S = 128$, {and $n_0=1$}. For nonlinear  directional PCA, the smoothing parameter was set to $\varepsilon_n = 2$, a choice guided by visual assessment of the roughness of the resulting estimated curves, and the criterion was evaluated at $m_R=60$ radii, without trimming (see below for the role of~$m_R$). The principal curves shown in Figures~\ref{fig:OldFaithful} and~\ref{fig:9}, and the projections onto them, are regularized with $\varepsilon_n=8$, with the curves held at the center as in Setting~2 of Section~\ref{simsec}. Both variables are measured in minutes and were not standardized. The competing methods, which do not require a sample size of the form $n_Rn_S+n_0$, were applied to the original observations.

\begin{figure}[b!]
\subfloat[Directional principal curves]{%
    \includegraphics[width=.48\linewidth]{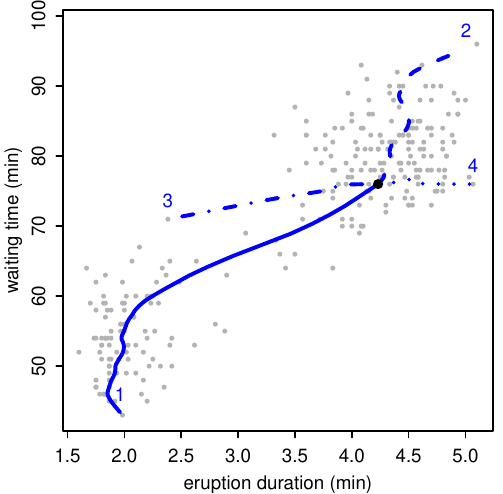}%
}\
\hfill
\subfloat[Dimension reduction]{%
    \includegraphics[width=.48\linewidth]{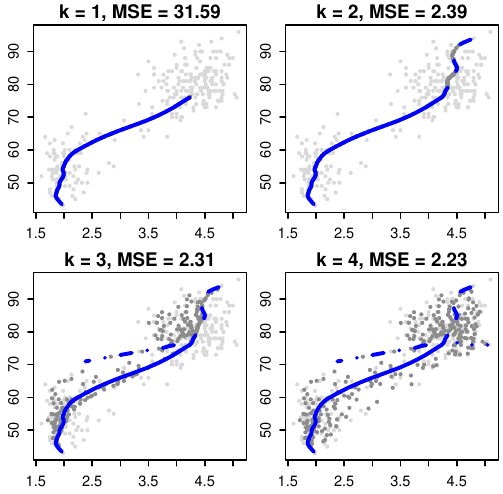}%
}
\caption{%
{Nonlinear directional PCA of the Old Faithful data. Subfigure~(1): the four directional principal curves, numbered and distinguished by their line types (plain,  dashed, ...) starting at the center (black dot). Subfigure~(2): the projections (dark grey) of the observations (light grey) onto the manifolds spanned by the first $k=1,\ldots,4$ directional principal curves (blue), together with the mean squared reconstruction errors. The axes show the same variables as in Subfigure~(1), and the points are the $272$ original observations.}}
\label{fig:9}
\end{figure}

We emphasize that our focus is on data visualization and representation through effective noise and dimension reduction, as discussed in Section \ref{sec:manifold}. Therefore, we opted not to conduct a comparative simulation study for two main reasons. Firstly, there is no universally applicable objective that applies to all nonlinear PCA extensions considered in our examples, apart from potentially using Wasserstein distances to assess the fidelity of the lower-dimensional projection (optimal curve). Second, since methods other than the specifically tuned nonlinear GS-PCA , unlike our approach,
 do not impose consistent constraints on the form and length of derived curves, the mean squared deviation from sample points to their corresponding projections on the curve is not universally valid as a measure of fit. For instance, in theory, any dataset could be perfectly fitted by a lengthy, space-filling curve;  {see e.g.  {\citep{mandelbrot1982fractal} or  \citep[Section 3]{GunsiliusSchennach2023}}}.%

Figure~\ref{fig:OldFaithful} compares the results obtained using the different methods. Both the autoencoder and classical PCA suggest an approximately linear relationship between eruption duration and waiting time. In contrast, the remaining approaches recover a more pronounced nonlinear, sigmoid-like structure. Among these, the directional approach exhibits a modest advantage: the first directional component primarily captures the regime associated with shorter waiting times and shorter eruptions, while the second component reflects the regime characterized by longer waiting times and longer eruptions.

{Figure~\ref{fig:9} shows all four directional principal curves together with the corresponding dimension reductions. The center lies in the larger cluster of long eruptions (at about $4.2$ minutes and $76$ minutes). The first two curves lead from the center into the two clusters, so that together they give a one-dimensional representation of the data along the sigmoid-like structure of Figure~\ref{fig:OldFaithful}. The third and fourth curves run from the center towards shorter and longer eruptions, respectively, at waiting times of about $70$ to $76$ minutes, that is, they describe the variation of the eruption duration that remains. Accordingly, the mean squared reconstruction error drops from $31.59$ for $k=1$ to $2.39$ for $k=2$, while the third and fourth components reduce it only slightly further ($2.31$ and $2.23$). Note that the reconstruction error with all four components is not zero, since the principal curves are regularized and the original observations are projected as new observations (the curves are estimated from the resampled data).

The order of the first directions in this example depends on the number~$m_R$ of radii at which the criterion is evaluated. With $m_R=n_R=20$, the second direction is orthogonal to the first one, so that the first two curves both lead into the cluster of short eruptions, and the reconstruction error with two components is hardly smaller than with one ($31.6$ for both). With $m_R=60$, the antipode of the first direction comes second, and this was the case for all six further resamples of the data that we tried (with~$m_R=40$, for four of the six). The criterion values of the two candidate directions are thus close, and the finer radial grid tips the balance towards the antipode. Since, for a fixed sample size, a finer radial grid does not necessarily improve the approximation of the curve lengths (see Section~\ref{sec:smooth}), this dependence should be kept in mind, and further investigation is needed on how to choose~$m_R$ in practice.

\subsubsection{\texorpdfstring{{The Hertzsprung--Russell diagram}}{The Hertzsprung--Russell diagram}}\label{sec:HR}

The Hertzsprung--Russell (HR) diagram is a classical tool in astronomy, displaying the absolute magnitude~$M_V$ of stars (a logarithmic measure of luminosity, with smaller values corresponding to brighter stars) against their color index $B-V$ (a proxy for surface temperature, with smaller values corresponding to hotter stars). We consider the stars of the Hipparcos catalogue \citep{Perryman1997} with parallaxes between 20 and 25 milliarcseconds (distances between 40 and 50 parsecs), as distributed with \cite{FeigelsonBabu2012}. Retaining the stars with a relative parallax error below 10\% yields a sample of size $n=2317$, with absolute magnitudes $M_V=V+5\log_{10}(\varpi/100)$, where $V$ denotes the apparent magnitude and $\varpi$ the parallax in milliarcseconds.

The data exhibit a dense core from which three branches of unequal lengths and densities emanate: the upper main sequence (hot and bright stars), the lower main sequence (cool dwarfs), and the subgiant and red giant branch, which leaves the main sequence at an acute angle towards the red clump ($B-V\approx 1$ and~$M_V\approx 0.5$). The two parts of the main sequence extend in opposite directions from the core, whereas the giant branch lies on one side only. The resulting point cloud is thus clearly non-elliptical and asymmetric, making it a natural example for assessing the ability of nonlinear principal components to recover structures that cannot be represented well by a single curve.

For nonlinear directional PCA, {the tuning parameters are chosen in advance as in Section~\ref{simsec}: the target grid is defined by $n_R=20$, $n_S=115$, and $n_0=17$, the upper trimming level is $\delta=0.1$, and the regularization parameter is set to $\varepsilon_n=1$, both for the criterion and for the principal curves. The principal directions are selected from $m_S=2415$ directions, so that $m_Rm_S\geq n$.} Since $B-V$ ranges over about 2 magnitudes and $M_V$ over about 11, all methods are applied to the standardized data, and the results are displayed in the original units, with the magnitude axis reversed, as is customary for HR diagrams. The other methods are used with their default settings.

Figure~\ref{fig:HR}(a) compares the results obtained with the different methods. For nonlinear directional PCA, the first two principal directions are antipodal, and the third one is orthogonal to them; the main sequence is thus described by a pair of antipodal directions, and the giant branch by a single direction. The first principal curve follows the lower main sequence, including its bend at the red end, while the second one runs along the upper main sequence. The third curve rises from the core through the subgiants and bends towards the red clump, thereby reproducing the shape of the giant branch. The fourth curve is short and describes the width of the main sequence. The center  $(B-V, M_V)=(0.66, 4.74)$ of the fit  practically coincides with the position $(0.65, 4.83)$ of the Sun.

To assess these findings, we classify the stars into four groups based on criteria that do not depend on any of the methods considered. Stars more than one magnitude brighter than a robust (loess) estimate of the main-sequence ridge line and with $B-V>0.5$ are classified as red giants and clump stars if $M_V<2$, and as subgiants otherwise; the remaining stars are assigned to the upper or the lower main sequence according to whether their color index is below or above that of the Sun ($B-V=0.65$), and five stars lying far below the main sequence are left aside. Figure~\ref{fig:HR}(b) shows the resulting groups along with the nonlinear directional principal curves. {Each star is attributed to the directional principal component with the largest score. The upper main sequence is then associated with the second component (85\% of its stars) and the lower main sequence with the first one (75\%, with a further 19\% attributed to the fourth component, whose curve runs next to it), while 87\% of the subgiants and all red giants and clump stars are attributed to the third component. These findings do not depend on the choice of the tuning parameters: for each of 18 combinations of grid sizes ($n_R\in\{20,30,40\}$, with $n_S=\lfloor n/n_R\rfloor$), regularization parameters ($\varepsilon_n\in\{0.5,1,2\}$), and trimming levels ($\delta\in\{0.05,0.1\}$), the three branches are described by three distinct principal curves, with between 79\% and 90\% of the upper main sequence, 64\% and 83\% of the lower main sequence, 77\% and 92\% of the subgiants, and 98\% and 100\% of the red giants and clump stars attributed to the corresponding component.}

In contrast, none of the competing methods represents the three branches separately. Classical PCA yields a single line along the main sequence, which misses the bend of the lower main sequence. The Hastie--Stuetzle principal curve follows the main sequence but ignores the giants, while the autoencoder, which has to cover two branches with a single curve, passes through the empty region between the main sequence and the giant branch. The first GS-PCA curve captures the main sequence; the second one  {runs from the core through the subgiants towards the red giants, but it passes to the right of the red clump, and at both ends it extends into regions that contain no stars (below the main sequence and beyond the red clump), so that it describes the giant branch only roughly.} This example illustrates the advantage of a directional approach when several branches of different lengths emanate from a common core.

\begin{figure}[t!]
\subfloat[All methods]{%
    \includegraphics[width=.48\linewidth]{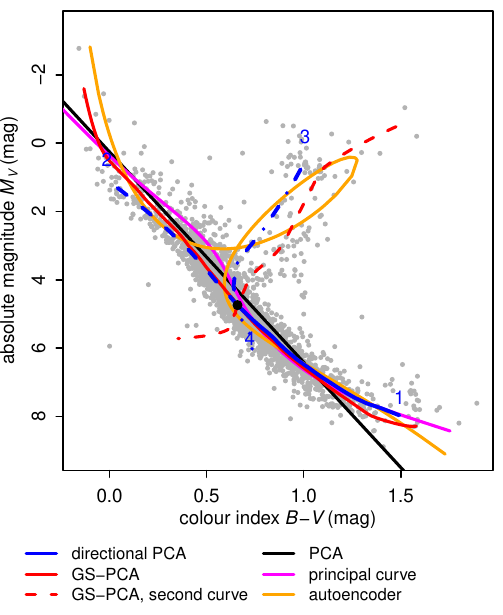}%
}\hfill
\subfloat[Stellar groups]{%
    \includegraphics[width=.48\linewidth]{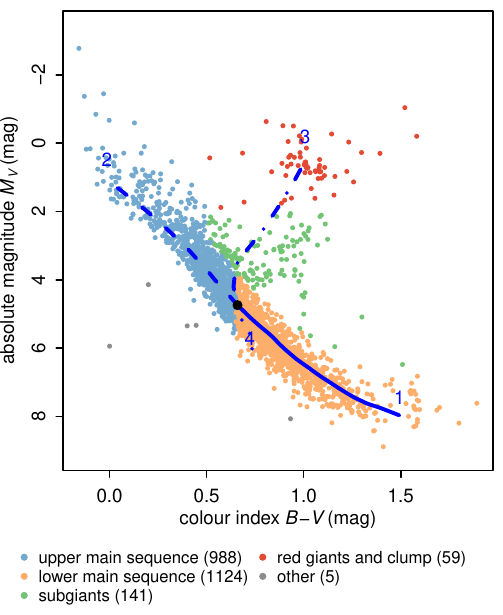}%
}
\caption{Hertzsprung--Russell diagram of the Hipparcos data. (a)~The four regularized nonlinear directional PCA curves (\textcolor{blue}{blue}, numbered with different line types), the two GS-PCA curves (\textcolor{red}{red}, solid and dashed), the first classical linear principal component (\textcolor{teal}{green}), the Hastie--Stuetzle principal curve (\textcolor{magenta}{magenta}), and the autoencoder (\textcolor{orange}{orange}); the dot marks the center. (b)~The nonlinear directional PCA curves, with the stars colored by group.}
\label{fig:HR}
\end{figure}

{\subsubsection{\texorpdfstring{{Gvessel Data}}{Gvessel Data}} \label{Gvessel}

The Gvessel North Atlantic Water Temperature dataset consists of multivariate oceanographic
measurements collected over nine consecutive days in May 2000 by the German research vessel
\emph{Gauss} in the North Atlantic. The data have been retrieved from the World Ocean Database and
have previously been used in the principal-curve literature as a benchmark example for data
compression and nonlinear representation; see, e.g.,  {\cite{einbeck2010data}}. In our analysis,
we focus on the three variables {salinity, temperature, and depth (\texttt{salg}, \texttt{tempg} and \texttt{depthg} in the dataset, $643$ observations)}, which capture the local hydrographic conditions along the vessel's
measurement track. The resulting point cloud exhibits a clearly non-elliptical structure, making it a
suitable example for assessing the ability of nonlinear principal
components to recover intrinsic low-dimensional features beyond classical (bidirectional) PCA. {The observations originate from vertical profiles: in the upper ocean, temperature and salinity decrease strongly with depth, whereas below about~$1500$~meters both are almost constant (see Figure~\ref{fig:Gvessel}). Note also that the variables have very different scales: the standard deviation of the depth ($1319$~meters) is more than $4000$ times that of the salinity ($0.29$).}}

\begin{figure}[t!]
\centering
\subfloat[Univariate approximations]{%
    \includegraphics[width=\linewidth]{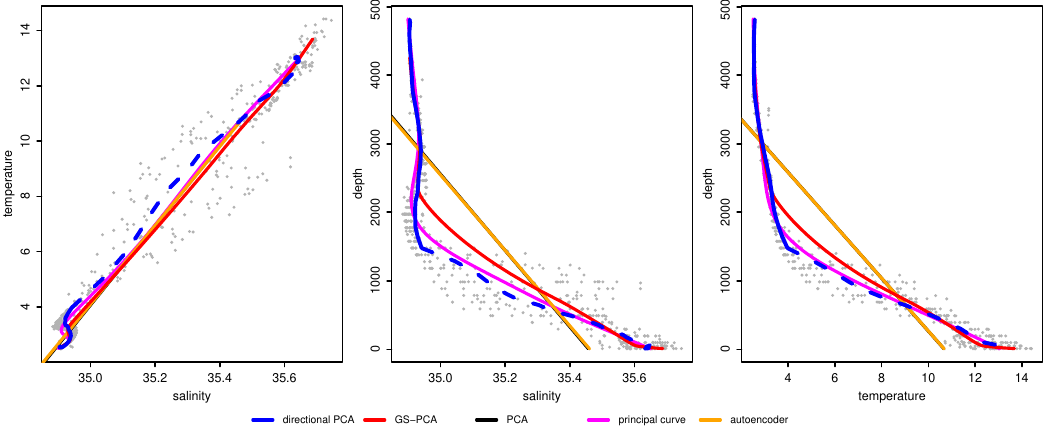}%
}\\
\subfloat[Dimension reduction]{%
    \includegraphics[width=\linewidth]{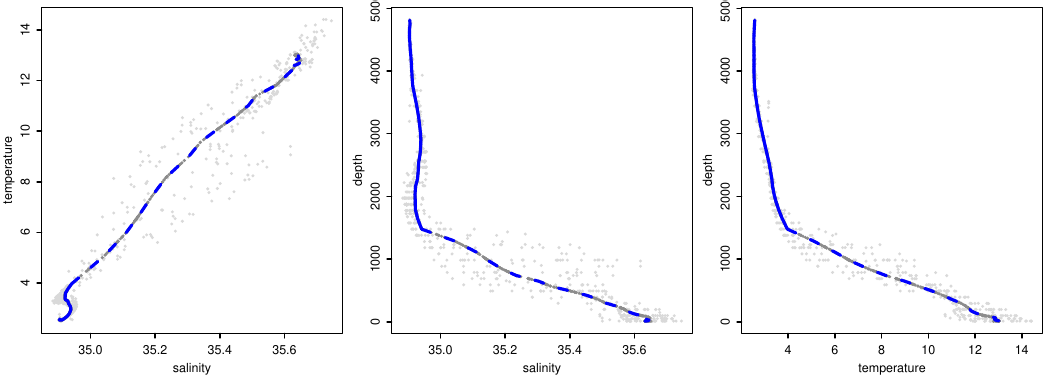}%
}
\caption{{Gvessel data, shown for the three pairs of variables. Subfigure~(1): the first two regularized nonlinear directional PCA curves (\textcolor{blue}{blue}), the first GS-PCA curve (\textcolor{red}{red}), the first classical linear principal component (black), the Hastie--Stuetzle principal curve (\textcolor{magenta}{magenta}), and the autoencoder (\textcolor{orange}{orange}), which almost coincides with the classical principal component. For directional PCA, solid and dashed (\textcolor{blue}{blue}) lines correspond to the first and second directional components, respectively. Subfigure~(2): the projections (dark grey) of the observations (light grey) onto the manifold spanned by the first two directional principal curves (\textcolor{blue}{blue}). The points are the $643$ original observations. An interactive three-dimensional version of Subfigure~(1), which can be rotated and zoomed, is available at \url{https://klauschn.github.io/DirectionalNonlinearPCA/gvessel_3d.html}.}}
\label{fig:Gvessel}
\end{figure}

In order to
apply nonlinear directional PCA, we work with a resampled dataset
obtained by sampling with replacement from the original observations to yield a sample of size
$n=4001${, and the ties created by the resampling were broken by adding a small amount of noise}. This resampling allows nonlinear  directional PCA to be implemented using target grid
sizes $n_R=40$, $n_S=100$, and $n_0=1$. For nonlinear  directional PCA, the smoothing parameter was set to~$\varepsilon_n=2$,  based on visual inspection of the roughness and stability of the resulting
estimated principal curves{; the criterion is evaluated at $m_R=n_R$ radii, without trimming. The competing methods were applied to the $643$ original observations.

Directional PCA, like classical PCA, the principal curve, and the autoencoder, was applied to the variables in their original units. The depth then dominates the analysis, and the transport essentially orders the observations by their depth. Since the depth is the natural coordinate of a vertical profile, we consider this appropriate here. We also applied directional PCA to the standardized variables; the resulting principal directions, however, depended strongly on the smoothing parameter and on~$m_R$. GS-PCA, in contrast, was applied to the standardized variables, with its default grid of normal quantiles, since in the original units the estimated Jacobians are not positive definite at almost all grid points.}

As shown in Figure~\ref{fig:Gvessel}, nonlinear  directional PCA {and the principal curve follow the vertical profile closely, and GS-PCA captures its overall shape but cuts across the bend of the profile at depths between about $500$ and $2300$~meters}, whereas the autoencoder and linear PCA, which almost coincide, fail to recover it and provide only an essentially uninformative low-dimensional representation.

The second directional principal direction is the antipode of the first one, and the values of the criterion for the remaining four directions are negligible (below one thousandth of the first value). The data are thus essentially one-dimensional, and the first two directional principal curves together form a single curve through the center, which lies at a depth of about $1480$~meters (with a temperature of $4.0^\circ$C and a salinity of $34.94$). The first curve leads from the center down to the deep water at about $4800$~meters, and the second one up through the thermocline to the warm and saline surface water. The projection onto the manifold spanned by these two curves (Subfigure~(2) of Figure~\ref{fig:Gvessel}) therefore parametrizes the trivariate observations by a single parameter, their position along the profile.

Unlike the parameter of a principal curve, the two directional components also have a direct interpretation. For all but $16$ of the $643$ observations, either the first or the second component is zero: observations below the center have a positive first component, and observations above the center a positive second component. The first component thus describes the deep water, in which temperature and salinity hardly change, and the second one the stratified upper ocean. Note that the center, as the center-outward median, is determined by the method and not by the physics of the ocean; here it lies, however, close to the depth below which temperature and salinity are almost constant. The example thus illustrates, as the Old Faithful data in Section~\ref{sec::realdata}, how the one-sided components separate two regimes of the data, while the representation as a whole is comparable to that of the principal curve.

{\section{Conclusion and outlook} \label{ConclSec}
Prompted by the observation that classical PCA is inherently linear and bidirectional (a component is identified only up to its sign) and therefore may fail to adequately represent nonlinear and/or asymmetric patterns, this paper develops a new framework for principal component analysis that is both \emph{directional} and \emph{nonlinear}. By replacing the traditional variance-maximization objective with a one-sided criterion based on the expected arc lengths of ``curvilinear projections,'' we obtain a nonlinear and directional analogue of PCA.  %
This modification yields, in dimension $d$, up to $2d$ ordered principal curves, allowing the method to capture one-sided or skewed structures that are masked  by the bidirectional nature of standard PCA. Numerical illustrations in low dimensions indicate that the resulting  principal curves offer improved interpretability and, in  asymmetric settings, a more faithful reconstruction of the data geometry.

 Based on the concepts of   center-outward distribution and quantile functions (${\textbf F}_{\pm}$ and ${\textbf Q}_{\pm}$, respectively,  defined through measure transportation), our principal curves %
 are obtained as images by the center-outward quantile map ${\textbf Q}_{\pm}$ of radial directions ${\textbf u}$ in the unit ball. The resulting concept preserves the core intuition of PCA---identifying the directions $\textbf u$ that maximize the expected variations of a projection of the variable of interest $\textbf X$. However, instead of the expected  squared modulus of a linear projection of $\textbf X$ to the straight line   $\{ t{\textbf u}\vert\, t\in{\mathbb R}\}$, we consider the expected squared arc length of a ``curvilinear projection'' of $\textbf X$ to the curve~${\textbf Q}_{\pm}(\{t{\textbf u}\vert\, t\in{\mathbb R}^+\})$  that can capture curved and non-elliptical structures. Examples illustrate that this approach recovers meaningful nonlinear structures (such as banana-shaped distributions) in a way that classical PCA cannot, while enjoying the same natural orthogonal equivariance properties, hence behaving consistently under rotations and symmetries of the data space.

A key application of PCA is %
 \emph{dimension reduction}. In the linear case, the PCA-based reduction procedure consists of projecting the $d$-dimensional observations on the $k$-dimensional linear space spanned by the first~$k<d$ principal components. Our nonlinear dimension reduction similarly projects the observations, in an adequate curvilinear way, to the manifold obtained as the positive span of the first $k<2d$ principal curves.

While preserving the intuitive features of traditional principal components, the directional nonlinear approach we are proposing offers a flexible and much more informative analysis of asymmetric and non-elliptical data, thus paving the way for improved implementations of the many applications, whether data-analytic or inferential, of principal components---data visualization, pattern recognition, identification of clusters, trends, and outliers; removal of redundant information; principal component regression in presence of high-dimensional regressors;  accelerated learning and image processing;   efficient data compression;~etc.

\appendix

\section*{Appendix}
\addcontentsline{toc}{section}{Appendix}

\section{Proofs}\label{sec:proofs}

\subsection{Proof of Proposition~\ref{prop:invariance}}

  (a) Statements  (i) and (ii). Since  $\textbf{F}^{\textbf X}_\pm$ and $\textbf{F}^{{\boldsymbol{\mathcal O}}\textbf X}_\pm$ are  \textcolor{black}{monotone}  transports pushing the distributions $\rm{P}^{\textbf X}\in{\mathcal P}^d_\pm$ of $\textbf{X}$ and~$\rm{P}^{{\boldsymbol{\mathcal O}}{\textbf X}}\in~\!{\mathcal P}^d_\pm$ of ${\boldsymbol{\mathcal O}}{\textbf X}$ to the uniform~$\rm{U}_d$ over~${\mathbb S}_d$, the claims follow from the general equivariance properties of such   transports: see Proposition 2.2 in  {\cite{HHH20}}.\smallskip

 \noindent  (b) Statement (iii). Result from (i) and (ii), and the orthogonal invariance of the $L_2$ norm, further imply that,
    for any~$\textbf{u}\in{\mathcal S}_d$ and $r\in (0,1)$,
    $L^2\Big(\big(\textbf{F}^{{\boldsymbol{\mathcal O}}\textbf X}_{\pm}\big)^{-1}(r\textbf{u})\Big)
    =L^2\Big({\boldsymbol{\mathcal O}}\big(\textbf{F}^{\textbf X}_{\pm}\big)^{-1}(r{\boldsymbol{\mathcal O}}^\prime \textbf{u})\Big)
    =L^2\Big(\big(\textbf{F}_{\pm}^{{\textbf X}}\big)^{-1}(r{\boldsymbol{\mathcal O}}^\prime\textbf{u})\Big)$. Let~${\textbf U}^{{\textbf X}}_{\pm ,i}$ and~${\textbf U}^{{\boldsymbol{\mathcal O}}{\textbf X}}_{\pm ,i}$ stand for the $i$th  principal directions of $\textbf X$ and ${\boldsymbol{\mathcal O}}{\textbf X}$, respectively.  Applying  this in the definition of  principal curves yields
    $$\mathcal{L}_{\pm,i}^{\boldsymbol{\mathcal O}{\textbf X}}
    =\{\big(\textbf{F}^{\boldsymbol{\mathcal O}{\textbf X}}_{\pm}\big)^{-1}(r
    {\textbf U}^{{\boldsymbol{\mathcal O}}{\textbf X}}_{\pm ,i}
     |r\in(0,1)\}
    =\{{\boldsymbol{\mathcal O}}\big(\textbf{F}^{\textbf X}_{\pm}\big)^{-1}(r{\textbf U}^{{\textbf X}}_{\pm ,i})|r\in(0,1)\}
    \eqqcolon {\boldsymbol{\mathcal O}}\mathcal{L}_{\pm ,i}^{\textbf X},\quad  i=1,\dots,{2d}.$$ %

\noindent  (c) Statement (iv).  \textcolor{black}{From the definitions of $P^{{\boldsymbol{\mathcal O}}{\textbf X}}_{\pm,i}$ and $P^{{\textbf X}}_{\pm,i}$ (see \eqref{eq:nonlinear-projection+}), we obtain}
\begin{align*}
P^{{\boldsymbol{\mathcal O}}{\textbf X}}_{\pm,i}&=\int_{0}^{\Big\|\textbf{F}_{\pm}^{{\boldsymbol{\mathcal O}}\textbf{X}}(\boldsymbol{\mathcal O}\textbf{X})\Big\|}\left\langle\frac{\frac{\mathrm{d}}{\mathrm{d}t}(\textbf{F}_{\pm}^{{\boldsymbol{\mathcal O}}\textbf{X}})^{-1}(t{\textbf U}^{{\boldsymbol{\mathcal O}}{\textbf X}}_{\pm ,i})}{\Big\|\frac{\mathrm{d}}{\mathrm{d}t}(\textbf{F}_{\pm}^{{\boldsymbol{\mathcal O}}\textbf{X}})^{-1}(t{\textbf U}^{{\boldsymbol{\mathcal O}}{\textbf X}}_{\pm ,i})\Big\|},\frac{\mathrm{d}}{\mathrm{d}t}(\textbf{F}_{\pm}^{{\boldsymbol{\mathcal O}}\textbf{X}})^{-1}\Bigg(t\frac{\textbf{F}_{\pm}^{{\boldsymbol{\mathcal O}}\textbf{X}}({\boldsymbol{\mathcal O}}\textbf{X})}{\Big\|\textbf{F}_{\pm}^{{\boldsymbol{\mathcal O}}\textbf{X}}({\boldsymbol{\mathcal O}}\textbf{X})\Big\|}\Bigg)\right\rangle\mathrm{d}t\\
&=\int_{0}^{\Big\|\textbf{F}_{\pm}^{\textbf{X}}(\textbf{X})\Big\|}\left\langle{\boldsymbol{\mathcal O}}\frac{\frac{\mathrm{d}}{\mathrm{d}t}(\textbf{F}_{\pm}^{\textbf{X}})^{-1}(t{\textbf U}^{{\textbf X}}_{\pm ,i})}{\Big\|\frac{\mathrm{d}}{\mathrm{d}t}(\textbf{F}_{\pm}^{\textbf{X}})^{-1}(t{\textbf U}^{{\textbf X}}_{\pm ,i})\Big\|},{\boldsymbol{\mathcal O}}\frac{\mathrm{d}}{\mathrm{d}t}(\textbf{F}_{\pm}^{\textbf{X}})^{-1}\Bigg(t\frac{\textbf{F}_{\pm}^{\textbf{X}}(\textbf{X})}{\Big\|\textbf{F}_{\pm}^{\textbf{X}}(\textbf{X})\Big\|}\Bigg)\right\rangle\mathrm{d}t
=P^{{\textbf X}}_{\pm,i},\quad i=1,\dots,2d,%
\end{align*}%
which completes the proof.\hfill$\square$

\subsection{Proof of Theorem \ref{Thm:Directions_Consistency}}

The proof of Theorem \ref{Thm:Directions_Consistency} relies on  five intermediate results, one theorem (Theorem~\ref{Argmax}) and four lemmas (Lemmas~\ref{Argmax}--\ref{lem:uniform_convergence_M_n}). Theorem~\ref{Argmax} is a special case of the standard result on the consistency of M-estimators (see, e.g., Corollary 3.2.3 in  \citet{vanderVaart1996}) and involves an indexing set~$H$ which, for the sake of generality, we assume to be a compact subset of some metric space with distance function $d(\cdot ,\, \cdot)$. In the subsequent lemmas %
 and the proof of Theorem \ref{Thm:Directions_Consistency}, however, $H$ is the unit hypersphere~${\mathcal S}_{d-1}$, which is a compact subset of ${\mathbb R}^d$ equipped with the Euclidean norm $\Vert\cdot\Vert$. %

{\begin{Thm}\label{Argmax} Let $M_n\coloneqq \{M_n(h):h\in H\}$, $n\geq 1$ be a sequence of real-valued random processes defined over some  probability space~$(\Omega, {\mathcal F}, {\mathrm P})$ and  indexed on a compact subset $H$ of a metric space with distance function $d(\cdot ,\, \cdot)$, and denote by $M$
 a deterministic mapping from $H$ to $\mathbb R$ such that
  $$
\sup_{h\in H}|M_n(h)-M(h)|=o_{ \rm P}(1)\quad\text{ as }\quad n\to\infty.$$

Assume that   $h\mapsto M(h)$ is %
 continuous on $H$ and admits a %
  unique maximizer %
  {$\hat{h}\coloneqq \argmax_{h\in H}M(h)$}.\linebreak Let   $\hat{h}_n$, $n\in{\mathbb N}$ be a sequence of $H$-valued random variables,  defined over the same probability space $(\Omega, {\mathcal F}, {\mathrm P})$ as $M_n$, such that
  $$M_n(\hat{h}_n) \geq \sup_{h\in H}M_n(h)-o_{\rm P}(1)\quad\text{ as $n\to\infty$.}$$
   Then,  $d(\hat{h}_n,\, \hat{h})=o_{\rm P}(1)$ as $n\to\infty$.
\end{Thm}
}

\begin{proof}[Proof of Theorem~\ref{Argmax}]
The result follows from  Corollary 3.2.3 (i) in~\cite{vanderVaart1996}. To show this, we only need to verify the assumption on the maximum $\hat h\in H$ being well separated in the sense that, for every open set $O\subset H$ containing $\hat h$,
$
\sup_{h\notin O}M(h)<M(\hat{h}).%
$
This, however, is a direct consequence of the uniqueness of $\hat h\in H$, the  continuity of $h\mapsto M(h)$, and the compactness of $H$.
\end{proof}

In the proof of  Theorem \ref{Thm:Directions_Consistency}, %
$H$ will be the unit sphere ${\mathcal S}_{d-1}\subset{\mathbb R}^d$, equipped with the Euclidean norm~$\Vert\cdot\Vert$ and the distance~$d({\bf u}_1,\,{\bf u}_2) = \Vert {\bf u}_1 - {\bf u}_2 \Vert$.
  To apply Theorem %
   \ref{Argmax} in this proof, %
   we first establish the uniform convergence, in probability,  of the sample objective function $M_n^{\delta_1^{(n)}\!,\delta_2^{(n)}}: H={\mathcal S}_{d-1}\to{\mathbb R} $,  $\textbf{u}\mapsto M_n^{\delta_1^{(n)}\!,\delta_2^{(n)}}(\textbf{u})$ in~{\eqref{MEmpirical}}.} The proof for that convergence is based on Lemmas~\ref{lem:Figalli}--\ref{lem:uniform_convergence_M_n}.

To start with, recall the following result \cite[Theorem 1.1]{Fig18} by Figalli.
\begin{Lemma}\label{lem:Figalli}
Under Assumption \ref{AssA1}, $r{\textbf u}\mapsto\textbf{Q}_\pm (r{\textbf u})$ is a diffeomorphism of class
$C^{1,\alpha}_{\mathrm{loc}}(K)$ over any compact subset $K$ of the punctured unit ball $\mathbb{S}_d\setminus\{\textbf{0}\}$.
\end{Lemma}

 Next, Lemma~\ref{lem:bounds} provides %
 bounds for various quantities arising in the estimation procedure.

\begin{Lemma}
\label{lem:bounds}
Fix $\delta_1,\delta_2\in (0,1)$ such that $\delta_1<1-\delta_2$. For $\textbf{u}\in{\mathcal S}_{d-1}$ and $r\in [\delta_1,1-\delta_2]$,
let %
$
\gamma_{\textbf{u}}(r) \coloneqq  \textbf{Q}_\pm(r\textbf{u}),
$
and  define the (trimmed at the origin) arc–length function
 $
L^{(\delta_1)}(\textbf{Q}_\pm(r\textbf{u})) \coloneqq  \int_{\delta_1}^r \left\|\gamma_{\textbf{u}}'(t)\right\|\,\mathrm{d}t .
$
Then, under Assumption \ref{AssA1},%
\begin{enumerate}
\item[(i)] for every $\delta_1,\delta_2%
$, there exists a constant $C_{\delta_1,\delta_2}<\infty$ not depending on $\textbf{u}$  %
such that, for   all~$r,s\in[\delta_1,1-\delta_2]$, and $\textbf{u}\in{\mathcal S}_{d-1}$,
    \begin{enumerate}
        \item[(ia)] $ \|\gamma_{\textbf{u}}(r)\| \leq C_{\delta_1,\delta_2},\quad\|\gamma_{\textbf{u}}'(r)\| \leq C_{\delta_1,\delta_2}$;
        \item[(ib)] $\|\gamma_{\textbf{u}}(r)-\gamma_{\textbf{u}}(s)\|
    \leq C_{\delta_1,\delta_2} |r-s|,\quad  \|\gamma_{\textbf{u}}'(r)-\gamma_{\textbf{u}}'(s)\|
    \leq C_{\delta_1,\delta_2} |r-s|^\alpha
    $;
    \item[(ic)] $
   \Big|L^{(\delta_1)}(\textbf{Q}_\pm(r\textbf{u}))\Big| \leq C_{\delta_1,\delta_2}$, and
  $\Big|\frac{\mathrm{d}}{\mathrm{d}r}L^{(\delta_1)}(\textbf{Q}_\pm(r\textbf{u}))\Big| \leq C_{\delta_1,\delta_2}
    $.
    \end{enumerate}
    \item[(ii)] letting $\pi_{\delta_1,\delta_2}\coloneqq \{\delta_1 =  r_{n_{\delta_1}}<\dots<r_{n_{\delta_2}}=1-\delta_2\}$ characterize a   partition
    of $[\delta_1,1-\delta_2]$ with mesh size~$\|\pi_{\delta_1,\delta_2}\|\coloneqq {\max_{n_{\delta_1}\leq l\leq n_{\delta_2}}}(r_l-r_{l-1})$ and  defining   the
    corresponding polygonal length as
    $$
    L_n^{(\delta_1)}(\textbf{Q}_\pm(r\textbf{u}))
    \coloneqq  \sum_{{l=n_{\delta_1}+1}}^{k}
       \|\gamma_{\textbf{u}}(r_l)-\gamma_{\textbf{u}}(r_{l-1})\|
    $$
    where $k$ is such that $r_k\leq r<r_{k+1}$, we have,
 for all
    $\textbf{u}\in{\mathcal S}_{d-1}$ and  $r\in[\delta_1,1-\delta_2]$,
    \begin{enumerate}
        \item[(iia)] $
    0\leq L_n^{(\delta_1)}(\textbf{Q}_\pm(r\textbf{u}))\leq C_{\delta_1,\delta_2}
    $;
    \item[(iib)] $
    \big|L_n^{(\delta_1)}(\textbf{Q}_\pm(r\textbf{u})) - L^{(\delta_1)}(\textbf{Q}_\pm(r\textbf{u}))\big|
    \leq \left(1+\frac{2}{1+\alpha}\right)C_{\delta_1,\delta_2}\|\pi_{\delta_1,\delta_2}\|^\alpha$;
    \item[(iic)] $\big|(L_n^{(\delta_1)}(\textbf{Q}_\pm(r\textbf{u})))^2 - (L^{(\delta_1)}(\textbf{Q}_\pm(r\textbf{u})))^2\big|
    \leq 2C_{\delta_1,\delta_2}^2{\left(1+\frac{2}{1+\alpha}\right)}\|\pi_{\delta_1,\delta_2}\|^\alpha
    $.
    \end{enumerate}
\end{enumerate}
\end{Lemma}

\begin{proof}[\bf Proof of Lemma~\ref{lem:bounds}]

Statement {\it (i)}. Fixing $\delta_1,\delta_2\in (0,1)$, $\delta_1<1-\delta_2$, consider the compact annulus~$
K_{\delta_1,\delta_2} \coloneqq  \overline{\mathbb{S}_d(1-\delta_2)}\setminus \mathbb{S}_d(\delta_1),
$
which contains all  points of the form $r\textbf{u}$ with
$r\in[\delta_1,1-\delta_2]$ and $\textbf{u}\in{\mathcal S}_{d-1}$. It follows from  Lemma~\ref{lem:Figalli}  that  $\textbf{Q}_\pm$ and~$\nabla\textbf{Q}_\pm$ are continuous, hence bounded, on   $ K_{\delta_1,\delta_2}$. Therefore, there exists $C_{\delta_1,\delta_2}<\infty$ such that
$$
\sup_{  \delta_1\leq r\leq 1-\delta_2}\|\gamma_\textbf{u}(r)\|\leq C_{\delta_1,\delta_2}<\infty\quad\text{and}\quad \sup_{\delta_1\leq\|\textbf{x}\|\leq 1-\delta_2}\|\nabla\textbf{Q}_\pm(\textbf{x})\|_\mathrm{op}\leq C_{\delta_1,\delta_2} < \infty,
$$
where the constant $C_{\delta_1,\delta_2}$ does not depend on $\textbf{u}\in{\mathcal S}_{d-1}$ and  $\|\cdot\|_\mathrm{op}$ is the operator norm of the %
Jacobian~$\nabla\textbf{Q}_\pm$. For each $\textbf{u}\in{\mathcal S}_{d-1}$ and $r\in[\delta_1,1-\delta_2]$,
the chain rule and the Cauchy-Schwarz inequality yield
$$
\|\gamma_{\textbf{u}}'(r)\|
\leq \|\nabla\textbf{Q}_\pm(r\textbf{u})\|_\mathrm{op}\leq C_{\delta_1,\delta_2};
$$
statement {\it (ia)} follows.

As a consequence of {\it (ia)},  $\gamma'_\textbf{u}$ is (uniformly in $\textbf{u}$) bounded so that  {$\textbf{r}\mapsto \gamma_\textbf{u}(r)$} %
is  Lipschitz with  Lipschitz constant not depending on $\textbf{u}$. Hence,
 $
\|\gamma_\textbf{u}(r)-\gamma_\textbf{u}(s)\|\leq C_{\delta_1,\delta_2}|r-s|$ for all  $\textbf{u}\in{\mathcal S}_{d-1}$. On the other hand, since $\nabla\textbf{Q}_\pm$ is
$\alpha$-Hölder continuous on the compact annulus
$K_{\delta_1,\delta_2}$, there exists a constant~$C_{\delta_1,\delta_2}<\infty$ not depending on $\textbf{u}$ such that,
for all $r,s\in[\delta_1,1-\delta_2]$,
\begin{align*}
\|\gamma'_{\textbf{u}}(r)-\gamma'_{\textbf{u}}(s)\|=
\|[\nabla\textbf{Q}_\pm(r\textbf{u})
-\nabla\textbf{Q}_\pm(s\textbf{u})]\textbf{u}\|\leq
\|\nabla\textbf{Q}_\pm(r\textbf{u})
-\nabla\textbf{Q}_\pm(s\textbf{u})\|_{\rm op}\leq
C_{\delta_1,\delta_2}|r-s|^\alpha.
\end{align*}
Thus, $\gamma'_{\textbf{u}}$ is, uniformly in $\textbf{u}\in{\mathcal S}_{d-1}$,
$\alpha$-Hölder continuous; {\it (ib)} follows.

Turning to {\it (ic)}, by the fundamental theorem of calculus,
$$
L^{(\delta_1)}(\textbf{Q}_\pm(r\textbf{u}))
= \int_{\delta_1}^r \|\gamma_{\textbf{u}}'(t)\|\,\mathrm{d}t
\quad\text{and}\quad
\frac{\mathrm{d}}{\mathrm{d}r} L^{(\delta_1)}(\textbf{Q}_\pm(r\textbf{u})) = \|\gamma_{\textbf{u}}'(r)\|.
$$
The uniform bound on $\|\gamma_{\textbf{u}}'(r)\|$ in \unaAdded{\it (ia)} implies
$$
\left|L^{(\delta_1)}(\textbf{Q}_\pm(r\textbf{u}))\right|\leq C_{\delta_1,\delta_2}r\leq C_{\delta_1,\delta_2}<\infty \quad\text{and}\quad \left|\tfrac{\mathrm{d}}{\mathrm{d}r}L^{(\delta_1)}(\textbf{Q}_\pm(r\textbf{u}))\right|\leq C_{\delta_1,\delta_2}<\infty,
$$
which completes the proof of {\it (ic)}.

\bigskip

Statement {\it (ii)}.  %
The Lipschitz bound from {\it (ib)} yields
$$
L_n^{(\delta_1)}(\textbf{Q}_\pm(r\textbf{u}))
\leq \sum_{{l=n_{\delta_1}+1}}^{n_{\delta_2}}
   \|\gamma_{\textbf{u}}(r_l)-\gamma_{\textbf{u}}(r_{l-1})\| \leq  \sum_{{l=n_{\delta_1}+1}}^{n_{\delta_2}}
   C_{\delta_1,\delta_2}(r_l-r_{l-1})\leq C_{\delta_1,\delta_2}{(1-\delta_1-\delta_2)}\leq C_{\delta_1,\delta_2},
$$
hence {\it (iia)}. Using the fact that
 $
\gamma_{\textbf{u}}(r_l)-\gamma_{\textbf{u}}(r_{l-1})
= \int_{r_{l-1}}^{r_l} \gamma_{\textbf{u}}'(t)\,\mathrm{d}t
$
and the inequality $\big|\|A\|-\|B\|\big|\leq \|A-B\|$, we get
\begin{align*}
\int_{r_{l-1}}^{r_l}\|\gamma_{\textbf u}'(t)\|\mathrm{d}t
&-\left\|\gamma_{\textbf u}(r_l)-\gamma_{\textbf u}(r_{l-1})\right\| \\
&=\int_{r_{l-1}}^{r_l}\|\gamma_{\textbf u}'(t)\|\mathrm{d}t
-\left\|\int_{r_{l-1}}^{r_l}\gamma_{\textbf u}'(t),\mathrm{d}t\right\| \\
&=\int_{r_{l-1}}^{r_l}\|\gamma_{\textbf u}'(t)\|\mathrm{d}t
\pm \|\gamma_\textbf{u}'(r_{l-1})\|(r_l-r_{l-1})-\left\|\int_{r_{l-1}}^{r_l}\gamma_{\textbf u}'(t),\mathrm{d}t\right\| \\
&\leq \int_{r_{l-1}}^{r_l}\left(\|\gamma_{\textbf u}'(t)\|-\|\gamma_\textbf{u}'(r_{l-1})\|\right)\mathrm{d}t
+\left\|\gamma_\textbf{u}'(r_{l-1})(r_l-r_{l-1})-\int_{r_{l-1}}^{r_l}\gamma_{\textbf u}'(t)\mathrm{d}t\right\|\\
&\leq \int_{r_{l-1}}^{r_l}\|\gamma_{\textbf u}'(t)-\gamma_\textbf{u}'(r_{l-1})\|\mathrm{d}t
+\left\|\int_{r_{l-1}}^{r_l}\left(\gamma_\textbf{u}'(r_{l-1})-\gamma_{\textbf u}'(t)\right)\mathrm{d}t\right\|
\leq 2\int_{r_{l-1}}^{r_l}\|\gamma_{\textbf u}'(t)-\gamma_\textbf{u}'(r_{l-1})\|\mathrm{d}t,
\end{align*}
 where the last inequality follows from  Jensen's  inequality and the convexity of the norm.
Still from {\it (ib)},
$$
\left\|\gamma_{\textbf{u}}'(t)-\gamma_{\textbf{u}}'(r_{l-1})\right\|
\leq C_{\delta_1,\delta_2} (t-r_{l-1})^\alpha,
\quad t\in[r_{l-1},r_l],
$$
hence
$$
{0\leq} \int_{r_{l-1}}^{r_l}\|\gamma_{\textbf{u}}'(t)\|\mathrm{d}t
-
\left\|\gamma_{\textbf{u}}(r_l)-\gamma_{\textbf{u}}(r_{l-1})\right\|
\le
2C_{\delta_1,\delta_2}\int_{r_{l-1}}^{r_l} (t-r_{l-1}) ^\alpha\mathrm{d}t
=
\frac{2}{1+\alpha}C_{\delta_1,\delta_2}(r_l-r_{l-1})^{\alpha+1}.
$$
Let $k$ be such that $r_k\leq r<r_{k+1}$. Then
\begin{align*}
\left|L^{(\delta_1)}(\textbf Q_\pm(r\textbf u))
-L_n^{(\delta_1)}(\textbf Q_\pm(r\textbf u))\right|&\leq
\left|L^{(\delta_1)}(\textbf Q_\pm(r_k\textbf u))
-L_n^{(\delta_1)}(\textbf Q_\pm(r_k\textbf u))\right|
+
\int_{r_k}^{r}\|\gamma'_{\textbf u}(t)\|\,\mathrm dt\\
&\leq
2\frac{C_{\delta_1,\delta_2}}{1+\alpha}
\|\pi_{\delta_1,\delta_2}\|^\alpha
+
C_{\delta_1,\delta_2}\|\pi_{\delta_1,\delta_2}\|
\leq
C_{\delta_1,\delta_2}
\left(1+\frac{2}{1+\alpha}\right)
\|\pi_{\delta_1,\delta_2}\|^\alpha,
\end{align*}
hence {\it (iib)}, with a  bound that does not depend on $\textbf{u}$, $r$, nor the choice of the points in the  partition $\pi_{\delta_1,\delta_2}$. %

As for {\it (iic)}, observe that
\begin{align}\nonumber
\Big|\left(L^{(\delta_1)}(\textbf{Q}_\pm(r\textbf{u}))\right)^2 - \left(L_n^{(\delta_1)}(\textbf{Q}_\pm(r\textbf{u}))\right)^2\Big|=\Big|L^{(\delta_1)}(\textbf{Q}_\pm(r\textbf{u}))& -  L_n^{(\delta_1)}(\textbf{Q}_\pm(r\textbf{u}))\Big| \\
&\times\Big|L^{(\delta_1)}(\textbf{Q}_\pm(r\textbf{u})) + L_n^{(\delta_1)}(\textbf{Q}_\pm(r\textbf{u}))\Big|.\label{eq:a^2-b^2}
\end{align}
From {\it (iia)} and {\it (ic)}, $L_n^{(\delta_1)}(\textbf{Q}_\pm(r\textbf{u}))%
$ and $L^{(\delta_1)}(\textbf{Q}_\pm(r\textbf{u}))%
$ both are bounded by $C_{\delta_1,\delta_2}$ which, in combination with {\it (iib)} and \eqref{eq:a^2-b^2},  yields
\begin{align*}
\Big|\left(L^{(\delta_1)}(\textbf{Q}_\pm(r\textbf{u}))\right)^2 - \left(L_n^{(\delta_1)}(\textbf{Q}_\pm(r\textbf{u}))\right)^2\Big|&\leq {2C_{\delta_1,\delta_2}^2{\left(1+\frac{2}{1+\alpha}\right)}}\|\pi_{\delta_1,\delta_2}\|^{\alpha},
\end{align*}
reducing  to
\begin{align*}
\Big|\left(L^{(\delta_1)}(\textbf{Q}_\pm(r\textbf{u}))\right)^2 - \left(L_n^{(\delta_1)}(\textbf{Q}_\pm(r\textbf{u}))\right)^2\Big|&\leq 2C_{\delta_1,\delta_2}^2{\left(1+\frac{2}{1+\alpha}\right)}n_{\delta_2}^{-\alpha}
\end{align*}
for  equidistant grids. This completes the proof of  {\it (iic)} and Lemma~\ref{lem:bounds}.
\end{proof}

The next lemma establishes the continuity of  expected trimmed arc lengths  with respect to trimming levels.  Let
 $\displaystyle
M^{\delta_1,\delta_2}(\textbf u)
\coloneqq
\int_{\delta_1}^{1-\delta_2}
\left(L^{(\delta_1)}(\textbf Q_\pm(r\textbf u))\right)^2\,\mathrm d r
$
with $\displaystyle L^{(\delta_1)}(\textbf Q_\pm(r\textbf u))
\coloneqq
\int_{\delta_1}^r
\|\gamma_{\textbf u}'(t)\|\,\mathrm d t $, %
$ \textbf u\in\mathcal S_{d-1}.
 $

\begin{Lemma}\label{lem:M_delta_continuity}

Assume that the conditions of Lemma~\ref{lem:bounds} hold. Then, for any sequences
 $\delta_1^{(n)}\downarrow\varepsilon_0$ and~$
\delta_2^{(n)}\downarrow\delta_0$ as $n\to\infty$
with $0<\varepsilon_0\leq 1-\delta_0<1$, it holds that
$$
\sup_{\textbf u\in\mathcal S_{d-1}}
\left|
M^{\delta_1^{(n)},\delta_2^{(n)}}(\textbf u)
-
M^{\varepsilon_0,\delta_0}(\textbf u)
\right|
\to 0\qquad\text{as $n\to\infty$} .
$$
\end{Lemma}

\begin{proof}[\bf Proof of Lemma~\ref{lem:M_delta_continuity}]
Set
$a_n\coloneqq \delta_1^{(n)}$ and
 $b_n\coloneqq \delta_2^{(n)}$.
Then
\begin{equation}\label{integrals}
M^{a_n,b_n}(\textbf u)
=
\int_{a_n}^{1-b_n}
\left(L^{(a_n)}(\textbf Q_\pm(r\textbf u))\right)^2\,\mathrm d r
\quad
\text{and}
\quad
M^{\varepsilon_0,\delta_0}(\textbf u)
=
\int_{\varepsilon_0}^{1-\delta_0}
\left(L^{(\varepsilon_0)}(\textbf Q_\pm(r\textbf u))\right)^2\,\mathrm d r .
\end{equation}

Since $a_n\downarrow\varepsilon_0$ and $b_n\downarrow\delta_0$, for  $n$ sufficiently large,
$a_n\geq \varepsilon_0/2$ and $b_n\geq \delta_0/2$.
Hence,   $r$  in the integrals~\eqref{integrals} belongs to
$[\varepsilon_0/2,\,1-\delta_0/2]$.
 By Lemma~\ref{lem:bounds} {\it (ic)}, there exists a  constant $C_0\coloneqq C_{\varepsilon_0/2,\delta_0/2}<\infty$
such that
$$
\sup_{\textbf u\in\mathcal S_{d-1}}
\sup_{r\in[\varepsilon_0/2,\,1-\delta_0/2]}
L^{(\varepsilon_0)}(\textbf Q_\pm(r\textbf u))
\leq C_0,
\quad
\quad
\sup_{\textbf u\in\mathcal S_{d-1}}
\sup_{r\in[\varepsilon_0/2,\,1-\delta_0/2]}
L^{(a_n)}(\textbf Q_\pm(r\textbf u))
\leq C_0,
$$
and
$$
\sup_{\textbf u\in\mathcal S_{d-1}}
\sup_{r\in[\varepsilon_0/2,\,1-\delta_0/2]}
\|\gamma_{\textbf u}'(r)\|
\leq C_0 .
$$

We first compare the two length functionals on the common interval $[a_n,\,1-b_n].$  For $r\in[a_n,1-b_n]$,
$$
L^{(\varepsilon_0)}(\textbf Q_\pm(r\textbf u))
=
\int_{\varepsilon_0}^{r}\|\gamma_{\textbf u}'(t)\|\,\mathrm d t \quad \text{and}\quad
L^{(a_n)}(\textbf Q_\pm(r\textbf u))
=
\int_{a_n}^{r}\|\gamma_{\textbf u}'(t)\|\,\mathrm d t .
$$
Therefore,
$
\left|
L^{(\varepsilon_0)}(\textbf Q_\pm(r\textbf u))
-
L^{(a_n)}(\textbf Q_\pm(r\textbf u))
\right|
=
\int_{\varepsilon_0}^{a_n}
\|\gamma_{\textbf u}'(t)\|\,\mathrm d t
$
and, consequently,
$$
\sup_{\textbf u\in\mathcal S_{d-1}}
\sup_{r\in[a_n,1-b_n]}
\left|
L^{(\varepsilon_0)}(\textbf Q_\pm(r\textbf u))
-
L^{(a_n)}(\textbf Q_\pm(r\textbf u))
\right|
\leq
C_0 |a_n-\varepsilon_0|.
$$
Recalling that $x^2-y^2=(x-y)(x+y)$, we get
\begin{align}\nonumber
&\sup_{\textbf u\in\mathcal S_{d-1}}
\int_{a_n}^{1-b_n}
\left|
\left(L^{(a_n)}(\textbf Q_\pm(r\textbf u))\right)^2
-
\left(L^{(\varepsilon_0)}(\textbf Q_\pm(r\textbf u))\right)^2
\right|\,\mathrm d r\\
&\quad \leq
\sup_{\textbf u\in\mathcal S_{d-1}}
\int_{a_n}^{1-b_n}
\left|
L^{(a_n)}(\textbf Q_\pm(r\textbf u))
-
L^{(\varepsilon_0)}(\textbf Q_\pm(r\textbf u))
\right|\left(
L^{(a_n)}(\textbf Q_\pm(r\textbf u))
+
L^{(\varepsilon_0)}(\textbf Q_\pm(r\textbf u))
\right)\,\mathrm d r \nonumber\\
&\quad\leq
\left(C_0 |a_n-\varepsilon_0|\right)(2C_0)(1-b_n-a_n)
\leq
2C_0^2 |a_n-\varepsilon_0|.\label{bound1}
\end{align}

It remains to control the difference between the domains of integration. The contributions of the  lower and upper endpoints satisfy
\begin{equation}\label{bound2}
\sup_{\textbf u\in\mathcal S_{d-1}}
\int_{\varepsilon_0}^{a_n}
\left(L^{(\varepsilon_0)}(\textbf Q_\pm(r\textbf u))\right)^2\,\mathrm d r
\leq
C_0^2 |a_n-\varepsilon_0|
\end{equation}
and
\begin{equation}\label{bound3}
\sup_{\textbf u\in\mathcal S_{d-1}}
\int_{1-b_n}^{1-\delta_0}
\left(L^{(\varepsilon_0)}(\textbf Q_\pm(r\textbf u))\right)^2\,\mathrm d r
\leq
C_0^2 |b_n-\delta_0|,
\end{equation}
respectively. Combining the three bounds \eqref{bound1}--\eqref{bound3} yields
$$
\sup_{\textbf u\in\mathcal S_{d-1}}
\left|
M^{a_n,b_n}(\textbf u)
-
M^{\varepsilon_0,\delta_0}(\textbf u)
\right|
\leq
3C_0^2 |a_n-\varepsilon_0|
+
C_0^2 |b_n-\delta_0|
$$
where the right-hand side converges to zero since $a_n\downarrow\varepsilon_0$ and $b_n\downarrow\delta_0$, as $n\to\infty$. The claim follows.
\end{proof}

Based on Lemmas  \ref{lem:bounds} and \ref{lem:M_delta_continuity} , we now prove the uniform (in ${\textbf u}\in{\mathcal S}_{d-1}$) convergence of the sample objective function to the population one.

\begin{Lemma}\label{lem:uniform_convergence_M_n}
  Under the assumptions and with the notation of Theorem~\ref{Thm:Directions_Consistency},
    define, for   $\delta_1,\delta_2>0$  such that~$\delta_1<1-\delta_2$,
    \begin{equation}\label{def1}
M_n^{\delta_1,\delta_2}:{\mathcal S}_{d-1}\to \mathbb{R}, \, \textbf{u}\mapsto M_n^{\delta_1,\delta_2}(\textbf{u})\coloneqq \frac{1}{m_R+1%
}\sum_{j=m_{\delta_1}}^{m_{\delta_2}} \left(L_n^{(\delta_1)}\left(\overline{\textbf{Q}}_\pm^{(n)}(r_j{\textbf u})\right)\right)^2
\end{equation}
with   $m_{\delta_1}\coloneqq \lfloor\delta_1(m_R+1)\rfloor$ and   $m_{\delta_2}\coloneqq \lfloor(1-\delta_2)(m_R+1)\rfloor$. Considering the partition $\pi_{\delta_1\delta_2}\coloneqq \{r_{m_{\delta_1}}\leq ~\!\cdots\leq~\!r_{m_{\delta_2}}\}$  of~$(\delta_1,1-\delta_2)$, which is contained in the partition $\pi\coloneqq \{r_0<r_1<\cdots<r_{m_R}\}$ where~$r_j\coloneqq j/(m_R+1)$ for~$j=0,\dots,m_R$, let %
    \begin{equation}\label{def2}
M^{\delta_1,\delta_2}: {\mathcal S}_{d-1}\to \mathbb{R}, \, \textbf{u}\mapsto M^{\delta_1,\delta_2}(\textbf{u})\coloneqq \int_{\delta_1}^{1-\delta_2}
\left(L^{(\delta_1)}(\textbf Q_\pm(r\textbf u))\right)^2\mathrm \mathrm d r.%
\end{equation}
Then, for  $1>\delta_1^{(n)}\downarrow\varepsilon_0$ and  $1>\delta_2^{(n)}\downarrow\delta_0$, where $0<\varepsilon_0$ and $\delta_0<1$ are as in Assumption \ref{AssA2},
$$
\big\|M_n^{\delta_1^{(n)},\delta_2^{(n)}}-M^{\varepsilon_0,\delta_0}\big\|_\infty\coloneqq \sup_{\textbf{u}\in\mathcal{S}_{d-1}}\big \vert M_n^{\delta_1^{(n)},\delta_2^{(n)}}(\textbf{u})-M^{\varepsilon_0,\delta_0}(\textbf{u})\big\vert = o_{\rm P} (1)\quad\text{as $n\to\infty$}.
$$
\end{Lemma}

\begin{proof}[\bf Proof of Lemma \ref{lem:uniform_convergence_M_n}]

For any $\delta_1,\delta_2>0$, $\pi_{\delta_1\delta_2}$ is a subpartition of $\pi$;  both $\pi_{\delta_1\delta_2}$ and $\pi$ have  mesh size~$(m_R+1)^{-1}$.  Observe that $M_n^{\delta_1\delta_2}$ as defined in~\eqref{def2} depends on the partition $\pi$, a dependence which, for the sake of simplicity, we do not reflect in the notation.
Fix~$\delta_1\in [\varepsilon_0,1)$ and $\delta_2\in [\delta_0,1)$ such that $\delta_1<1-\delta_2$ and  assume, {for simplicity of the exposition},  that there  exist $j=m_{\delta_1}\in\{1,\dots,m_R\}$ and $k=m_{\delta_2}\in\{1,\dots,m_R\}$ such that $r_{m_{\delta_1}}=\delta_1$ and $r_{m_{\delta_2}}=1-\delta_2$. Iin the general case, the discrepancy between the trimming endpoints and the nearest gridpoints is of order $O(m_R^{-1})$: the resulting additional terms are therefore absorbed into the existing $O(m_R^{-1})$ remainder and do not affect the proof.  The triangle inequality implies
 \begin{align*}
\Big\vert M_n^{\delta_1,\delta_2}(\textbf{u})-M^{\delta_1,\delta_2}(\textbf{u})\Big\vert =&\left|\frac{1}{m_R+1}\sum_{j=m_{\delta_1}}^{m_{\delta_2}}\left(L_n^{(\delta_1)}\left((\overline{\textbf{Q}}_\pm^{(n)}(r_j{\textbf u})\right)\right)^2   -\int_{\delta_1}^{1-\delta_2}\left(L^{(\delta_1)}\left(\textbf{Q}_\pm
(r{\textbf u})\right)\right)^2
  \mathrm{d}r\right |\nonumber\\
  \leq & \left|\frac{1}{m_R+1}\sum_{j=m_{\delta_1}}^{m_{\delta_2}}\left(L_n^{(\delta_1)}\left((\overline{\textbf{Q}}_\pm^{(n)}(r_j{\textbf u})\right)\right)^2 - \frac{1}{m_R+1}\sum_{j=m_{\delta_1}}^{m_{\delta_2}}\left(L_n^{(\delta_1)}\left(\textbf{Q}_\pm(r_j{\textbf u})\right)\right)^2\right |%
  \\
  &+ \left|\frac{1}{m_R+1}\sum_{j=m_{\delta_1}}^{m_{\delta_2}}\left(L_n^{(\delta_1)}\left(\textbf{Q}_\pm(r_j{\textbf u})\right)\right)^2-\frac{1}{m_R+1}\sum_{j=m_{\delta_1}}^{m_{\delta_2}}\left(L^{(\delta_1)}\left(\textbf{Q}_\pm(r_j{\textbf u})\right)\right)^2\right |%
  \\
  &+\left| \frac{1}{m_R+1}\sum_{j=m_{\delta_1}}^{m_{\delta_2}}\left( L^{(\delta_1)}\left(\textbf{Q}_\pm(r_j{\textbf u})\right)\right)^2 -\int_{\delta_1}^{1-\delta_2}\left(L^{(\delta_1)}\left(\textbf{Q}_\pm(\rho{\textbf u})\right)\right)^2
  \mathrm{d}r\right|.%
  \\
  \eqqcolon& A_1 + A_2 + A_3,\ \text{ say.}
\end{align*}

\medskip

First consider $A_3$.
For fixed $\textbf{u}\in{\mathcal S}_{d-1}$,
\begin{equation}\label{Riemann}
\frac{1}{m_R+1}\sum_{j=m_{\delta_1}}^{m_{\delta_2}} \left(L^{(\delta_1)}\left(\textbf{Q}_\pm(r_j\textbf{u})\right)\right)^2
\end{equation}
is the Riemann sum  associated with the equidistant grid $\pi_{\delta_1,\delta_2}$, with mesh size $(m_R+1)^{-1}$, for the integral~\eqref{def2} with inte\-grand~$r\mapsto g_\textbf{u}(r)\coloneqq \left(L^{(\delta_1)}\Big(\textbf{Q}_\pm(r\textbf{u})\right)\Big)^2$ defining $M^{\delta_1,\delta_2}(\textbf{u})$.
By the chain rule, for~$r\in[\delta_1,1-\delta_2]$ and $\textbf{u}\in{\mathcal S}_{d-1}$
\begin{equation}\label{g'}
g_\textbf{u}'(r)=2L^{(\delta_1)}(\textbf{Q}_\pm(r\textbf{u}))\frac{\mathrm{d}}{\mathrm{d}r}L^{(\delta_1)}(\textbf{Q}_\pm(r\textbf{u})).
\end{equation}
Plugging the bounds from Lemma~\ref{lem:bounds} {\it (ib)} into \eqref{g'}  yields
\begin{equation}\label{eq:g'bound}
\sup_{\textbf{u}\in{\mathcal S}_{d-1}}\sup_{r\in[\delta_1,1-\delta_2]}|g_\textbf{u}'(r)|\leq 2C_{\delta_1,\delta_2}^2.
\end{equation}
Plugging further  \eqref{eq:g'bound} into the Riemann sum \eqref{Riemann}, we obtain
\begin{equation}\label{eq:3.6_delta_bound}
    \sup_{\textbf{u}\in{\mathcal S}_{d-1}} \left|\frac{1}{m_R+1}\sum_{j=m_{\delta_1}}^{m_{\delta_2}} \left(L^{(\delta_1)}\left(\textbf{Q}_\pm(r_j\textbf{u})\right)\right)^2 - \int_{\delta_1}^{1-\delta_2}\left[\left(L^{(\delta_1)}\left(\textbf{Q}_\pm(r\textbf{u})\right)\right)^2\right]\mathrm{d}r\right|   \leq C_{\delta_1,\delta_2}^2m_R^{-1},
\end{equation}
where we have used that interval length is $1-\delta_1-\delta_2\leq 1$ and, since by  construction $\pi_{\delta_1,\delta_2}\subset\pi$ is an equidistant grid, $\|\pi_{\delta_1,\delta_2}\|=(m_R+1)^{-1} \leq m_R^{-1}$.

\medskip

As for  $A_2$, the triangle inequality and inequality {\it (iic)} of Lemma \ref{lem:bounds}  yield
\begin{align*}
 A_2&\leq \frac{1}{m_R+1}\sum_{j=m_{\delta_1}}^{m_{\delta_2}}\left|\left(L_n^{(\delta_1)}\left(\textbf{Q}_\pm(r_j{\textbf u})\right)\right)^2-\left(L^{(\delta_1)}\left(\textbf{Q}_\pm(r_j{\textbf u})\right)\right)^2\right |\\
    &\leq \sup_{0 <  \delta_1\leq r\leq 1-\delta_2}\left|\left(L_n^{(\delta_1)}\left(\textbf{Q}_\pm(r{\textbf u})\right)\right)^2-\left(L^{(\delta_1)}\left(\textbf{Q}_\pm(r{\textbf u})\right)\right)^2\right | \\
    &\leq 2C_{\delta_1,\delta_2}^2{\left(1+\frac{2}{1+\alpha}\right)}\|\pi_{\delta_1,\delta_2}\|^\alpha\leq 2C_{\delta_1,\delta_2}^2{\left(1+\frac{2}{1+\alpha}\right)}m_R^{-\alpha}.
\end{align*}
That  bound does not depend on $\textbf{u}$, so that it also holds for the $\sup$ over ${\textbf u}\in{\mathcal S}_{d-1}$:
\begin{equation}\label{eq:3.5_delta_bound}
    \sup_{\textbf{u}\in{\mathcal S}_{d-1}}\left|\frac{1}{m_R+1}\sum_{j=m_{\delta_1}}^{m_{\delta_2}}\left(L_n^{(\delta_1)}\left(\textbf{Q}_\pm(r_j{\textbf u})\right)\right)^2-\frac{1}{m_R+1}\sum_{j=m_{\delta_1}}^{m_{\delta_2}}\left(L^{(\delta_1)}\left(\textbf{Q}_\pm(r_j{\textbf u})\right)\right)^2\right |\leq 2C_{\delta_1,\delta_2}^2{\left(1+\frac{2}{1+\alpha}\right)}m_R^{-\alpha}.
\end{equation}

Turning to $A_1$,  the triangle inequality entails
\begin{align}
|L_n^{(\delta_1)}\left((\overline{\textbf{Q}}_\pm^{(n)}(r_j{\textbf u})\right) - L_n^{(\delta_1)}\left(\textbf{Q}_\pm(r_j{\textbf u})\right)|&=\Bigg|\sum_{l=m_{\delta_1}}^j\!\!\left(\Vert\overline{\textbf{Q}}_\pm^{(n)}(r_l{\textbf u})-\overline{\textbf{Q}}_\pm^{(n)}(r_{l-1}{\textbf u})\Vert-\Vert{\textbf{Q}}_\pm(r_l{\textbf u})-{\textbf{Q}}_\pm(r_{l-1}{\textbf u})\Vert\right)\Bigg|\nonumber\\
&\leq \sum_{\unaAdded{l}=m_{\delta_1}}^{j}\!\!\left(\Vert\overline{\textbf{Q}}_\pm^{(n)}(r_l{\textbf u})-{\textbf{Q}}_\pm(r_l{\textbf u})\Vert+\Vert\overline{\textbf{Q}}_\pm^{(n)}(r_{l-1}{\textbf u})-{\textbf{Q}}_\pm(r_{l-1}{\textbf u})\Vert\right).\label{eq:L_n expansion}
\end{align}%
Now,  for every $r\textbf{u}\in K_{\delta_1,\delta_2}\coloneqq \overline{\mathbb{S}(1-\delta_2)}\setminus\mathbb{S}(\delta_1)$, and  $\textbf{x}\coloneqq \overline{\textbf{Q}}_\pm^{(n)}(r\textbf{u})$, we have
\begin{align*}
    \|\overline{\textbf{Q}}_\pm^{(n)}(r\textbf{u})-\textbf{Q}_\pm(r\textbf{u})\|&= \|\overline{\textbf{Q}}_\pm^{(n)}(\overline{\textbf{F}}_\pm^{(n)}(\textbf{x}))-\textbf{Q}_\pm(\overline{\textbf{F}}_\pm^{(n)}(\textbf{x}))\|\\
    &=\|\textbf{x}-\textbf{Q}_\pm(\overline{\textbf{F}}_\pm^{(n)}(\textbf{x})-{\textbf{F}}_\pm(\textbf{x})+{\textbf{F}}_\pm(\textbf{x}))\|
    \\
    &
    =\|\textbf{Q}_\pm(\textbf{F}_\pm(\textbf{x}))-\textbf{Q}_\pm(\overline{\textbf{F}}_\pm^{(n)}(\textbf{x})-{\textbf{F}}_\pm(\textbf{x})+{\textbf{F}}_\pm(\textbf{x}))\|.
\end{align*}
Observe that $\|\overline{\textbf{F}}_\pm^{(n)}(\textbf{x})\|=\|r\textbf{u}\|= r\in[\delta_1, 1-\delta_2]$. On the other hand,
\begin{align*}
\|{\textbf{F}}_\pm(\textbf{x})\|&\leq
\|\overline{\textbf{F}}_\pm^{(n)}(\textbf{x})\|+\|{\textbf{F}}_\pm(\textbf{x})-\overline{\textbf{F}}_\pm^{(n)}(\textbf{x})\|\leq 1-(\delta_2-s_n)
\end{align*}
and
\begin{align*}
\|{\textbf{F}}_\pm(\textbf{x})\|&\geq
\|\overline{\textbf{F}}_\pm^{(n)}(\textbf{x})\|-\|{\textbf{F}}_\pm(\textbf{x})-\overline{\textbf{F}}_\pm^{(n)}(\textbf{x})\|\geq \delta_1-s_n.
\end{align*}
Moreover, for every $t\in[0,1]$, setting
$z_t\coloneqq (1-t)\textbf{F}_\pm(\textbf x)
+t\overline{\textbf F}^{(n)}_\pm(\textbf x)$, we have,
since $\|\textbf F_\pm(\textbf x)-\overline{\textbf F}^{(n)}_\pm(\textbf x)\|\le s_n,$
$$
\|z_t\|\ge \|\textbf F_\pm(\textbf x)\|-s_n
\geq \delta_1-2s_n\quad\text{ and }\quad \|z_t\|\le \|\textbf F_\pm(\textbf x)\|+s_n
\leq 1-(\delta_2-2s_n).
$$
Hence, the   segment   $[\textbf F_\pm(\textbf x), \overline{\textbf F}^{(n)}_\pm(\textbf x)]$ is contained in
$K_{\delta_1-2s_n,\delta_2-2s_n}$. Therefore, by the mean value theorem and Lemma~\ref{lem:bounds},
\begin{equation}\label{eq:uniform_convergence_of_Qn}
\|\overline{\textbf Q}^{(n)}_\pm(r\textbf u)-\textbf Q_\pm(r\textbf u)\|
\leq
C_{\delta_1-2s_n,\delta_2-2s_n}
\|\textbf F_\pm(\textbf x)-\overline{\textbf F}^{(n)}_\pm(\textbf x)\|
\leq
C_{\delta_1-2s_n,\delta_2-2s_n}s_n
\end{equation}
for $n$ large enough  that $0<\delta_i-2s_n<1$, $i=1,2$.%

\medskip

Plugging \eqref{eq:uniform_convergence_of_Qn} into \eqref{eq:L_n expansion}, we get

\begin{equation*}
\left|L_n^{(\delta_1)}\left(\overline{\textbf{Q}}_\pm^{(n)}(r_j{\textbf u})\right) - L_n^{(\delta_1)}\left(\textbf{Q}_\pm(r_j{\textbf u})\right)\right|\leq
2\sum_{l=m_{\delta_1}}^{j}C_{\delta_1-2s_n,\delta_2-2s_n}s_n\leq 2m_RC_{\delta_1-2s_n,\delta_2-2s_n}s_n,
\end{equation*}
that is,
\begin{equation}\label{eq:L_n of U_n and Q}
L_n^{(\delta_1)}\left(\overline{\textbf{Q}}_\pm^{(n)}(r_j{\textbf u})\right)\leq   L_n^{(\delta_1)}\left(\textbf{Q}_\pm(r_j{\textbf u})\right) +2 m_RC_{\delta_1-2s_n,\delta_2-2s_n}s_n.
\end{equation}

In view of Lemma \ref{lem:bounds} {\it (iia)},
$\displaystyle\sup_{ \delta_1\leq r\leq 1-\delta_2} L_n^{(\delta_1)}\left(\textbf{Q}_\pm(r{\textbf u})\right)\leq C_{\delta_1,\delta_2}$, a bound that does not depend on $\textbf{u}$. Plugging this bound   into \eqref{eq:L_n of U_n and Q}, we get a uniform (in $\textbf{u}\in {\mathcal S}_{d-1}$) bound on the polygonal length of the sample curves:
$$
\sup_{\textbf{u}\in{\mathcal S}_{d-1}}\sup_{ \delta_1\leq r\leq 1-\delta_2}L_n^{(\delta_1)}(\overline{\textbf{Q}}_\pm^{(n)}(r{\textbf u}))\leq C_{\delta_1,\delta_2} +2C_{\delta_1-2s_n,\delta_2-2s_n}m_Rs_n.
$$
Using the identity $a^2-b^2=(a-b)(a+b)$, we further have
\begin{align}\label{L^2 q_n-Q bound}
\left|\left(L_n^{(\delta_1)}\left(\overline{\textbf{Q}}_\pm^{(n)}(r_j{\textbf u})\right)\right)^2\right. &- \left.\left(L_n^{(\delta_1)}\left(\textbf{Q}_\pm(r_j{\textbf u})\right)\right)^2\right|\nonumber\\
&= \left|L_n^{(\delta_1)}(\overline{\textbf{Q}}_\pm^{(n)}(r_j{\textbf u})) - L_n^{(\delta_1)}\left(\textbf{Q}_\pm(r_j{\textbf u})\right)\ \right|\left(L_n^{(\delta_1)}(\overline{\textbf{Q}}_\pm^{(n)}(r_j{\textbf u})) + L_n^{(\delta_1)}\left(\textbf{Q}_\pm(r_j{\textbf u})\right)\ \right)\nonumber\\
&\leq 2m_RC_{\delta_1-2s_n,\delta_2-2s_n}s_n(C_{\delta_1,\delta_2} +2C_{\delta_1-2s_n,\delta_2-2s_n}m_Rs_n),
\end{align}
\vspace{-12mm}

\noindent a bound which again is uniform in $\textbf{u}\in {\mathcal S}_{d-1}$. The triangle inequality finally gives
\begin{align}\label{eq:3.4 delta bound}
    \sup_{\textbf{u}\in{\mathcal S}_{d-1}}\Bigg| \frac{1}{m_R+1}&\sum_{j=m_{\delta_1}}^{m_{\delta_2}}\left(L_n^{(\delta_1)}\left((\overline{\textbf{Q}}_\pm^{(n)}(r_j{\textbf u})\right)\right)^2 - \frac{1}{m_R+1}\sum_{j=m_{\delta_1}}^{m_{\delta_2}}\left(L_n^{(\delta_1)}\left(\textbf{Q}_\pm(r_j{\textbf u})
  \right)\right)^2\Bigg |\nonumber\\
  &\hspace{15mm}\leq  {2m_RC_{\delta_1-2s_n,\delta_2-2s_n}s_n(2C_{\delta_1,\delta_2} +2C_{\delta_1-2s_n,\delta_2-2s_n}m_Rs_n).}
\end{align}

Combining \eqref{eq:3.6_delta_bound}, \eqref{eq:3.5_delta_bound}, and \eqref{eq:3.4 delta bound}, we get
\begin{align}\label{eq:M_delta_-Mn_delta}
\sup_{\textbf{u}\in{\mathcal S}_{d-1}}
\left|M_n^{\delta_1,\delta_2}(\textbf{u})
-M^{\delta_1,\delta_2}(\textbf{u})\right|
\leq{}&
{
C_{\delta_1,\delta_2}^2m_R^{-1}
+2C_{\delta_1,\delta_2}^2
\left(1+\frac{2}{1+\alpha}\right)m_R^{-\alpha}
}
\nonumber\\
&{
+2m_RC_{\delta_1-2s_n,\delta_2-2s_n}s_n
\left(
2C_{\delta_1,\delta_2}
+2C_{\delta_1-2s_n,\delta_2-2s_n}m_Rs_n
\right).
}
\end{align}

Take now $1>\delta_2=\delta_2^{(n)}\downarrow\delta_0$ and $1>\delta_1=\delta_1^{(n)}\downarrow\varepsilon_0$. As previously discussed, $s_n\to 0$ a.s.\ as $n\to\infty$. For~$n$ large enough, thus,  $s_n\leq \min\{\delta_0/4,\varepsilon_0/4\}$,
hence  $\delta_1^{(n)}-2s_n\geq \varepsilon_0/2$, $\delta_2^{(n)}-2s_n\geq \delta_0/2$.
From the proof of Lemma \ref{lem:bounds}, it is straightforward to see that $(\delta_1,\delta_2)\mapsto C_{\delta_1,\delta_2}$ is a decreasing function in both its arguments. Therefore, $C_{\delta_1',\delta_2'}$, with $\delta_i'\leq \delta_i$, $i=1,2$, can be substituted for $C_{\delta_1,\delta_2}$ in the inequalities of  in Lemma \ref{lem:bounds}.  It follows that we can bound \eqref{eq:M_delta_-Mn_delta} with  constants that depend only on $\delta_0,\varepsilon_0$:
$$
\sup_{\textbf{u}\in{\mathcal S}_{d-1}}\left|M_n^{\delta_1^{(n)},\delta_2^{(n)}}(\textbf{u})-M^{\delta_1^{(n)},\delta_2^{(n)}}(\textbf{u}) \right|\leq  {C_{\varepsilon_0,\delta_0}^2m_R^{-1}+2C_{\varepsilon_0,\delta_0}^2{\left(1+\frac{2}{1+\alpha}\right)}m_R^{-\alpha}+4m_RC_{\varepsilon_0/2,\delta_0/2}^2s_n(1 +2m_Rs_n)}.
$$
By assumption, $m_R$ is growing slow enough that $m_Rs_n=o_{\rm P}(1)$ %
 as $n\to\infty$, so that
\begin{equation}\label{eq:M_n-M delta bound}
\sup_{\textbf{u}\in{\mathcal S}_{d-1}}\left|M_n^{\delta_1^{(n)},\delta_2^{(n)}}(\textbf{u})-M^{\delta_1^{(n)},\delta_2^{(n)}}(\textbf{u}) \right|=o_{\rm P}(1).
\end{equation}
Applying  the triangle inequality again,
\begin{align*}
\sup_{\textbf{u}\in{\mathcal S}_{d-1}}\left|M_n^{\delta_1^{(n)},\delta_2^{(n)}}(\textbf{u})-M^{\varepsilon_0,\delta_0}(\textbf{u}) \right|\leq \sup_{\textbf{u}\in{\mathcal S}_{d-1}}\left|M_n^{\delta_1^{(n)},\delta_2^{(n)}}(\textbf{u})-M^{\delta_1^{(n)},\delta_2^{(n)}}(\textbf{u}) \right|+\!\!\sup_{\textbf{u}\in{\mathcal S}_{d-1}}\left|M^{\delta_1^{(n)},\delta_2^{(n)}}(\textbf{u})-M^{\varepsilon_0,\delta_0}(\textbf{u}) \right|
\end{align*}
The claim then follows from Lemmas~\ref{lem:M_delta_continuity} and \eqref{eq:M_n-M delta bound}.
\end{proof}

We are now equipped to prove Theorem \ref{Thm:Directions_Consistency}.

\begin{proof}[\bf Proof of Theorem~\ref{Thm:Directions_Consistency}]

Let  $M_n \coloneqq M_n^{\delta_1^{(n)},\delta_2^{(n)}}$ and $M \coloneqq M^{\varepsilon_0,\delta_0}.$ By Lemma~\ref{lem:uniform_convergence_M_n},\vspace{-4mm}
\begin{equation}\label{eq:uniform_Mn_M_theorem}
\|M_n-M\|_\infty
\coloneqq \sup_{\textbf u\in\mathcal S_{d-1}}
|M_n(\textbf u)-M(\textbf u)|
=o_{\rm P}(1).
\end{equation}\vspace{-8mm}

Moreover, $M$ is continuous on the compact set $\mathcal S_{d-1}$. The population principal direction~$\textbf{U}_{\pm,1}$ is defined in~\eqref{defU1''} as the maximizer over $\mathcal{S}_{d-1}$ of
$\displaystyle M^0({\textbf u})
\coloneqq
\mathbb{E}_{r\sim\mathcal{U}(0,1]}
\left[
L^2\left(\textbf{Q}_\pm(r{\textbf u})\right)
\right],$
provided that this expectation is finite for all ${\textbf u}$.  Assumption~\ref{AssA2}, however, implies that $\textbf{U}_{\pm,1}$ then is also the maximizer of the trimmed objective function $M^{\varepsilon_0,\delta_0}$. In case $M^0({\textbf u})$ is infinite for some ${\textbf u}$, convergence is to the maximizer of the trimmed objective $M^{\varepsilon_0,\delta_0}$. This distinction, however, has no impact on finite-$n$ practical situations.

Define therefore $\hat{\textbf h}\coloneqq \textbf U_{\pm,1}$ and let further $\hat{\textbf h}_n
\coloneqq\textbf{U}_{\pm,1}^{(n)}$ be the maximizer of  $M_n^{\delta_1^{(n)},\delta_2^{(n)}}(\textbf{u})$ over the grid~$\mathfrak{G}^{\mathcal S}_m=\{\textbf u_1,\ldots,\textbf u_{m_S}\}
\subset\mathcal S_{d-1}$.  It remains to verify that $\hat{\textbf h}_n$ is an approximate
maximizer of $M_n$ over the whole unit sphere.

By construction of the grid,\vspace{-6mm}
\begin{equation}\label{eq:grid_convergence}
\rho_n
\coloneqq
\sup_{\textbf h\in\mathcal S_{d-1}}
\min_{\textbf u\in\mathfrak G_{m}^{\mathcal S}}
\|\textbf h-\textbf u\|
\longrightarrow 0
\qquad\text{as }m_S\to\infty .
\end{equation}
For each $n\in\mathbb{N}$, choose $\textbf u_n^\ast\in\mathfrak G_{m}^{\mathcal S}$ such that
$\displaystyle \|\textbf u_n^\ast-\hat{\textbf h}\|
=
\min_{\textbf u\in\mathfrak G_{m}^{\mathcal S}}
\|\textbf u-\hat{\textbf h}\|.$ Then \eqref{eq:grid_convergence} implies
$\|\textbf u_n^\ast-\hat{\textbf h}\|\leq \rho_n\to 0$.

The  continuity of $M$ entails\vspace{-2mm} %
\begin{equation}\label{eq:M_grid_population}
M({\textbf u}_n^\ast)\to M(\hat{\textbf h}) \quad \text{as $n\to\infty$}. %
\end{equation}
Since $\hat{\textbf h}_n$ maximizes $M_n$ $\mathrm{P}$-a.s.\ over $\mathfrak G_{m}^{\mathcal S}$, $M_n(\hat{\textbf h}_n) \geq M_n(\textbf u_n^\ast)$,   %
giving further
\begin{equation}\label{eq:3.1temp1}
    0 \leq
\sup_{\textbf h\in\mathcal S_{d-1}}[M_n(\textbf h)
-
M_n(\hat{\textbf h}_n)]
\leq
\sup_{\textbf h\in\mathcal S_{d-1}}[M_n(\textbf h)
-
M_n(\textbf u_n^\ast)],\quad {\mathrm{P}-a.s.}
\end{equation}
Lemma~\ref{lem:sup-ineq} with
$A=\mathcal S_{d-1}$ and the fact that $\hat{\textbf h}$ maximizes $M$ yield\vspace{-4mm}
\begin{equation}\label{eq:3.1temp2}
\sup_{\textbf h\in\mathcal S_{d-1}}M_n(\textbf h)
\leq
M(\hat{\textbf h})+\|M_n-M\|_\infty,\quad  {\mathrm{P}-a.s.}
\end{equation}
On the other hand $ M_n(\textbf u_n^\ast)-
M(\textbf u_n^\ast)\geq - |M_n(\textbf u_n^\ast)-
M(\textbf u_n^\ast)|\geq  -\|M_n-M\|_\infty$ {$\mathrm{P}$-a.s.}, giving
\begin{equation}\label{eq:3.1temp3}
 M_n(\textbf u_n^\ast)\geq
M(\textbf u_n^\ast) -\|M_n-M\|_\infty \quad {\mathrm{P}-a.s.}
\end{equation}
Plugging \eqref{eq:3.1temp2} and \eqref{eq:3.1temp3} into \eqref{eq:3.1temp1} then implies
\begin{align}
0\leq
\sup_{\textbf h\in\mathcal S_{d-1}}M_n(\textbf h)
-
M_n(\hat{\textbf h}_n)\leq
M(\hat{\textbf h})-M(\textbf u_n^\ast)
+
2\|M_n-M\|_\infty
=
o_{\rm P}(1)\quad\text{as $n\to\infty$},
\label{eq:approx-maximizer}
\end{align}
where the last equality follows from
\eqref{eq:M_grid_population} and
\eqref{eq:uniform_Mn_M_theorem}. Hence, $M_n(\hat{\textbf h}_n)
\geq
\sup_{\textbf h\in\mathcal S_{d-1}}M_n(\textbf h)
-o_{\rm P}(1)$ as~$n\to\infty$.

 All conditions of Theorem~\ref{Argmax}, thus, are satisfied, with
$$\text{%
 $H=\mathcal S_{d-1}$, \ $d({\textbf u} ,\  {\textbf v}) = \Vert {\textbf u} - {\textbf v}\Vert$, \ %
 $M_n=M_n^{\delta_1^{(n)}\!, \delta_2^{(n)}}\!\!$, \ %
 $M=M^{\varepsilon_0,\delta_0}$, \
 $%
 \hat{\textbf h}_n=\textbf U_{\pm,1}^{(n)}$, \ and \
$ \hat{\textbf h}=\textbf U_{\pm,1}$.
}.$$ %
 It follows that
$%
 \|\textbf U_{\pm,1}^{(n)}-\textbf U_{\pm,1}\|
 =o_{\rm P}(1)$,
 which completes the proof.
\end{proof}

\color{black}

\subsection{Proof of Theorem~\ref{Thm:Directions_Consistency_i}}
We next discuss Theorem \ref{Thm:Directions_Consistency_i} and the consistency of the subsequent directions $\textbf{U}_{\pm,2}^{(n)},\ldots, \textbf{U}_{\pm,2d}^{(n)}$.
A key difficulty  is that the sets of feasible solutions  involved in %
definitions  \eqref{defUi''} and \eqref{defUin'} now depend on the previous principal directions and,  therefore, are random in the empirical case. As a consequence,  neither the standard M-estimation consistency results (Theorem~\ref{Argmax}), nor the existing generalizations~\citep{cox2022}
 can be used to obtain consistency in our setting. In this section, we therefore develop (Theorem~\ref{thm:Argmax-compact-random}) a consistency result for M-estimators when the feasible sets are random and converge, in probability and with respect to the Hausdorff distance,  to their population counterparts.%

That consistency result is of independent interest, as it applies broadly to sequential estimation procedures in which constraints are generated by previously estimated quantities. In particular, it provides a general framework for establishing   consistency results in deflation-based methods such as PCA and ICA. For the sake of generality, we prove Theorem~\ref{thm:Argmax-compact-random} in the general case of a compact subset~$H$ of some metric space although, in  Theorem~\ref{Thm:Directions_Consistency_i}, we only need it for $H={\mathcal S}_{d-1}$.%

\subsubsection{{A Consistency result for M-estimators over random feasible sets}}\label{SecA31}

This section involves real random variables, %
 random Borel subsets\footnote{See, e.g., Section~1.1 in  {\cite{molchanov2018random}} for a precise definition of random subsets of a compact set $H\subset{\mathbb R}^d$.}  $\Omega\ni\omega\mapsto  \Lambda(\omega)$ of~$\mathbb R^p$, and stochastic processes $\{\Omega\ni\omega\mapsto\{ {M}({\textbf u};\,\omega)\vert {\textbf u}\in H\}$  indexed on a compact subset~$H$ of
 a metric space
  (equivalently, random functions from $H$ to $\mathbb R$), all defined over some common probability space~$(\Omega, {\mathcal F}, {\mathrm P})$. %
  Lighter notation~$\Lambda$ and~$M({\textbf u})$ (and $\Lambda_n$, $C$, $C_n$,  $M_n({\textbf u})$,
  $\ldots$, $n\in{\mathbb N}$) is used whenever no confusion is possible. %

\begin{Thm}%
\label{thm:Argmax-compact-random}
Let $M_n$, $n\in{\mathbb N}$ denote a sequence of real-valued stochastic processes defined over some common probability space $(\Omega, {\mathcal F}, {\mathrm P})$  indexed on \textcolor{black}{a  compact subset $H$ of some metric space  with distance~$d(\cdot,\,\cdot)$.} %
Let  further $H\ni \textbf{u}\mapsto M(\textbf{u})$ be a continuous (deterministic) function with domain~$H$, and denote by $\Lambda$ and~$\Lambda_n $, $n\in\mathbb{N}$,  where~$\Lambda_n$  is random,
 non-empty closed ${\mathrm{(P}}$-a.s.\  so for $\Lambda_n${\rm )} subsets of $H$.
Assume that %
\begin{itemize}
\item[(A1)]%
 $\sup_{\textbf{{\textbf u}}\in H}|M_n(\textbf{u})-M({\textbf u})| = o_{\mathrm P}(1)\ \text{as \ }n\to\infty$.
\item[(A2)]   $ d_H(\Lambda_n,\Lambda) = o_{\mathrm P}(1)\quad \text{as \ }n\to\infty$, where $ d_H(\Lambda_n,\Lambda)$ stands for the Hausdorff distance between $\Lambda_n$ and~$\Lambda$.
\item[(A3)] %
the maximizer
$\displaystyle\hat{\textbf{u}} \coloneqq  \arg\max_{\textbf{u}\in\Lambda} M(\textbf{\textbf u})$ of $M$ over $\Lambda$ is $\mathrm P$-a.s.\  unique.
\end{itemize}
Let  $\hat{\textbf{u}}_n\in\Lambda_n$
be an approximate maximizer of $M_n$ over $\Lambda_n$ in the sense that  {\begin{itemize}
    \item[(A4)] $r_n\coloneqq \sup_{\textbf{u}\in\Lambda_n} [M_n(\textbf{u})]  - M_n(\hat{ \textbf{u}}_n) ] =~\!o_{\mathrm P}(1)$ as~$n\to\infty$. %
\end{itemize}}
Then,
{$d(\hat{\textbf{u}}_n,\,  \hat{\textbf{u}}) =o_{\mathrm P}(1)$}
  as $n\to\infty$.
\end{Thm}%

The proof of Theorem \ref{thm:Argmax-compact-random} relies on the following three auxiliary lemmas (same notation and same assumptions as in Theorem~\ref{thm:Argmax-compact-random}).

\begin{Lemma}%
\label{lem:separation}
Let $C\subset H$ be compact and let $f:C\to\mathbb R$ be continuous and admit (over $C$)
 a unique maxi\-mizer~${\hat{\bf h}_C}\coloneqq    {\argmax}_{{\bf h}\,\in\, C}f({\bf h})$.  %
Then, under the assumptions of Theorem~\ref{thm:Argmax-compact-random}, for every $\varepsilon>0$,
 $$
\Delta_\varepsilon
\coloneqq
f({\hat{\bf h}_C})
-
\sup\{f({\bf h}):{\bf h}\in C,\  {d({\bf h},\,  {\hat{\bf h}_C})}\geq\varepsilon\} %
\;>\;0.
$$
\end{Lemma}

\begin{proof}%
The set $\{{\bf h}\in C:\,{d({\bf h},\, {\hat{\bf h}_C}})\ge\varepsilon\}$ is compact.
By continuity, $f$ attains a maximum on this set.
Uniqueness of the maximizer implies that this maximum is strictly less than $f({\hat{\bf h}_C})$.
\end{proof}

\begin{Lemma}
\label{lem:sup-ineq}
Let %
 $A\subseteq H$ be a non-empty %
  subset of~$H$. Then, under the assumptions of Theorem~\ref{thm:Argmax-compact-random}, letting $\displaystyle\|M_n-M\|_\infty\coloneqq  \sup_{\textbf{u}\in H}|M_n(\textbf{u})-M(\textbf{u})|$,
\begin{align}
\sup_{\textbf{u}\in A} M(\textbf{u})
-
\|M_n-M\|_\infty
\leq \sup_{\textbf{u}\in A} M_n(\textbf{u})
\leq\sup_{\textbf{u}\in A} M(\textbf{u})
+
\|M_n-M\|_\infty
\label{eq:sup-ineq1}
\quad \text{$\mathrm P$-a.s.,\  \ $n\in{\mathbb N}$} \end{align}
\end{Lemma}

\begin{proof}%
For every $\textbf{u}\in A$,
\begin{align*}
 M(\textbf{u}) -  \Vert M_n - M \Vert_\infty &\leq M(\textbf{u}) - |M_n(\textbf{u})-M(\textbf{u})| \leq M_n(\textbf{u}) \\
 &\leq M(\textbf{u}) + |M_n(\textbf{u})-M(\textbf{u})| \leq M(\textbf{u}) + \Vert M_n - M \Vert_\infty\  {\mathrm P}-\text{a.s.}
\end{align*}
Taking $\sup_{\textbf{u}\in A}$ in this double inequality yields the desired result~\eqref{eq:sup-ineq1}.
\end{proof}

\begin{Lemma}%
\label{lem:sup-convergence-deterministic}
{Let $C_n$, $n\in{\mathbb N}$ be a sequence of nonempty random compact subsets of $H$, and let $C\subseteq H$ be a compact set such that
$d_H(C_n,C) =  {o_{\mathrm P}(1)}$  as~$n\to\infty$ ($d_H$ the Hausdorff distance associated with $d(\cdot ,\,\cdot)$).} %
Then, under the assumptions of Theorem~\ref{thm:Argmax-compact-random},
 $$
\Big|\sup_{\textbf{u}\in C_n} M_n(\textbf{u}) -  \sup_{\textbf{u}\in C} M(\textbf{u})\Big| = o_{\mathrm P}(1)
\quad\text{ as $n\to\infty$.}$$
\end{Lemma}

\begin{proof}
Let
 $
\varpi_M(\delta)\coloneqq \sup_{{\textbf{x},\textbf{y}\in H}:\, {d(\textbf{x},\, \textbf{y})}\leq\delta}|M(\textbf{x})-M(\textbf{y})|
$
denote the {(deterministic)} modulus of continuity of~$M$.  %
Since ${\textbf u}\mapsto M({\textbf u})$ is continuous, and thus uniformly continuous, on the compact~$H$,
 {$\varpi_M(\delta)\to 0$ as~$\delta\downarrow 0.$}
For any nonempty compacts $A,B\subset H$,
\begin{equation}
\label{eq:sup-det-ineq}
\Big|\sup_{\textbf{u}\in A} M(\textbf{u}) - \sup_{\textbf{u}\in B} M(\textbf{u})\Big| \leq \varpi_M (d_H(A,B)).%
\end{equation}
Indeed, let $\delta=d_H(A,B)$. For any $\textbf{a}\in A$ there exists $\textbf{b}\in B$ such that  {$d(\textbf{a},\, \textbf{b})\leq\delta$}, hence
$$
M(\textbf{a})\leq M(\textbf{b})+\varpi_M(\delta)\leq \sup_{\textbf{u}\in B} M(\textbf{u})+\varpi_M(\delta).%
 $$
Taking supremum over $\textbf{a}\in A$ in this inequality yields
$
\displaystyle\sup_{\textbf{u}\in A} M(\textbf{u})\leq \sup_{\textbf{u}\in B} M(\textbf{u})+\varpi_M(\delta).%
$
Exchanging the roles of $A$ and $B$ similarly yields  $
\displaystyle\sup_{\textbf{u}\in B} M(\textbf{u})\leq \sup_{\textbf{u}\in A} M(\textbf{u})+\varpi_M(\delta)$, hence %
$
\Big|\sup_{\textbf{u}\in A} M(\textbf{u})- \sup_{\textbf{u}\in B} M(\textbf{u})\Big|\leq\varpi_M(\delta).%
$

Defining $\displaystyle S_n\coloneqq \sup_{\textbf{u}\in C_n} M_n(\textbf{u})$ and $\displaystyle S\coloneqq \sup_{\textbf{u}\in C} M(\textbf{u})$, and
applying Lemma \ref{lem:sup-ineq} with $A=C_n$, we obtain
$$
S_n\leq \sup_{\textbf{u}\in C_n}M(\textbf{u})+\|M_n-M\|_\infty \quad{\mathrm P}\text{-a.s.},\,  n\in{\mathbb N},$$
hence, subtracting $\displaystyle S=\sup_{\textbf{u}\in C}M(\textbf{u})$ from  both sides, %
\begin{equation}\label{eq:S_n-S_lower1}
S_n - S \leq \Big(\sup_{\textbf{u}\in C_n} M(\textbf{u}) - \sup_{\textbf{u}\in C} M(\textbf{u})\Big) + \|M_n-M\|_\infty \quad{\mathrm P}\text{-a.s.},\,  n\in{\mathbb N}.
\end{equation}
Similarly, by Lemma \ref{lem:sup-ineq},
$
S_n\geq  \sup_{\textbf{u}\in C_n}M(\textbf{u})-\|M_n-M\|_\infty$, ${\mathrm P}$-a.s.,
which implies  %
\begin{equation}\label{eq:S_n-S_lower2}
S - S_n \leq \Big(\sup_{\textbf{u}\in C} M(\textbf{u}) - \sup_{\textbf{u}\in C_n} M(\textbf{u})\Big) + \|M_n-M\|_\infty \quad{\mathrm P}\text{-a.s.},\,  n\in{\mathbb N}.
\end{equation}
Piecing together \eqref{eq:S_n-S_lower1} and \eqref{eq:S_n-S_lower2} yields
\begin{equation}\label{eq:S_n-S_final_bound}
    |S_n - S | \leq \Big|\sup_{\textbf{u}\in C} M(\textbf{u}) - \sup_{\textbf{u}\in C_n} M(\textbf{u})\Big| +\|M_n-M\|_\infty\leq \varpi_Md_H(C_n,C)+\|M_n-M\|_\infty \quad{\mathrm P}\text{-a.s.},\,  n\in{\mathbb N}
\end{equation}
where the last inequality follows from \eqref{eq:sup-det-ineq}.

By assumption, $d_H(C_n,C)=o_{\mathrm P}(1)$ as $n\to\infty$. The continuity of $\varpi_M$ at $0$ and the  continuous mapping theorem imply that  $\varpi_Md_H(C_n,C)=o_{\mathrm P}(1)$  as $n\to\infty$, and the claim %
 follows from the assumption that~$\| M_n-~\!M\|_\infty$ is $o_{\mathrm P}(1)$ as $n\to\infty$.
\end{proof}

\begin{proof}[\bf Proof of Theorem~\ref{thm:Argmax-compact-random}]
For all $\varepsilon>0$,  let
$
F_\varepsilon\coloneqq { \{\textbf{u}\in H:d({\textbf u},\, \hat{\textbf u})}
\ge\varepsilon\}.
$
The proof consists of showing that, for all $\varepsilon >0$, $\mathrm{P}(\hat{\textbf{u}}_n\in F_\varepsilon) =  \mathrm{P}({d(\hat{{\textbf u}}_n,\, \hat{{\textbf u}})}\ge \varepsilon)
\to 0$ as $n\to\infty$.

As a closed subset  of a compact set,  $\Lambda\subseteq H$ is compact.
 \unaAdded{Letting $\delta_n = \delta_n(\omega)\coloneqq d_H(\Lambda_n(\omega),\Lambda)$, con\-sider~$E^{(n)}_\varepsilon\coloneqq \{\omega \vert \delta_n(\omega )\le \varepsilon/2\}$   and
observe that, by~(A2), $\mathrm P(E^{(n)}_\varepsilon)\to 1$ as $n\to~\!\infty$  {for any $\varepsilon>0$. }}

\unaAdded{
Suppose first that %
 $\Lambda\cap F_{\varepsilon/2}=\varnothing$. Then  $\textbf{v}\in\Lambda$ implies~$d(\textbf{v},\hat{\textbf{u}})<\varepsilon/2$. {On $E_\varepsilon^{(n)}$, for every $\textbf{u}\in\Lambda_n$,} %
 there exists $\textbf{v}\in\Lambda$ such that $d(\textbf{u},\textbf{v})\leq \delta_n\leq \varepsilon/2$,  hence  $d(\textbf{u},\hat{\textbf{u}})\leq d(\textbf{u},\textbf{v})+d(\textbf{v},\hat{\textbf{u}})<\varepsilon$, so that {$\Lambda_n\cap F_\varepsilon=\varnothing$, on $E_\varepsilon^{(n)}$}. %
 In particular, on $E^{(n)}_\varepsilon$,  $\hat{\textbf{u}}_n\notin F_\varepsilon$, so that $\mathrm{P}(\hat{\textbf{u}}_n\in F_\varepsilon)\leq \mathrm{P}(E_\varepsilon^{(n)}{}^c)\to 0$ as~$n\to\infty$, and the result follows. We can thus simply assume $\Lambda\cap F_{\varepsilon/2}\neq\varnothing$.
}

 By (A3) and Lemma~\ref{lem:separation}
(applied to~$M$~on~$\Lambda$), %
\begin{equation}
\label{eq:gap}
M(\hat{\textbf{u}})-\sup_{\textbf{u}\in \Lambda\cap F_{\varepsilon/2}} M(\textbf{u})
\eqqcolon\Delta_{\varepsilon/2}>0.
\end{equation}

Denote by $\varpi_M(\delta)\coloneqq \sup\{|M({\textbf x})-M({\textbf y})|:\, {d({\textbf x},\, {\textbf y})}\le \delta\}$  the modulus of continuity of $M$ on $H$.
Since~$M$ is  %
 continuous on  $H$ (which is compact), $\varpi_M(\delta)\to 0$ %
   as $\delta\downarrow 0$
Putting  $C_n=\Lambda_n$ and $C=\Lambda$ in Lemma~\ref{lem:sup-convergence-deterministic} and using  \eqref{eq:a^2-b^2} and \eqref{integrals}, we obtain
\begin{equation}
\label{eq:global-sup}
A_n\coloneqq \sup_{\textbf{u}\in\Lambda_n} M_n(\textbf{u})  -  \sup_{\textbf{u}\in\Lambda} M(\textbf{u})=\sup_{\textbf{u}\in\Lambda_n} M_n(\textbf{u})  -M(\hat{\textbf{u}}) = o_{\mathrm P}(1)\quad \text{as $n\to\infty$}.
\end{equation}

\unaAdded{
Define $G^{(n)}_\varepsilon:=\{\omega\vert\Lambda_n(\omega)\cap F_{\varepsilon}\neq \varnothing\}$.  On $(G^{(n)}_\varepsilon)^c$, since $\hat{\textbf{u}}_n\in\Lambda_n$, necessarily $\hat{\textbf{u}}_n\notin F_{\varepsilon}$.  Hence, $\{\hat{\textbf{u}}_n\in~\!F_{\varepsilon}\}\subseteq~\!G^{(n)}_\varepsilon$ a.s.,  and  it is enough to
 show that~$\mathrm{P}(\{\hat{\textbf{u}}_n\in F_\varepsilon\}\cap G^{(n)}_\varepsilon)\to 0$. }

\unaAdded{For $\omega\in G_\varepsilon^{(n)}$, let} %
$B_n(\omega)\coloneqq \sup_{\textbf{u}\in\Lambda_n\cap F_\varepsilon} M_n(\textbf{u})$;  similarly define
$B\coloneqq \sup_{\textbf{u}\in\Lambda\cap F_{\varepsilon/2}} M(\textbf{u})$.
\unaAdded{On $G_\varepsilon^{(n)}$, for every $\textbf{u}\in\Lambda_n\cap F_\varepsilon ,%
$,} %
there ${\mathrm P}$-a.s. exists $\textbf{v}\in\Lambda$ such that  ${d(\textbf{u},\, \textbf{v})} \leq d_H(\Lambda_n,\Lambda)\leq \delta_n$.  Moreover, by the definition of $\varpi_M$,
\begin{equation}\label{eq:mod_cont_uv}
    M(\textbf{u})\leq M(\textbf{v})+\varpi_M(\delta_n) \quad {\mathrm P}\text{-a.s.}, \ n\in{\mathbb N}.
\end{equation}

On $E_\varepsilon^{(n)}\unaAdded{\cap G_\varepsilon^{(n)}}$, %
 it holds that
$
{d(\textbf{v},\, \hat{\textbf{u}})} \geq {d(\textbf{u},\, \hat{\textbf{u}})}-{d(\textbf{u},\, \textbf{v})}
\geq \varepsilon-\delta_n\geq \varepsilon/2 \quad {\mathrm P}\text{-a.s.},
$
implying
\begin{equation}\label{eq:v_on_En}
\textbf{v}\in\Lambda\cap F_{\varepsilon/2} \quad {\mathrm P}\text{-a.s.}
\end{equation}
Using \eqref{eq:v_on_En} in \eqref{eq:mod_cont_uv}, we get that, on $E_\varepsilon^{(n)}\unaAdded{\cap G_\varepsilon^{(n)}}$, %
\begin{equation}\label{eq:u-v1}
M(\textbf{u})\le M(\textbf{v})+\varpi_M(\delta_n)\leq \sup_{\textbf{u}\in\Lambda\cap F_{\varepsilon/2}} M(\textbf{u})+\varpi_M(\delta_n)=B+\varpi_M(\delta_n)    \quad {\mathrm P}\text{-a.s.},\ n\in{\mathbb N};
\end{equation}
 taking the supremum over $\textbf{u}\in\Lambda_n\cap F_\varepsilon$ yields, still on $E_\varepsilon^{(n)}\unaAdded{\cap G_\varepsilon^{(n)}}$,
\begin{equation}
\label{eq:Mn-far-on-En}
\sup_{\textbf{u}\in\Lambda_n\cap F_\varepsilon} M(\textbf{u})
\le B+\varpi_M(\delta_n),
\quad {\mathrm P}\text{-a.s.}
\end{equation}
Applying Lemma~\ref{lem:sup-ineq} with $A=\Lambda_n\cap F_\varepsilon$ yields, on $E_\varepsilon^{(n)}\unaAdded{\cap G_\varepsilon^{(n)}}$,
\begin{equation}
\label{eq:Bn-on-En}
B_n
\leq \sup_{\textbf{u}\in\Lambda_n\cap F_\varepsilon} M(\textbf{u})+\|M_n-M\|_\infty\leq
B+\varpi_M(\delta_n)+\|M_n-M\|_\infty  \quad {\mathrm P}\text{-a.s.,} \ n\in{\mathbb N}
\end{equation}
where the second inequality follows from \eqref{eq:Mn-far-on-En}.
Fix   $\eta>0$. It follows from \eqref{eq:Bn-on-En} that\vspace{-4mm}
\begin{align}%
    \mathrm{P}\left(\{B_n > B+\eta\}\unaAdded{\cap G_\varepsilon^{(n)}}\right)
&\leq
 \mathrm{P}(E_\varepsilon^{(n)}{}^c)
+
 \mathrm{P}\left(\{B_n > B+\eta\}\unaAdded{\cap G_\varepsilon^{(n)}}\cap E_\varepsilon^{(n)}\right)\nonumber\\
&\leq
 \mathrm{P}(E_\varepsilon^{(n)}{}^c)
+
 \mathrm{P}\left(\{\varpi_M(\delta_n)+\|M_n-M\|_\infty>\eta\}\cap E_\varepsilon^{(n)}\right)\nonumber\\
&\leq
 \mathrm{P}(E_\varepsilon^{(n)}{}^c)
+
\mathrm{P}\left(\varpi_M(\delta_n)+\|M_n-M\|_\infty>\eta\right)\qquad n\in{\mathbb N}.  %
\label{eq:eliminate_E_n}
\end{align}
Since $ \mathrm{P}(E_\varepsilon^{(n)}{}^c)=o(1)$, $\delta_n =o_{\mathrm P} (1)$, and $\|M_n-M\|_\infty = o_{\mathrm P} (1)$ as $n\to\infty$, and  since $\varpi_M$ is %
a deterministic function continuous at $0$,
the right-hand side in \eqref{eq:eliminate_E_n} tends to zero. %
\unaAdded{Thus, for every $\eta>0$
\begin{equation}\label{eq:final1}
  \mathrm{P}\left(\{B_n > B+\eta\}\cap G_\varepsilon^{(n)}\right)\to 0\quad\text{as}\quad n\to\infty.
  \end{equation}
}

From \eqref{eq:global-sup}, recalling that, \unaAdded{on $G_\varepsilon^{(n)}$}, $
B_n\coloneqq \sup_{\textbf{u}\in\Lambda_n\cap F_\varepsilon} M_n(\textbf{u})$ and
$B\coloneqq \sup_{\textbf{u}\in\Lambda\cap F_{\varepsilon/2}} M(\textbf{u}),
$ we obtain
\begin{equation*}%
    \sup_{{\textbf u}\in\Lambda_n} M_n({\textbf u})
=
M(\hat{{\textbf u}})+o_{\mathrm P}(1)\quad\text{as $n\to\infty$},
\end{equation*}
hence, in view of  \eqref{eq:gap},
$
M(\hat{{\textbf u}})
=
\sup_{{\textbf u}\in\Lambda\cap F_{\varepsilon/2}} M({\textbf u})
+
\Delta_{\varepsilon/2}$ ${\mathrm P}$-a.s.
Therefore, \unaAdded{still on $G_\varepsilon^{(n)}$},  \vspace{-4mm}
\begin{equation}\label{eq:A41}
    \sup_{{\textbf u}\in\Lambda_n} M_n({\textbf u})
=
\sup_{{\textbf u}\in\Lambda\cap F_{\varepsilon/2}} M({\textbf u})
+
\Delta_{\varepsilon/2}
+
o_{\mathrm P}(1) \quad\text{as $n\to\infty$}.
\end{equation}

By \eqref{eq:final1}, \unaAdded{on $G_\varepsilon^{(n)}$,}
\begin{equation}\label{eq:A42}
\sup_{{\textbf u}\in\Lambda_n\cap F_\varepsilon} M_n({\textbf u})
\leq
\sup_{{\textbf u}\in\Lambda\cap F_{\varepsilon/2}} M({\textbf u})
+
o_{\mathrm P}(1),\quad n\to\infty.
\end{equation}
\eqref{eq:A41} and \eqref{eq:A42} yield, still \unaAdded{on $G_\varepsilon^{(n)}$},
 $
\sup_{{\textbf u}\in\Lambda_n} M_n({\textbf u})
-
\sup_{{\textbf u}\in\Lambda_n\cap F_\varepsilon} M_n({\textbf u})
\geq
\Delta_{\varepsilon/2}
+
o_{\mathrm P}(1).
$
for   $n\to\infty$.
Consequently,\vspace{-4mm}
\begin{equation}\label{eq:semi_final}
 \mathrm{P}\left(\left\{
\sup_{{\textbf u}\in\Lambda_n} M_n({\textbf u})
-
\sup_{{\textbf u}\in\Lambda_n\cap F_\varepsilon} M_n({\textbf u})
\unaAdded{\leq}
\Delta_{\varepsilon/2}/2\right\}\unaAdded{\cap G_\varepsilon^{(n)}}
\right)\to \unaAdded{0}\quad \text{as $n\to\infty$}.
\end{equation}
Now, assume that $\hat{{\textbf u}}_n\in F_\varepsilon$.
Since $\hat{{\textbf u}}_n\in\Lambda_n$,
$
M_n(\hat{{\textbf u}}_n)
\leq
\sup_{{\textbf u}\in\Lambda_n\cap F_\varepsilon} M_n({\textbf u})
$ ${\mathrm P}$-a.s.
On the other hand, by assumption,
 $
M_n(\hat{{\textbf u}}_n)
\geq
\sup_{{\textbf u}\in\Lambda_n} M_n({\textbf u})-r_n.
$   ${\mathrm P}$-a.s.
Combining the two inequalities gives \vspace{-4mm}
$$
\sup_{{\textbf u}\in\Lambda_n} M_n({\textbf u})
-
\sup_{{\textbf u}\in\Lambda_n\cap F_\varepsilon} M_n({\textbf u})
\leq r_n \quad  {\mathrm P}-\text{a.s.}
$$
Therefore,  \vspace{-4mm}
$$
\{\hat{{\textbf u}}_n\in F_\varepsilon\}
\cap
\Big\{
\sup_{{\textbf u}\in\Lambda_n} M_n({\textbf u})
-
\sup_{{\textbf u}\in\Lambda_n\cap F_\varepsilon} M_n({\textbf u})
>
\Delta_{\varepsilon/2}/2
\Big\}
\subseteq
\{r_n>\Delta_{\varepsilon/2}/2\}   \quad  {\mathrm P}-\text{a.s.}.
$$
\unaAdded{As $\{\hat{\textbf{u}}_n\in F_\varepsilon\}\subseteq G_\varepsilon^{(n)}$, this further implies} %
\begin{align*}
 \mathrm{P}(\hat{{\textbf u}}_n\in F_\varepsilon)&\unaAdded{=\mathrm{P}(\{\hat{{\textbf u}}_n\in F_\varepsilon\}\cap G_\varepsilon^{(n)})}
\\
&\le
 \mathrm{P}(r_n>\Delta_{\varepsilon/2}/2)
+
 \mathrm{P}\!\left(\left\{
\sup_{{\textbf u}\in\Lambda_n} M_n({\textbf u})
-
\sup_{{\textbf u}\in\Lambda_n\cap F_\varepsilon} M_n({\textbf u})
\le
\Delta_{\varepsilon/2}/2\right\}\unaAdded{\cap G_\varepsilon^{(n)}}
\right)\eqqcolon P_1 + P_2.
\end{align*}
Since $r_n$ is $o_{\rm P}(1)$ by (A4) and since $P_2 = o(1)$ by \eqref{eq:semi_final}, we obtain that
 $
 \mathrm{P}(\hat{{\textbf u}}_n\in F_\varepsilon)\to 0
 $ as $n\to\infty$.
As this holds for any $\varepsilon>0$, we   conclude that
 $
{d(\hat{{\textbf u}}_n,\, \hat{{\textbf u}})} = o_{\mathrm P}(1)%
$. The theorem follows.
\end{proof}

\subsubsection{\bf Proof of Theorem \ref{Thm:Directions_Consistency_i}}\label{SecA32}

We now turn to the proof of Theorem \ref{Thm:Directions_Consistency_i}. Recall that here $H={\mathcal S}_{d-1}$ and $d(\cdot ,\,\cdot)$ coincides with the Euclidean norm $\Vert\cdot - \cdot\Vert$. The following two lemmas establish the desired convergence of the sample feasible sets. These   lemmas and the proof of  Theorem~\ref{Thm:Directions_Consistency_i} involve maximization problems with various feasible sets $\Lambda$,  of the form  $\Lambda^k({\bf u})\coloneqq \Lambda^k_1({\bf u})\cup\Lambda^k_2({\bf u})$,
 $k\in\{1,\ldots,2d-1\}$, $n\in{\mathbb N}$, with ${\bf u}=({\textbf u}_1,\ldots, {\textbf u}_k)\in{\mathbb R}^{dk}$,
{
\begin{equation}\label{defLambda}
\Lambda^k_{1}({\bf u}) \coloneqq\left\{
\textbf{v}\in\mathcal{S}_{d-1}\!:
\textbf{v}' %
\textbf{u}_j= 0,\ j=1,\dots,k
\right\} \ \ \text{and}\ \ %
\Lambda^k_{2} ({\bf u})\coloneqq \left\{
       -{\textbf u}_i\!:\!
       -{\textbf u}_i\
       \!\notin\!
       \{{\textbf u}_1,\ldots,{\textbf u}_k\},
       \ \  i=1,\ldots,k
    \right\}.
\end{equation}}
\begin{Lemma}\label{lem:orthogonality_convergence}
For all $n\in\mathbb{N}$ and  $k\in\{1,\ldots,  {2d-1}\}$, denote by  ${\bf u}\coloneqq \{\textbf{u}_1,\dots,\textbf{u}_k\}$
 a deterministic $k$-tuple and
 by~${\bf u}^{(n)}\coloneqq \{\textbf{u}^{(n)}_1,\dots,\textbf{u}^{(n)}_k\}$ a random $k$-tuple of points on $\mathcal{S}_{d-1}$
 such that, for all $i\neq j=1\dots k$,
$$
\textbf u_i'\textbf u_j\in\{0,-1\},
\quad
(\textbf u_i^{(n)})'\textbf u_j^{(n)}\in\{0,-1\},\quad\text{and}\quad
\|\textbf u_i^{(n)}-\textbf u_i\|=o_{\rm P}(1), \ i=1,\ldots,k, \ \text{as $n\to\infty$.}$$
 Then, \unaAdded{$\Lambda^k_{1}({\bf u}^{(n)})$ is a compact random set and
 \begin{itemize}
     \item[(i)] if $\Lambda_1^k(\textbf{u})\neq \varnothing$
     \begin{equation}\label{A8}d_H(\Lambda^k_{1}({\bf u}^{(n)}),\Lambda^k_{1}({\bf u})) =o_{\rm P} (1)\quad\text{ as $n\to\infty$.} \end{equation}
     \item[(ii)]  if $\Lambda_1^k(\textbf{u})=\varnothing$ (equivalently, if $\text{\rm span}(\textbf{u}_1,\dots,\textbf{u}_k) = \mathbb{R}^d$), $\mathrm{P}(\Lambda_1^k(\textbf{u}^{(n)})=\varnothing)\to 1$, as $n\to\infty$.
 \end{itemize}
 }
\end{Lemma}

\begin{proof}[\bf Proof of Lemma \ref{lem:orthogonality_convergence}]
First, note that the set $\Lambda^k_{1}({\bf u}^{(n)})=\bigcap_{j=1}^k \{\textbf{v}\in{\mathcal S_{d-1}}:\textbf{v}'\textbf{u}^{(n)}_{j}=0\}$, as the (possibly empty) intersection of finitely many closed random subsets of $\mathcal S_{d-1}$, is ($\mathrm P$-a.s.) closed, hence, since~$\mathcal S_{d-1}$ itself is compact, $\mathrm P$-a.s.\ compact, for every $n\in\mathbb N$ and~$k\in\{1,\ldots,2d-1\}$. Random set measurability of~$\Lambda_1^k(\textbf{u}^{(n)})$ follows from \citet[Example 1.7]{molchanov2018random}.

Consider, for $1\leq i\neq j\leq k$, the events
\begin{equation}\label{defE_n}
E^{(n)}_{i,j}({\bf u})\coloneqq \{\omega :\, ({\bf u}^{(n)}_i(\omega))^{\prime}{\bf u}^{(n)}_j(\omega) = {\bf u}_i^{\prime}{\bf u}_j\}\quad\text{and}\quad E^{(n)}({\bf u})\coloneqq \bigcap_{1\leq i\neq j\leq k}E^{(n)}_{i,j}({\bf u}).
\end{equation}
On $E^{(n)}({\bf u})$, antipodality and orthogonality among the $k$-tuples ${\bf u}^{(n)}$ and $\bf u$ coincide, i.e., for $\omega\in E^{(n)}({\bf u})$ and any $1\leq i\neq j\leq k$, $({\bf u}^{(n)}_i(\omega))^{\prime} {\bf u}^{(n)}_j(\omega) = 0$ iff ${\bf u}_i^{\prime}{\bf u}_j= 0$ and $({\bf u}^{(n)}_i(\omega))^{\prime}{\bf u}^{(n)}_j(\omega) = -1$ iff ${\bf u}_i^{\prime}{\bf u}_j= -1$.

Since $\|\textbf u_i^{(n)}-\textbf u_i\|$ is~$o_{\rm P}(1)$, also $(\textbf u_i^{(n)})'\textbf u_j^{(n)}=\textbf u_i'\textbf u_j+o_{\rm P}(1)$ for $1\leq i\neq j\leq k$ and, because both inner products only can take values $0$ and~$-1$, ${\rm P}(E_{i,j}^{(n)}(\textbf{u}))\to1$ for all pairs $(i,j)$. The number of such pairs being finite,~${\rm P}(E^{(n)}(\textbf{u}))\to 1$ as $n\to\infty$.

Let ${\cal J}\coloneqq\{j:\ \textbf{u}_i\neq-\textbf{u}_j\ \text{for all}\ i<j\}$, which keeps one index of each antipodal pair. Since all inner products ${\bf u}_i^\prime{\bf u}_j$ take values in $\{0,-1\}$, $\{\textbf{u}_j:j\in{\cal J}\}$ is a family of $|\cal J|$ orthonormal vectors and each ${\bf u}_i$, $i=1,\ldots,k$, equals $\pm\textbf{u}_j$ for some $j\in{\cal J}$, so that $\Lambda_1^k(\textbf{u})=\{\textbf{v}\in{\cal S}_{d-1}:\textbf{v}'\textbf{u}_j=0,\ j\in{\cal J}\}$. On $E^{(n)}(\textbf{u})$, since the inner products among the ${\bf u}^{(n)}$'s coincide with those among the $\bf u$'s, $\{\textbf{u}_j^{(n)}:j\in{\cal J}\}$ is orthonormal as well, and~$\Lambda_1^k(\textbf{u}^{(n)})=\{\textbf{v}\in{\cal S}_{d-1}:\textbf{v}'\textbf{u}_j^{(n)}=0,\ j\in{\cal J}\}$. Let further $\varepsilon_n\coloneqq\big(\sum_{j\in{\cal J}}\|\textbf{u}^{(n)}_j-\textbf{u}_j\|^2\big)^{1/2}=o_{\rm P}(1)$.

(i) Assume~$\Lambda_1^k(\textbf{u})\neq \varnothing$. On $E^{(n)}(\textbf{u})\cap\{\varepsilon_n<1\}$, let $\textbf{v}\in\Lambda_1^k(\textbf{u}^{(n)})\neq\varnothing$ and take $\textbf{w}\coloneqq\textbf{v}-\sum_{j\in{\cal J}}(\textbf{v}'\textbf{u}_j)\textbf{u}_j$. Then $\textbf{w}'\textbf{u}_j=0$ for $j\in{\cal J}$ and, since $\textbf{v}'\textbf{u}_j=\textbf{v}'(\textbf{u}_j-\textbf{u}^{(n)}_j)$,
$$
\|\textbf{v}-\textbf{w}\|=\Big(\sum_{j\in{\cal J}}(\textbf{v}'\textbf{u}_j)^2\Big)^{1/2}\leq\varepsilon_n,\quad \|\textbf{w}\|\geq \|\textbf{v}\| - \|\textbf{v}-\textbf{w}\|\geq 1-\varepsilon_n>0.
$$
Thus $\tilde{\textbf{v}}\coloneqq\textbf{w}/\|\textbf{w}\|\in\Lambda_1^k(\textbf{u})$, and
$\|\textbf{v}-\tilde{\textbf{v}}\|\leq\|\textbf{v}-\textbf{w}\|+\big|\|\textbf{v}\|-\|\textbf{w}\|\big|\leq2\|\textbf{v}-\textbf{w}\|\leq2\varepsilon_n$.
Exchanging the roles of $\textbf{u}$ and $\textbf{u}^{(n)}$ gives %
$d_H(\Lambda_1^k({\bf u}^{(n)}),\Lambda_1^k({\bf u}))\leq2\varepsilon_n$ on $E^{(n)}(\textbf{u})\cap\{\varepsilon_n<1\}$. As $\mathrm{P}(E^{(n)}(\textbf{u}))\to 1$ and $\varepsilon_n=o_{\rm P}(1)$ as~$n\to\infty$, \eqref{A8} follows.

(ii) If $\Lambda_1^k(\textbf{u})=\varnothing$, then $\{\textbf u_j:j\in{\cal J}\}$ is an orthonormal basis of $\mathbb R^d$. %
On $E^{(n)}(\textbf u)$, so is $\{\textbf u_j^{(n)}:j\in{\cal J}\}$, implying $\Lambda_1^k(\textbf{u}^{(n)})=\varnothing$. Therefore, $\mathrm{P}(\Lambda_1^k(\textbf{u}^{(n)})=\varnothing)\geq \mathrm{P}(E^{(n)}(\textbf{u}))\to 1$ as $n\to\infty$.
\end{proof}

\begin{Lemma}\label{lemma:convergence_of_antipodal_sets}
For any $k\leq 2d-1$ and $n\in{\mathbb N}$, let ${\bf u}^{(n)}$ %
 and  ${\bf u}$ %
satisfy the same conditions as in Lemma~\ref{lem:orthogonality_convergence}.
Then,
$\Lambda^k_2({\bf u}^{(n)})$
is a random compact set, and
\begin{itemize}
    \item[(i)] \unaAdded{if $\Lambda_2^{k}(\textbf{u})\neq \varnothing$, $d_H(\Lambda^k_2({\bf u}^{(n)}),\Lambda^k_2({\bf u}))=o_{\mathrm P}(1)$ as~$n\to\infty$;}
    \item[(ii)]  \unaAdded{if $\Lambda_2^{k}(\textbf{u})= \varnothing$, $\mathrm{P}(\Lambda_2^k(\textbf{u}^{(n)})=\varnothing)\to 1$ as $n\to\infty$.}
\end{itemize}
\end{Lemma}

\begin{proof}[\bf Proof of Lemma~\ref{lemma:convergence_of_antipodal_sets}]
For $n\in\mathbb{N}$, $\Lambda^k_2({\bf u}^{(n)})$ is, $\rm P$-a.s., a finite set of at most $k$ random points.  \unaAdded{Therefore, it is $\rm P$-a.s.\ compact}; its \unaAdded{random-set measurability} follows from Example 1.5 in {\cite{molchanov2018random}}.

Recall from the proof of Lemma~\ref{lem:orthogonality_convergence}, that, for $E^{(n)}({\bf u})$ defined in \eqref{defE_n}, \unaAdded{${\rm P}(E^{(n)}({\bf u})) \to 1$\linebreak  as $n\to~\!\infty$.}
On~$E^{(n)}({\bf u})$, %
 antipodality and orthogonality  among
$\{\textbf{u}_1^{(n)},\ldots,\textbf{u}_k^{(n)}\}$
coincide with  antipodality and ortho\-gonality   among~$\{\textbf{u}_1,\ldots,\textbf{u}_k\}$.  %
The index set
$
{\cal I}\coloneqq \{i:\,-\textbf{u}_i\notin\{\textbf{u}_1,\ldots,\textbf{u}_k\}\}\subseteq\{1,\dots,k\}
$
is such that %
${\Lambda^k_2({\bf u})}=\{-\textbf{u}_i:i\in {\cal I}\}$. On $E^{(n)}({\bf u})$, the same index set characterizes ${\Lambda^k_2({\bf u}^{(n)})}=\{-\textbf{u}_i^{(n)}:i\in\mathcal{I}\}$.

\unaAdded{(i) If $\Lambda_2^k(\textbf{u})\neq\varnothing$, then $\mathcal{I}\neq\varnothing$ and, on $E^{(n)}(\textbf{u}),$ $\Lambda^k_2({\bf u}^{(n)})\neq \varnothing$. } Therefore, on  $E^{(n)}({\bf u})$,
$$
d_H({\Lambda^k_2({\bf u}^{(n)})},{\Lambda^k_2({\bf u})})
\leq
\max_{i\in {\cal I}}
\|\textbf u_i^{(n)}-\textbf u_i\|.
$$
\unaAdded{Hence, for every $\delta>0$,
\begin{align*}
\mathrm{P}(d_H({\Lambda^k_2({\bf u}^{(n)})},{\Lambda^k_2({\bf u})})>\delta)&\leq \mathrm{P}(E^{(n)}(\textbf{u})^c)+\mathrm{P}(E^{(n)}(\textbf{u})\cap \{d_H({\Lambda^k_2({\bf u}^{(n)})},{\Lambda^k_2({\bf u})})>\delta\})\\
&\leq
\mathrm{P}(E^{(n)}(\textbf{u})^c)+\mathrm{P}(\max_{i\in {\cal I}}
\|\textbf u_i^{(n)}-\textbf u_i\|> \delta)\to 0\quad\text{as}\quad n\to\infty,
\end{align*}
since $\max_{i\in {\cal I}}
\|\textbf u_i^{(n)}-\textbf u_i\| =o_{\mathrm P}(1)$ and $\mathrm P(E^{(n)}({\bf u}))\to1$ as $n\to\infty$.}%

\unaAdded{(ii) If $\Lambda_2^k(\textbf{u})=\varnothing$, then $\mathcal{I}=\varnothing$. Hence, on $E^{(n)}(\textbf{u})$, $\Lambda_2^k(\textbf{u}^{(n)})=\varnothing$, which implies
$$\mathrm{P}(\Lambda_2^k(\textbf{u}^{(n)})=\varnothing)\geq \mathrm{P}(E^{(n)}(\textbf{u}))\to 1\quad \text{ as $n\to\infty$.}$$
\vspace{-21mm}}

\end{proof}

We now turn to the proof of Theorem~\ref{Thm:Directions_Consistency_i}. %

\begin{proof}[\bf Proof of Theorem~\ref{Thm:Directions_Consistency_i}]
We proceed by induction. The consistency of
$\textbf U_{\pm,1}^{(n)}$ is established in Theorem~\ref{Thm:Directions_Consistency}.
Suppose, for some $k\in\{1,\ldots,2d-1\}$, that  $\|\textbf U_{\pm,i}^{(n)}-\textbf U_{\pm,i}\|=o_{\rm P}(1)$,  $i=1,\ldots,k$. Then, using Theorem~\ref{thm:Argmax-compact-random}, we show that
$\|\textbf U_{\pm,k+1}^{(n)}-\textbf U_{\pm,k+1}\|=o_{\rm P}(1)$.

As in Theorem~\ref{Thm:Directions_Consistency}, let $M_n(\textbf u)\coloneqq     M_n^{\delta_1^{(n)},\delta_2^{(n)}}(\textbf u)$ and $M(\textbf u)\coloneqq M^{\varepsilon_0,\delta_0}(\textbf u),$ $\textbf u\in\mathcal S_{d-1}.$   Lemma~\unaAdded{\ref{lem:uniform_convergence_M_n}}%
implies that~$\|M_n-M\|_\infty=o_{\rm P}(1)$ on $\mathcal{S}_{d-1}$.

Consider, for $k=1,\ldots, 2d-1$, the population feasible sets ${\Lambda^k({\bf U}^k_\pm)=\Lambda^k_1({\bf U}^k_\pm)}\cup{\Lambda^k_2({\bf U}^k_\pm)}$ (as defined in~\eqref{defLambda}),
where ${\bf U}^k_\pm\coloneqq ({\bf U}_{\pm ,1},\ldots,{\bf U}_{\pm , k})$, and their sample counterparts ${\Lambda^{k}({\bf U}^{(n)k}_\pm)}={\Lambda^{k}_{1}({\bf U}^{(n)k}_{\pm}) }\cup{\Lambda^{k}_2({\bf U}^{(n)k}_{\pm}) }$ where~${\bf U}^{(n)k}_\pm\coloneqq ({\bf U}^{(n)}_{\pm ,1},\ldots,{\bf U}^{(n)}_{\pm ,k})$.
\unaAdded{Lemmas~\ref{lem:orthogonality_convergence} and \ref{lemma:convergence_of_antipodal_sets}, together with $$
d_H(\Lambda^{k}({\bf U}^{(n)k}_{\pm}) ,\Lambda^k({\bf U}^k_\pm))%
\leq \max\left\{d_H(\Lambda^k_1({\bf U}^k_\pm), {\Lambda^{k}_{1}({\bf U}^{(n)k}_{\pm}) }),d_H(\Lambda^k_2({\bf U}^k_\pm), {\Lambda^{k}_{2}({\bf U}^{(n)k}_{\pm}) }) \right\},
$$
yield $d_H(\Lambda^{k}({\bf U}^{(n)k}_{\pm}) ,\Lambda^k({\bf U}^k_\pm))=o_{\mathrm{P}}(1)$
as~$n\to~\!\infty$.}

By definition, $\textbf U_{\pm,k+1}$ is the unique maximizer of $M$
over $\Lambda^k({\bf U}^k_\pm)$, so that $M(\textbf U_{\pm,k+1}) = \sup_{\textbf u\in\Lambda^k({\bf U}^k_\pm)}M(\textbf u).$
It thus remains to verify that $\textbf U_{\pm,k+1}^{(n)}$ is an
approximate maximizer of $M_n$ over ${\Lambda^{k}({\bf U}^{(n)k}_\pm)}$. We proceed as in the proof of Theorem~\ref{Thm:Directions_Consistency}, now applying Theorem~\ref{thm:Argmax-compact-random} instead of Theorem~\ref{Argmax}.

Since $d_H(\Lambda^{k}({\bf U}^{(n)k}_{\pm}) ,\Lambda^k({\bf U}^k_\pm)) =o_{\rm P}(1)$ and
$\textbf U_{\pm,k+1}\in\Lambda^k({\bf U}^k_\pm)$, there exists
${\textbf v}^{(n)}\in\Lambda^{k}({\bf U}^{(n)k}_{\pm})$ such\linebreak  that~$
\|{\textbf v}^{(n)}-\textbf U_{\pm,k+1}\|
=o_{\rm P}(1).$ If ${\textbf v}^{(n)}\in\Lambda^{k}_2({\bf U}^{(n)k}_{\pm})$, set
${\textbf u}^{(n)}_*\coloneqq {\textbf v}^{(n)}$. Otherwise, by the weak convergence of the empirical distribution over the grid $ \mathfrak G_{m_{k+1}}^{\mathcal S}$ to a uniform distribution over
$\Lambda^{k}_1({\bf U}^{(n)k}_{\pm})$, there\linebreak exists~${\textbf u}^{(n)}_*\! \in\! \left(\Lambda^{k}_1({\bf U}^{(n)k}_{\pm})\cap \mathfrak G_{m_{k+1}}^{\mathcal S}
\right) \cup \Lambda^{k}_2({\bf U}^{(n)k}_{\pm})$
such that $\|{\textbf u}^{(n)}_*\!-{\textbf v}^{(n)}\|=o_{\rm P}(1)$, hence
$\|{\textbf u}^{(n)}_*\!-\textbf U_{\pm,k+1}\|=o_{\rm P}(1).$

Since $M$ is continuous on the compact set $\mathcal S_{d-1}$, it is uniformly
continuous, and therefore
\begin{equation}
M({\textbf u}^{(n)}_*)
-
M(\textbf U_{\pm,k+1})
=o_{\rm P}(1).
\label{eq:Thm32_grid}
\end{equation}
Furthermore, from Lemma~\ref{lem:sup-convergence-deterministic},
\begin{equation}
\Big|
\sup_{\textbf u\in{ \Lambda^{k}({\bf U}^{(n)k}_{\pm})}}\!\!\!\!M_n(\textbf u)
-\!\!\!\!
\sup_{\textbf u\in{\Lambda^k({\bf U}^k_\pm)}}\!\!\!\!M(\textbf u)
\Big|
=o_{\rm P}(1).
\label{eq:Thm32_sup}
\end{equation}
Since $\sup_{\textbf u\in{\Lambda^k({\bf U}^k_\pm)}}M(\textbf u)
=
M(\textbf U_{\pm,k+1})$   and  $\textbf U_{\pm,k+1}^{(n)}$ is obtained by maximizing $M_n$ over the discrete feasible set
$\displaystyle \left(
\Lambda^{k}_1({\bf U}^{(n)k}_{\pm})
\cap
\mathfrak G_{m_{k+1}}^{\mathcal S}
\right)
\cup
\Lambda^{k}_2({\bf U}^{(n)k}_{\pm})$, and since $\textbf{u}^*\in \displaystyle \left(
\Lambda^{k}_1({\bf U}^{(n)k}_{\pm})
\cap
\mathfrak G_{m_{k+1}}^{\mathcal S}
\right)
\cup
\Lambda^{k}_2({\bf U}^{(n)k}_{\pm})$, we obtain
\begin{align*}
0
&\leq
\Big|\!\!
\sup_{{\textbf u}\in
{
\Lambda^{k}
(
{\bf U}^{(n)k}_{\pm}
)
}}
\!\!\!\!M_n(\textbf u)
-
M_n(\textbf U_{\pm,k+1}^{(n)})
\Big|
=\!\!\!\!
\sup_{\textbf u\in{\Lambda^{k}({\bf U}^{(n)k}_{\pm})}}\!\!\!\!
M_n(\textbf u)
-
M_n(\textbf U_{\pm,k+1}^{(n)})\leq\!\!\!
\sup_{\textbf u\in{\Lambda^{k}({\bf U}^{(n)k}_{\pm})}}\!\!\!\!
M_n(\textbf u)
-
M_n({\textbf u}^{(n)}_*)
\\
&\leq
\Big|
\sup_{{\textbf u}\in
{
\Lambda^{k}
(
{\bf U}^{(n)k}_{\pm}
)
}}
\!\!\!\!M_n(\textbf u)
-
M(\textbf U_{\pm,k+1})
\Big|
+
\Big|
M(\textbf U_{\pm,k+1})-M({\textbf u}^{(n)}_*)
\Big|
+
\|M_n-M\|_\infty=o_{\rm P}(1),
\end{align*}
where the last equality follows from
\eqref{eq:Thm32_sup}, \eqref{eq:Thm32_grid}, and the fact that  \unaAdded{ $\|M_n-M\|_\infty=o_{\rm P}(1)$}. %
Hence,
$$
M_n(\textbf U_{\pm,k+1}^{(n)})
\geq
\sup_{{\textbf u}\in
{
\Lambda^{k}
(
{\bf U}^{(n)k}_{\pm}
)
}}
\!\!\!\!M_n(\textbf u)-o_{\rm P}(1)\quad\text{as $n\to\infty$}.
$$
With the uniform convergence
$\|M_n-M\|_\infty=o_{\rm P}(1)$, the uniqueness of
$\textbf U_{\pm,k+1}$ as the maximizer of $M$ over~$
{
\Lambda^{k}
(
{\bf U}^{k}_{\pm}
)
}$,  and the Hausdorff convergence
$d_H(
{
\Lambda^{k}
(
{\bf U}^{(n)k}_{\pm}
)
},
{
\Lambda^{k}
(
{\bf U}^{k}_{\pm}
)
})=o_{\rm P}(1)$, all
conditions of Theorem~\ref{thm:Argmax-compact-random}~are satisfied.
Therefore,
$\|\textbf U_{\pm,k+1}^{(n)}\!-\textbf U_{\pm,k+1}\|=o_{\rm P}(1).$
Theorem~\ref{Thm:Directions_Consistency_i} follows by induction\linebreak over~$k=~\!1,\ldots,2d-~\!1$.\end{proof}

\color{black}

\subsection{Proof of Theorem~\ref{Thm:principal_curves}}\label{SecA4}
Finally, we prove the consistency of the interpolated principal curves, uniformly over any compact subset of~$(0,1)$, as stated in Theorem \ref{Thm:principal_curves}.   We provide the proof for the first principal curve $\overline{\mathcal{L}}_{\pm, 1}^{(n)}$ only, as the results for the other ones follow along the same lines.   Fix $\delta\in(0,\unaAdded{1/2})$.

    By the triangle inequality,
\begin{align*}
    \sup_{r\in [\delta,1-\delta]}\|\mathcal{L}_{\pm, 1}(r)-\overline{\mathcal{L}}_{\pm, 1}^{(n)}(r)\|\leq \sup_{r\in [\delta,1-\delta]}\|\overline{\textbf{Q}}_\pm^{(n)}(r\textbf{U}_{\pm, 1}^{(n)})-{\textbf{Q}}_\pm(r\textbf{U}_{\pm, 1}^{(n)})\|+\sup_{r\in [\delta,1-\delta]}\|{\textbf{Q}}_\pm(r\textbf{U}_{\pm, 1}^{(n)})-{\textbf{Q}}_\pm(r\textbf{U}_{\pm, 1})\|,
\end{align*}
further yielding, for every $\varepsilon>0$,
\begin{align}\label{eq:Tm2_0}
    \mathrm{P}\big(\sup_{r\in [\delta,1-\delta]}\|\mathcal{L}_{\pm, 1}(r)-\overline{\mathcal{L}}_{\pm, 1}^{(n)}(r)\|>\varepsilon\big)&\leq \mathrm{P}\big(\sup_{r\in [\delta,1-\delta]}\|\overline{\textbf{Q}}_\pm^{(n)}(r\textbf{U}_{\pm, 1}^{(n)})-{\textbf{Q}}_\pm(r\textbf{U}_{\pm, 1}^{(n)})\|>\varepsilon/2\big)\nonumber\\
    &+ \mathrm{P}\big(\sup_{r\in [\delta,1-\delta]}\|{\textbf{Q}}_\pm(r\textbf{U}_{\pm, 1}^{(n)})-{\textbf{Q}}_\pm(r\textbf{U}_{\pm, 1})\|>\varepsilon/2\big).
\end{align}
For $\delta>0$, consider    the compact annulus ${K}_\delta=\{\textbf{x}\in\mathbb{R}^d:\delta\leq \|\textbf{x}\|\leq 1-\delta\}$. It follows from  {\cite{delBarrio2018}} that~$\displaystyle\sup_{\textbf{x}\in K_\delta}\|\overline{\textbf{Q}}_\pm^{(n)}(\textbf{x})-{\textbf{Q}}_\pm(\textbf{x})\|=o_{\rm P}(1)$,  further implying
\begin{equation}\label{eq:Tm2_1}
 \mathrm{P}(\sup_{r\in [\delta,1-\delta]}\|\overline{\textbf{Q}}_\pm^{(n)}(r\textbf{U}_{\pm, 1}^{(n)})-{\textbf{Q}}_\pm(r\textbf{U}_{\pm, 1}^{(n)})\|> \varepsilon/2)\leq \mathrm{P}(\sup_{\textbf{x}\in K_\delta}\|\overline{\textbf{Q}}_\pm^{(n)}(\textbf{x})-{\textbf{Q}}_\pm(\textbf{x})\|>\varepsilon/2)\to 0\  \ \text{as $n\to\infty$}.
\end{equation}

By Lemma~\ref{lem:bounds} (i),
$\textbf Q_\pm$ is $C^1(K_\delta)$,  hence %
continuous on the compact annulus $K_\delta$. Therefore,
$\textbf Q_\pm$ is uniformly continuous on $K_\delta$. Moreover, it follows from Theorem~\ref{Thm:Directions_Consistency} (for the first principal direction $\textbf U_{\pm,1}$; from Theorem~\ref{Thm:Directions_Consistency_i} for the subsequent ones $\textbf U_{\pm,i}$,  $i=2,\ldots, 2d$)
 that
$$
\sup_{r\in[\delta,1-\delta]}
\|r\textbf U_{\pm,1}^{(n)}-r\textbf U_{\pm,1}\|
\leq
\|\textbf U_{\pm,1}^{(n)}-\textbf U_{\pm,1}\|
=o_{\rm P}(1)\quad\text{as $n\to\infty$}
$$
 and, in view of  uniform continuity,
\begin{equation}\label{eq:Tm2_2}
\sup_{r\in[\delta,1-\delta]}
\|
\textbf Q_\pm(r\textbf U_{\pm,1}^{(n)})
-
\textbf Q_\pm(r\textbf U_{\pm,1})
\|
=o_{\rm P}(1)\quad\text{as $n\to\infty$}.
\end{equation}

Plugging \eqref{eq:Tm2_1} and \eqref{eq:Tm2_2} into \eqref{eq:Tm2_0} gives the desired statement \eqref{first statement}.

The second statement \eqref{second statement} is a direct consequence of the fact that  $\overline{\textbf Q}_\pm^{(n)}$ interpolates $\textbf{Q}_\pm^{(n)}$ on a compact subset of the domain.
\hfill$\square$

\section{Regularization of discrete principal curves}\label{sec:smooth}
The selection criterion, in \eqref{defU1n'} and \eqref{defUin'}, is based on a polygonal evaluation of empirical principal curve %
 lengths, that is, involves a linear interpolation of  $m_R$ values of  $\overline{\textbf{Q}}_\pm^{(n)}(r_j\textbf{u})$ for $m_S$ values of $\textbf u$. Depending on the sample size $n$, %
 {the approximation \ref{defLn} of \ref{defL} can be poor and biased. This difficulty is connected to the fact that the arc length is   only lower semicontinuous with respect to uniform convergence. Thus, if, e.g., a sequence of curves used in estimation converges uniformly to the limiting curve but has infinitely small, but rapid oscillations around it, the accumulated length induced by those oscillations can be quite large. Therefore, the radial mesh size $(m_R+1)^{-1}$ needs to be controlled relative to the estimation error $s_n$ defined in \eqref{GCrate}. In the worst case, the discrepancy between the polygonal lengths based on $\overline{\textbf{Q}}_\pm^{(n)}$ and $\textbf{Q}_\pm$ is of order~$m_Rs_n$:  see Theorem \ref{Thm:Directions_Consistency} and the discussion on its assumptions for details. Therefore, for a fixed sample size $n$, refining the radial grid doesn't necessarily improve the quality of approximation.
}

In practice, smoothing, for   each $i = 1. . . ,m_S$,  the $m_R$ interpolated values $\overline{\textbf{Q}}_\pm^{(n)}(r_j\textbf{u}_i)$, $j=1,\dots,m_R$ by penalizing the magnitude of the  squared sequential differences $\|\overline{\textbf{Q}}_\pm^{(n)}(r_{j+1}\textbf{u}_i)-\overline{\textbf{Q}}_\pm^{(n)}(r_{j}\textbf{u}_i)\|^2$, $j=1,\dots,m_R-1$ is advisable to overcome this difficulty. More precisely, writing
${\boldsymbol{\mathcal{L}}_{\pm}^{(n)}(\textbf{u}_{i})}\coloneqq (\overline{\textbf{Q}}_\pm^{(n)}(r_1{\textbf u}_{i}),\ldots,\overline{\textbf{Q}}_\pm^{(n)}(r_{m_R}{\textbf u}_{i}))$
 (a~${d\times m_R}$ matrix), and denoting by
${[\overline{\boldsymbol{\mathcal{L}}}_{\pm}^{(n)}(\textbf{u}_{i})]_l}$
 its ($(m_R)$-dimensional) $l$th row, $l=1,\dots,d$, we consider the   coordinatewise   {\it Tikhonov regularization}   \citep{tikhonov1963} of
 {\begin{align}\label{eq:Tikhono}
[\boldsymbol{\mathcal{L}}_{\pm;\varepsilon_n}^{(n)}(\textbf u_i)]_l
&:=
\argmin_{\textbf{v}\in\mathbb R^{m_R}}
\left\{
\left\|
[\boldsymbol{\mathcal{L}}_{\pm}^{(n)}(\textbf u_i)]_l-\textbf{v}
\right\|_{m_R}^2
+
\varepsilon_n
\sum_{j=1}^{m_R-1}
(v_{j+1}-v_j)^2
\right\}
\\
&=
\argmin_{\textbf{v}\in\mathbb R^{m_R}}
\left\{
\left\|
[\boldsymbol{\mathcal{L}}_{\pm}^{(n)}(\textbf u_i)]_l-\textbf{v}
\right\|_{m_R}^2
+
\varepsilon_n
\|\textbf L\textbf{v}\|_{m_R-1}^2
\right\}.
\end{align}
of~$\mathcal{L}_{\pm}^{(n)}(\textbf{u}_{i})$, where%
$$\textbf{L}\coloneqq \begin{pmatrix}
   -1 & 1 & 0 &\hdots &0\\
   0 &- 1 & 1 &\hdots &0\\
   \vdots&&\ddots&\ddots&\vdots \\
   0&&& -1 & 1
\end{pmatrix}%
$$
is an  $(m_R-1)\times m_R$ matrix and $\Vert\, .\, \Vert_{m_R}$ stands for the Euclidean norm in ${\mathbb R}^{m_{\!R}}$ (see, e.g.,  {\cite{Reichel2009}} and the references therein).  This  Tikhonov regularization  admits the closed form
\begin{equation}\label{eq:Tikhono_solution}
\boldsymbol{\mathcal{L}}_{\pm;\,\varepsilon_n}^{(n)}(\textbf{u}_{i})={\boldsymbol{\mathcal{L}}_{\pm}^{(n)}(\textbf{u}_\unaAdded{i})}(\textbf{I}_{m_R}+\varepsilon_n\textbf{L}'\textbf{L})^{-1}
\end{equation}
where the regularization parameter $\varepsilon_n$ determines the amount of smoothing imposed on $\mathcal{L}_{\pm}^{(n)}$ and should be chosen to decrease with the sample size, not to interfere with the consistency results.

Smoothing the curves prior to differentiation is  common practice in functional data analysis since taking differences between closely spaced noisy observations magnifies the influence of  measurement errors. Thus, derivatives are usually estimated from smooth versions of the raw data: see, e.g., \citet[Section 3.2.6]{ramsayFDA}. Note that \eqref{defLn} does not  explicitly involve derivatives; however, the incre\-ments~$\overline{\textbf{Q}}_\pm^{(n)}(r_{j+1}\textbf{u}_i)-\overline{\textbf{Q}}_\pm^{(n)}(r_{j}\textbf{u}_i)$, $j=1,\dots,m_R-1$  play the role of discretized derivatives scaled by the grid resolution $(m_R+1)^{-1}$.

Finally, based on \eqref{eq:Tikhono_solution}, substitute %
$$
\overline{\textbf{Q}}_{\pm;\,\varepsilon_n}^{(n)}(r_j\textbf{u}_\unaAdded{i})\coloneqq ({\boldsymbol{\mathcal{L}}}_{\pm;\,\varepsilon_n}^{(n)}(r_j\textbf{u}_{i})) ,\quad i=1,\dots,m_S,\ \   j=1,\dots,m_R$$
for $\overline{\textbf{Q}}_{\pm}^{(n)}(r_j\textbf{u}_{i})$ in \eqref{defU1n'}-\eqref{eq:PrincipalCurves_sample} to obtain the \textit{regularized empirical principal directions} $\textbf{U}_{\pm,{k};\,\varepsilon_n}^{(n)}$, ${k}=1,\ldots,2d$ and,   as in Section \ref{sec31}, the associated empirical quantities. This   regularization penalizes the excessive local variation and, therefore, reduces the contribution of the high-frequency fluctuations of  polygonal length, providing more stable length estimates. %


\end{document}